\documentclass[twocolumn]{svjour3}          % twocolumn
\smartqed  % flush right qed marks, e.g. at end of proof

\usepackage{url}
\usepackage{amsmath,graphicx}
\usepackage{subfigure}
\usepackage{tikz}
\usetikzlibrary{arrows}
\usepackage{slashbox}
\usepackage[english]{babel} % English
\usepackage{graphicx,amssymb,amstext,amsmath}
\usepackage{float}
\usepackage{amsfonts}
\usepackage{bbm}
\usepackage{bm}
\usepackage{verbatim}
\usepackage{color}
\usepackage{flushend}
\usepackage{multirow}
\usepackage{makecell}
\usepackage{tabu} 

\usepackage[numbers]{natbib}

\usepackage{colortbl}
\def\x{{\mathbf x}}
\newcommand{\bI}{{\bf I}}

\usepackage[figuresright]{rotating}

\begin{document}

\title{Layered Adaptive Importance Sampling%\thanks{Grants or other notes
}


\author{
L. Martino$^\star$ \and V. Elvira$^\dagger$ \and  D. Luengo$^\ddagger$ \and  J.  Corander$^\star$}
%\address{$^{\star}$ Dep. of Mathematics and Statistics, University of Helsinki,  Helsinki (Finland).\\
%$^{\dagger}$ Dep. of Signal Theory and Communic., Universidad Carlos III de Madrid, Legan\'es (Spain).\\
%$^\ddagger$ Dep. of Circuits and Systems Engineering, Universidad Polit\'ecnica de Madrid, Madrid (Spain). 
%}


\institute{$^\star$  Dep. of Mathematics and Statistics, University of Helsinki,  Helsinki (Finland). \\
           $^\dagger$ Dep. of Signal Theory and Communic., Universidad Carlos III de Madrid, Legan\'es (Spain). \\
            $^\ddagger$ Dep. of Circuits and Systems Engineering, Universidad Polit\'ecnica de Madrid, Madrid (Spain). 
            \vspace{-0.3cm}
}

\date{Received: date / Accepted: date \vspace{-0.1cm}}

\maketitle

\begin{abstract}
Monte Carlo methods represent the \textit{de facto} standard for approximating complicated  integrals involving multidimensional target distributions. In order to generate random realizations from the target distribution, Monte Carlo techniques use simpler proposal probability densities to draw candidate samples. The performance of any such method is strictly related to the specification of the proposal distribution, such that unfortunate choices easily wreak havoc on the resulting estimators. In this work, we introduce a \textit{layered} (i.e., hierarchical) procedure to generate samples employed within a Monte Carlo scheme. This approach ensures that an appropriate equivalent proposal density is always obtained automatically (thus eliminating the risk of a catastrophic performance), although at the expense of a moderate increase in the complexity. Furthermore, we provide a general unified importance sampling (IS) framework, where multiple proposal densities are employed and several IS schemes are introduced by applying the so-called deterministic mixture approach. Finally, given these schemes, we also propose a novel class of adaptive importance samplers using a population of proposals, where the adaptation is driven by independent parallel or interacting Markov Chain Monte Carlo (MCMC) chains. The resulting algorithms efficiently combine the benefits of both IS and MCMC methods.
\end{abstract}

%%%%%%%%%%%%%%%%%%
\section{Introduction}
\label{sec:1}
%%%%%%%%%%%%%%%%%%

%{\color{magenta} Evitar de entrar demasiado en detalles con formulas en la intro.}

Monte Carlo methods currently represent a maturing toolkit widely used throughout science and technology \citep{Doucet:MCSignalProcessing2005,Robert04,Wang:MCWirelessComm2002}.
%
%\citep{Fitzgerald01,Kotecha99,Luengo13}. 
Importance sampling (IS) and Markov Chain Monte Carlo (MCMC) methods are  well-known Monte Carlo (MC) techniques applied to compute integrals involving a high-dimensional target probability density function (pdf) $\bar{\pi}({\bf x})$.
In both cases, the choice of a suitable proposal density $q({\bf x})$ is crucial for the success of the Monte Carlo based approximation.
For this reason, the design of adaptive IS or MCMC schemes represents one of the most active research topics in this area, and several methods have been proposed in the literature \citep{Cappe04,CORNUET12,Craiu09,Haario2001,Luengo13}.

Since both IS and MCMC have certain intrinsic advantages and weaknesses, several attempts have been made to successfully marry the two approaches, producing hybrid techniques: IS-within-MCMC \citep{Andrieu10,Brockwell10,Liesenfeld08,Liu00,Neal11} or  MCMC-within-IS \citep{Beaujean13,Botev13,Chopin02,EBEB14,Mendes15,Neal01,Yuan13}.
To set the scene for such developments it is useful to recall briefly some of the main strengths of IS and MCMC, respectively.
For instance, one benefit of IS  is that it delivers a straightforward estimate of the normalizing constant of $\bar{\pi}({\bf x})$ \citep{Liang10,Robert04} (a.k.a. evidence or marginal likelihood), which is essential for several applications \citep{FRIEL11,Skilling06}.
In contrast, the estimation of the normalizing constant is highly challenging using MCMC methods, and several authors have investigated different approaches to overcome the obstacles related to the instability of the resulting estimators \citep{Botev08,CALDWELL14,Chib01,FRIEL11,Weinberg10}.
Furthermore, the application and the theoretical analysis of an IS scheme using an adaptive proposal pdf is easier than the theoretical analysis of the corresponding adaptive MCMC scheme, which is much more delicate \citep{Andrieu08}.

On the other hand, an appealing feature of MCMC algorithms is their explorative behavior.
For instance, the proposal function $q({\bf x}|{\bf x}_{t-1})$ can depend on the previous state of the chain ${\bf x}_{t-1}$ and foster movements between different regions of the target density.
For this reason, MCMC methods are usually preferred when no detailed information about the target $\bar{\pi}({\bf x})$ is available, especially in large dimensional spaces \citep{Andrieu:MCMachineLearning2003,Fitzgerald01}.
%
%{\color{black}Indeed, several authors have noted that this intrinsic explorative nature seems to safeguard the resulting MC estimators w.r.t. a rough tuning of the proposal $q({\bf x})$, explaining the wide success of MCMC methods \cite{Robert04,}.}
%A common feeling is that this intrinsic explorative nature seems to safeguard in some way the resulting Monte Carlo estimators with respect to a rough tuning of the proposal $q({\bf x})$, explaining the evident wide success of the MCMC methods.
%
%In this work, we provide a framework for explaining this common feeling about the MCMC based on {\it random walk} proposal densities.
%
Moreover, in order to amplify their explorative behavior several parallel MCMC chains can be run simultaneously \citep{Robert04,Liang10}.
This strategy facilitates the exploration of the state space, although at the expense of an increase in the computational cost.
Several schemes have been introduced to share information among the different chains \citep{Craiu09,O-MCMC,Smelly}, which further improves the overall convergence.

The main contribution of this work is the description and analysis of a hierarchical proposal procedure for generating samples, which can then be employed within any Monte Carlo algorithm.
In this hierarchical scheme, we consider two conditionally independent levels: the upper level is used to generate mean vectors for the proposal pdfs, which are then used in the lower level to draw candidate samples according to some MC scheme.
%
%Such an approach can be illustrated with an analogy of having a bag of potato chips/crisps layered on top of each other, such that their shapes and orientations may vary, mimicking a set of overlaid densities.
%
We show that the standard {\it Population Monte Carlo} (PMC) method \citep{Cappe04} can be interpreted as applying implicitly this hierarchical procedure.

The second major contribution of this work is providing a general framework for multiple importance sampling (MIS) schemes and their iterative adaptive versions.
We discuss several alternative applications of the so-called deterministic approach \citep{ElviraMIS15,Owen00,Veach95} for sampling a mixture of pdfs.
This general framework includes different MIS schemes used within adaptive importance sampling (AIS) techniques already proposed in literature, such as the standard PMC \citep{Cappe04}, the adaptive multiple importance sampling (AMIS) \citep{CORNUET12,Marin12}, and the adaptive population importance sampling (APIS) \citep{APIS15}.

Finally, we combine the general MIS framework with the hierarchical procedure for generating samples, introducing a new class of AIS algorithms.
More specifically, one or several MCMC chains are used for driving an underlying MIS scheme.
Each algorithm differs from the others in the specific Markov adaptation employed and the particular MIS technique applied for yielding the final Monte Carlo estimators.
This novel class of algorithms efficiently combines the main strengths of the IS and the MCMC methods, since it maintains an explorative behavior (as in MCMC) and can still easily estimate the normalizing constant (as in IS). 

We describe in detail the simplest possible algorithm of this class, called {\it random walk importance sampling}.
Moreover, we introduce two additional population-based variants that provide a good trade-off between performance and computational cost.
In the first variant, the mean vectors are updated according to independent parallel MCMC chains.
In the other one, an interacting adaptive strategy is applied. In both cases, all the adapted proposal pdfs collaborate to yield a single global IS estimator.
One of the proposed algorithms, called  {\it parallel interacting Markov adaptive importance sampling} (PI-MAIS), can be interpreted as parallel MCMC chains cooperating to produce a single global estimator, since the chains exchange statistical information to achieve a common purpose.

The rest of the paper is organized as follows.
Section \ref{PSsect} is devoted to the problem statement.
The hierarchical proposal procedure is then introduced in Section \ref{HMCsect}.
In Section \ref{LOCURASect2}, we describe a general framework for importance sampling schemes using a population of proposal pdfs, whereas Section \ref{SuperSect} introduces the adaptation procedure for the mean vectors of these proposal pdfs.
Numerical examples are provided in Section \ref{SIMUsect}, including comparisons with several benchmark techniques.
Different scenarios have been considered: a multimodal distribution, a nonlinear banana-shaped target, a high-dimensional example, and a localization problem in a wireless sensor network. 
Finally, Section \ref{sec:conclusions} contains some brief final remarks.

%%%%%%%%%%%%%%%%%%%%%%%%%%%
\section{Target distribution and related integrals}
\label{PSsect}
%%%%%%%%%%%%%%%%%%%%%%%%%%%

In this work, we focus on the Bayesian applications of IS and MCMC.
However, the algorithms described may also be used for approximating any target distribution that needs to be handled by simulation methods.
Let us denote the variable of interest as ${\bf x} \in \mathcal{X} \subseteq \mathbb{R}^{D_x}$, and let ${\bf y} \in \mathbb{R}^{D_y}$ be the observed data.
The posterior pdf is then given by
\begin{equation}
	\bar{\pi}({\bf x})=p({\bf x}| {\bf y})= \frac{\ell({\bf y}|{\bf x}) g({\bf x})}{Z({\bf y})},
\label{eq:posterior}
\end{equation}
where $\ell({\bf y}|{\bf x})$ is the likelihood function, $g({\bf x})$ is the prior pdf, and $Z({\bf y})$ is the model evidence or partition function.
In general, $Z({\bf y})$ is unknown, so we consider the corresponding unnormalized target,
\begin{equation}
\pi({\bf x})=\ell({\bf y}|{\bf x}) g({\bf x}).
\label{eq:target}
\end{equation}
Our goal is computing efficiently some integral measure w.r.t. the target pdf,
\begin{equation}
	I = \frac{1}{Z} \int_{\mathcal{X}} f({\bf x}) \pi({\bf x}) d{\bf x},
\label{eq:integral}
\end{equation}
where 
\begin{equation}
Z = \int_{\mathcal{X}} \pi({\bf x}) d{\bf x},
\label{eq:normConst}
\end{equation}
and $f$ is any square-integrable function (w.r.t. $\bar{\pi}({\bf x})$) of ${\bf x}$.\footnote{Note that, as both $\bar{\pi}({\bf x})$ and $Z$ depend on the observations ${\bf y}$, the use of $\bar{\pi}({\bf x}|{\bf y})$ and $Z({\bf y})$ would be more precise. However, since the observations are fixed, in the sequel we remove the dependence on ${\bf y}$ to simplify the notation.}
In this work, we address the problem of approximating  $I$ and $Z$ via Monte Carlo methods. 
Since drawing directly from $\bar{\pi}({\bf x}) \propto \pi({\bf x})$ is impossible in many applications, Monte Carlo techniques use a simpler proposal density $q({\bf x})$ to generate random candidates, testing or weighting them according to some suitable rule. 
Indeed, throughout the paper we focus on the combined use of several proposal pdfs, denoted as $q_1,\ldots, q_J$.

%%%%%%%%%%%%%%%%%%%%%%%%%%%%%%%%%%%%%%%%%%%%
\section{Hierarchical procedure for proposal generation} 
%%%%%%%%%%%%%%%%%%%%%%%%%%%%%%%%%%%%%%%%%%%%
\label{HMCsect}

The performance of MC methods depends on the discrepancy between the target, $\bar{\pi}({\bf x})\propto \pi({\bf x})$, and the proposal $q({\bf x})$.
Namely, the performance improves if $q({\bf x})$ is more similar (i.e., closer) to $\bar{\pi}({\bf x})$.
In general, tuning the parameters of the proposal is a difficult task that requires statistical information of the target distribution.
In this section, we deal with this important issue, focusing on the mean vector of the proposal pdf.
%, i.e., a parameter which determines the location or shift of the proposal density.
More specifically, we consider a proposal pdf defined by a mean vector ${\bm \mu}$ and covariance matrix ${\bf C}$, denoted as $q({\bf x}|{\bm \mu},{\bf C})=q({\bf x}-{\bm \mu}|{\bf C})$.
We propose the following hierarchical procedure for generating a set of samples that will be employed afterwards within some Monte Carlo technique:
\begin{enumerate}
\item For $j=1,\ldots,J:$
\begin{enumerate}
\item Draw a mean vector ${\bm \mu}_j \sim h({\bm \mu})$.
\item Draw ${\bf x}_j^{(m)}\sim q({\bf x}|{\bm \mu}_j,{\bf C})$ for $m=1,\ldots,M$.
\end{enumerate}
\item Use all the generated samples, ${\bf x}_j^{(m)}$ for $j=1,\ldots,J$ and $m=1,\ldots,M$, as candidates within some Monte Carlo method.
\end{enumerate}
%All the samples ${\bf x}_j^{(m)}$ are then used as candidates within a Monte Carlo technique.
%
Note that $h({\bm \mu})$ plays the role of a prior pdf over the mean vector of $q$ in this approach.
Hence, the pdf of each sample ${\bf x}_j^{(m)}$ can be expressed as
\begin{equation}
\label{EqPropIni}
\widetilde{q}({\bf x}|{\bf C})=\int_{\mathcal{X}}q({\bf x}|{\bm \mu},{\bf C})  h({\bm \mu}) d{\bm \mu},
\end{equation} 
i.e., the hierarchical procedure is equivalent to drawing directly ${\bf x}_j^{(m)} \sim \widetilde{q}({\bf x}|{\bf C})$ for all $j=1,\ldots,J$ and $m=1,\ldots,M$.
The density $\widetilde{q}$ is thus the {\it equivalent} proposal density of the whole hierarchical generating procedure.
Note also that the samples ${\bm \mu}_1,\ldots,{\bm \mu}_J$ are not directly used by the Monte Carlo estimator, since only the samples ${\bf x}_j^{(m)}$, for $j=1,\ldots,J$, $m=1,\ldots,M$, enter the actual estimator.
Hence, the computational cost per iteration of this hierarchical procedure is higher than the cost of a standard approach, However, it leads to substantial computational savings in terms of improved convergence towards the target, and thus a reduced number of iterations required, as shown later in the simulations.
Furthermore, note that the generation of the ${\bm\mu}_j$'s in the upper level is independent of the samples ${\bf x}_j^{(m)}$ drawn in the lower level, thus facilitating the theoretical analysis of the resulting algorithms, as discussed in Section \ref{TeoOtra}.\footnote{Note that, in the ideal case described here, each ${\bm \mu}_j$ is also independent of the other ${\bm \mu}$'s. However, in the rest of this work, we also consider cases where correlation among the mean vectors (${\bm \mu}_1,\ldots,{\bm \mu}_J$) is introduced.} %Below, we specifically clarify the difference with a standard adaptive Monte Carlo approach.

%%%%%%%%%%%%%%%%%%%%%%%%%%%%%%%%%%%%%%
%\subsection{{\color{black} Choice} of the prior pdf for the location parameters } 
%%%%%%%%%%%%%%%%%%%%%%%%%%%%%%%%%%%%%%
%\label{OptimalPrior}

%%%%%%%%%%%%%%%%%%%%%%%%%%%%%%%%%%%
\subsection{Optimal prior $h^*({\bm \mu})$} 
%%%%%%%%%%%%%%%%%%%%%%%%%%%%%%%%%%%%
\label{OptimalPrior}

Assuming that the parametric form of $q({\bf x}|{\bm \mu},{\bf C})$ and its covariance matrix ${\bf C}$ are fixed, we consider the problem of finding the optimal prior $h^{*}({\bm \mu}|{\bf C})$ over the mean vector ${\bm \mu}$.
%
%The optimal prior depends on the chosen covariance matrix ${\bf C}$ and,
Note that, since $q({\bf x}|{\bm \mu},{\bf C})=q({\bf x}-{\bm \mu}|{\bf C})$, we can write  
 \begin{equation}
 \label{EqQ2}
\widetilde{q}({\bf x}|{\bf C})=\int_{\mathcal{X}}q({\bf x}-{\bm \mu}|{\bf C}) h({\bm \mu}|{\bf C}) d{\bm \mu}.
\end{equation} 
regardless of the choice of the prior over the mean vectors in the upper level.
The desirable scenario is to have the equivalent proposal $\widetilde{q}({\bf x}|{\bf C})$ coinciding exactly with the target $\bar{\pi}({\bf x})$,\footnote{Given a function $f({\bf x})$,  the optimal proposal $q$ minimizing the variance of  the IS estimator is $\widetilde{q}({\bf x}|{\bf C}) \propto |f({\bf x})| \bar{\pi}({\bf x})$. However, in practical applications, we are often interested in computing expectations w.r.t. several $f$'s. In this context, a more appropriate strategy is to minimize the variance of the importance weights. In this case, the minimum variance is attained when $\widetilde{q}({\bf x}|{\bf C})= \bar{\pi}({\bf x})$ \cite{Doucet08tut}.} i.e., 
 {\color{black} 
\begin{equation}
\label{Condoptimal1}
\widetilde{q}({\bf x}|{\bf C})=\int_{\mathcal{X}}q({\bf x}-{\bm \mu}|{\bf C}) h^*({\bm \mu}|{\bf C}) d{\bm \mu} =\bar{\pi}({\bf x}),
\end{equation}
where $h^*({\bm \mu}|{\bf C})$ represents the optimal prior.} %This is the best case for the classical Monte Carlo approach, generating i.i.d. samples directly from ${\bar \pi}$.

\vspace*{12pt}

{\color{black}
%%%%%%%%%%%%%%%%%%%%%%%%%%%%%%%%%%%
\subsection{Asymptotically optimal choice of the prior $h({\bm \mu})$} 
%%%%%%%%%%%%%%%%%%%%%%%%%%%%%%%%%%%
\label{Prior2}

Since Eq. \eqref{Condoptimal1} cannot be solved analytically in general, in this section we relax that condition and look for an equivalent proposal $\widetilde{q}$ which fulfills \eqref{Condoptimal1} asymptotically as $J \to \infty$. }
%In order to obtain the optimal prior $h^*({\bm \mu})$ we have enforced Eq. \eqref{Condoptimal1}, i.e., $\widetilde{q}({\bf x}|{\bf C})=\bar{\pi}({\bf x})$. 
%
%In this section, we relax that condition and look for an equivalent proposal $\widetilde{q}$ which approximates the target $\bar{\pi}$, i.e.,
%\begin{equation}
%\label{Condoptimal2}
%\widetilde{q}({\bf x}|{\bf C})\approx\bar{\pi}({\bf x}).
%\end{equation}
%Throughout this section we first explain why the choice $h({\bm \mu}) = \bar{\pi}({\bm \mu})$ (i.e., using the target as the prior for the mean vectors in the upper level) allows us to fulfill Eq. \eqref{Condoptimal2}, and then discuss its practical implementation.
%%%%%%%%%%%%%%%%%%%%%%%%%%%%%%%%%%%%%%%%%%%
%\subsubsection{Motivation of the choice  $h({\bm \mu}) = \bar{\pi}({\bm \mu})$} 
%\label{Prior2}
%%%%%%%%%%%%%%%%%%%%%%%%%%%%%%%%%%%%%%%%%%%
For the sake of simplicity, let us set $M=1$.
Thus, we consider the generation of $J$ samples $\{{\bf x}_1,\ldots, {\bf x}_J\}$, drawn using the following hierarchical procedure:
\begin{enumerate}
	\item[(a)] Draw a mean vector ${\bm \mu}_j \sim h({\bm \mu})$.
	\item[(b)] Draw ${\bf x}_j\sim q({\bf x}|{\bm \mu}_j,{\bf C})$.
\end{enumerate}
Note that we are using $J$ different proposal pdfs,
$$
 q({\bf x}|{\bm \mu}_1,{\bf C}),\ldots,  q({\bf x}|{\bm \mu}_J,{\bf C}),
$$
to draw $\{{\bf x}_1,\ldots, {\bf x}_J\}$, with each ${\bf x}_j$ being drawn from the $j$-th proposal ${\bf x}_j\sim q({\bf x}|{\bm \mu}_j,{\bf C})$.
However, if the samples ${\bf x}_1,\ldots, {\bf x}_J$ are used altogether regardless of their order, then it can interpreted that they have been drawn from {\color{black} the following mixture using the deterministic mixture sampling scheme (see \citep[Chapter 9]{mcbookOwen}, \cite{ElviraMIS15}):}
\begin{equation}
\label{PHI_ForKernelEst}
\psi({\bf x}) = \frac{1}{J} \sum_{j=1}^J q({\bf x}|{\bm \mu}_j,{\bf C}).
\end{equation}
{\color{black} Note that, since ${\bm \mu}_j\sim h({\bm \mu})$, then $\psi({\bf x})$ is a Monte Carlo approximation of the integral in Eq. \eqref{Condoptimal1}, i.e., 
\begin{equation}
\label{Esto_CazzVic1}
\psi({\bf x}) \xrightarrow[J\rightarrow \infty]{a.s.} \widetilde{q}({\bf x}|{\bf C})=\int_{\mathcal{X}}q({\bf x}-{\bm \mu}|{\bf C}) h({\bm \mu}|{\bf C}) d{\bm \mu}.
\end{equation}
 Furthermore, if we choose $h({\bm \mu}) = \bar{\pi}({\bm \mu})$, i.e., ${\bm \mu}_j\sim  \bar{\pi}({\bm \mu})$, then $\psi({\bf x})$ is also a {\it kernel density estimator} of $ \bar{\pi}({\bf x})$, where the $q({\bf x}|{\bm \mu}_j,{\bf C})$ play the role of the kernel functions \cite{Wand94}. In general, this estimator has non-zero bias and variance, depending on the choice of $q$, ${\bf C}$ and the number of samples $J$. However, for a given value of $J$, there exists an optimal choice of ${\bf C}^*$ which provides the minimum Mean Integrated Square Error (MISE) estimator \cite{Wand94}. Using the optimal covariance matrix ${\bf C}^*$, it can be proved 
 \begin{equation}
 \label{Esto_CazzVic2}
\psi({\bf x})= \frac{1}{J} \sum_{j=1}^J q({\bf x}|{\bm \mu}_j,{\bf C}^*)\rightarrow \bar{\pi}({\bf x}),
\end{equation}
 pointwise as $J\rightarrow \infty$ \cite{Wand94}. Hence, the equivalent proposal density of the hierarchical approach converges to the target when $J\rightarrow \infty$. It is possible to show $||C^*||\rightarrow 0$ as $J\rightarrow \infty$, so that there is no contradiction between \eqref{Esto_CazzVic1} and \eqref{Esto_CazzVic2} since $q({\bf x}-{\bm \mu}|{\bf C}^*)$ becomes increasingly similar to $\delta({\bf x}-{\bm \mu})$, and thus $ \widetilde{q}({\bf x}|{\bf C}^*)\rightarrow  \bar{\pi}({\bf x})$ as $J\rightarrow \infty$.

% However, for any choice of ${\bf C}$, $\psi({\bf x})$ always shares several statistical information with the target $\bar{\pi}({\bf x})$. For instance, it is straightforward to show that the expected value of $\psi({\bf x})$ is the empirical mean,
%\begin{equation}
%\int_{\mathcal{X}} {\bf x} \psi({\bf x}) d{\bf x}=\frac{1}{J}\sum_{i=1}^J {\bm \mu}_i \xrightarrow[J\rightarrow \infty]{} \int_{\mathcal{X}} {\bf x} \bar{\pi}({\bf x}) d{\bf x},
%\end{equation}
%i.e.,  the unbiased estimator of the expected value of $\bar{\pi}$, when ${\bm \mu}_i \sim \bar{\pi}({\bf x})$ for all $i=1,\ldots,J$.
  }

%The performance of the corresponding Monte Carlo method, where such a hierarchical procedure is applied, depends on how closely $\Phi({\bf x})$ resembles $\bar{\pi}({\bf x})$.
 %This means that a good configuration of location parameters ${\bm \mu}_{n}$, $n=1,\ldots,N$, is around the modes of the target. 
%

%%%%%%%%%%%%%%%%%%%%%%%%%%%%%%%%%%%%%%%%%%%
\subsection{Practical implementation} 
%%%%%%%%%%%%%%%%%%%%%%%%%%%%%%%%%%%%%%%%%%%
\label{Prior4}

As explained in Section \ref{Prior2}, $h({\bm \mu}) = \bar{\pi}({\bm \mu})$ is a suitable choice from a kernel density estimation point of view.
However, sampling directly from $\bar{\pi}({\bm \mu})$ is unfeasible from a practical point of view (otherwise, we would not require any MC algorithm).
Therefore, we propose applying another sampling method, such as an MCMC algorithm, to obtain the samples $\{{\bm \mu}_{1},\ldots, {\bm \mu}_{J}\}\sim \bar{\pi}({\bm \mu})$.
More specifically, starting from an initial ${\bm \mu}_{0}$, we generate a sequence 
$$
{\bm \mu}_{j} \sim K({\bm \mu}_{j}|{\bm \mu}_{j-1}),  \quad j=1,\ldots,J,
$$ 
where $K$ is the kernel of the MCMC technique used.
%, leaving invariant $\bar{\pi}$.\footnote{Clearly, the samples within the burn-in period should be discarded.}
With the choice $h({\bm \mu}) = \bar{\pi}({\bm \mu})$,  the two levels of the sampler play different roles:
\begin{itemize}
\item The upper level attends the need for {\it exploration} of the state space, providing $\{{\bm \mu}_{1},\ldots, {\bm \mu}_{J}\}$.
\item The lower level is devoted to the {\it approximation} of local features of the the target, using $\{{\bf x}_1,\ldots, {\bf x}_J\}$.  
\end{itemize}
 In general, the two levels require their own tuning of the parameters of the corresponding proposals.

%{\color{magenta}
%describir aqu' la  $K_n()$ luego poner la tabla con los tres casos....es decir que nosotros controlamos la parte de arriba mientras otros no...tipo PMC.....
%}

%The use of the MCMC step facilitates the movement of the means towards the high-probability regions of the target, regardless of the choice of the initial parameters.

%%%%%%%%%%%%%%%%%%%%%%%%%%%%%%%%%%%%%%%
\subsection{Relationship with other adaptive MC schemes}
%%%%%%%%%%%%%%%%%%%%%%%%%%%%%%%%%%%%%%%

\label{RelationSect}
In contrast to the hierarchical approach described previously, in standard adaptive MC approaches \cite{Bugallo15,Haario2001,Luengo13} the parameter ${\bm \mu}_n$ is determined by a deterministic function, 
$$
\gamma: \mathbb{R}^{M\times D_x\times (n-1)}\rightarrow \mathbb{R}^{D_x},
$$
 of the previously generated samples (assuming to generate $M$ samples from each proposal), 
$$
{\bf X}_{j-1}=[{\bf x}_{1}^{(1)},\ldots,{\bf x}_{1}^{(M)},\ldots, {\bf x}_{j-1}^{(1)},\ldots,{\bf x}_{j-1}^{(M)}],
$$ 
namely,
 \begin{equation}
{\bm \mu}_j=\gamma({\bf X}_{j-1}).
 \end{equation}
 Although $\gamma$ is a deterministic function, the sequence $\{{\bm \mu}_j\}_{j=1}^J$ is generated according to a conditional pdf, $K({\bm \mu}_j|{\bm \mu}_{1},\ldots,{\bm \mu}_{j-1})$, since ${\bf X}_{j-1}$ is random. Unlike in the hierarchical scheme, in standard adaptive MC approaches, the sequence $\{{\bm \mu}_j\}_{j=1}^J$  typically converges to a fixed vector.
 
  In the standard PMC method \citep{Cappe04}  the sequence of mean vectors ${\bm \mu}_j$'s is also generated depending on the previous ${\bf x}$'s but, in this case, the final distribution is unknown and it is not a fixed vector, in general (for further details see Appendix \ref{PMClocura2}). Similar considerations also apply for Sequential Monte Carlo (SMC) schemes \cite{Moral06,Fearnhead13,Schafer13} where the adaptation is performed using a combination of resampling and MCMC steps. Other interesting and related techniques are the Particle MCMC (P-MCMC) \citep{Andrieu10} and the Sequentially Interacting MCMC (SI-MCMC) \citep{Brockwell10} methods. In this case, IS approximations of the target are used to build better proposal pdfs, employed within MCMC steps. Both methods are also able to provide efficient estimators of $Z$. However, unlike in PMC, SMC, P-MCMC and SI-MCMC, in the proposed hierarchical approach each ${\bm \mu}_j$ is always chosen independently of ${\bf X}_{j-1}$ and it is distributed according to $h({\bm \mu})$, decided in advance by the user. Moreover, the means ${\bm \mu}_1,\ldots,{\bm \mu}_j$ are not involved in the resulting estimators. Related observations are provided in Section \ref{TeoOtra}  and Table \ref{FantasticTable1negativo}.
 
 %Certain connections with the random-walk Metropolis Hastings (RWMH) are discussed in Appendix \ref{HierachInterp_app}.
%Finally, note that the complete generating procedure in Eq. \eqref{EqPropIni} can be also interpreted as a data augmentation approach \citep{Liang10,Robert04}, but we wish to emphasize the role of the prior over the mean vectors played by $h({\bm \mu})$, for reasons that will become apparent later. 
\vspace{0.3cm}

\section{Generalized Multiple Importance Sampling} 
%%%%%%%%%%%%%%%%%%%%%%%%%%%%%%%%%%%
%%%%%%%%%%%%%%%%%%%%%%%%%%%%%%%%%%%
%%%%%%%%%%%%%%%%%%%%%%%%%%%%%%%%%%%
\label{LOCURASect2}

So far, we have introduced a hierarchical procedure to generate candidates for an MC technique, adapting the mean vectors of a set of proposal densities. In this section, we provide a general framework for multiple importance sampling (MIS) techniques using a population of proposal densities, which embeds various alternative schemes proposed in the literature \cite{ElviraMIS15}. First, we consider several alternatives of static MIS, and then we focus on the corresponding adaptive MIS samplers.

%%%%%%%%%%%%%%%%%%%%%%%%%%%%%%%%
\subsection{Generalized Static Multiple Importance Sampling}
%%%%%%%%%%%%%%%%%%%%%%%%%%%%%%%%
\label{LOCURASect2_1}
As we have already highlighted, finding a good proposal pdf, $q({\bf x})$, is critical and is in general very challenging \citep{Owen00}.
An alternative strategy consists in using a population of proposal pdfs.
This approach is often known in the literature as multiple importance sampling (MIS)  \citep{mcbookOwen,Owen00,Veach95,ElviraMIS15}.
Consider a set of $J$ proposal pdfs, 
$$
q_1({\bf x}), \ldots, q_J({\bf x}),
$$ 
with heavier tails than the target $\pi$, and let us assume that $M$ samples are drawn from each of them, i.e., 
$$
{\bf x}_j^{(m)} \sim q_j({\bf x}),  \quad j=1,\ldots, J, \quad m=1,\ldots,M.
$$
%The aim of drawing exactly $M$ samples from each proposal is ensuring that all the proposals have an equal participation in the estimation.
%
In this scenario, the weights associated to the samples can be obtained following at least one of these two strategies:
\begin{itemize}
\item[(a)] {\it Standard} MIS (S-MIS): 
\begin{equation} 
 w_j^{(m)}= \frac{\pi({\bf x}_j^{(m)})}{{q_j({\bf x}_j^{(m)})}},
  \label{is_weights_static}
\end{equation}
for $j=1,...,J$  and $m=1,\ldots,M$,
\item[(b)] {\it Deterministic mixture MIS} (DM-MIS)  \citep{Owen00,Veach95}:
\begin{equation} 
	w_j^{(m)}=\frac{\pi({\bf x}_j^{(m)})}{\psi({\bf x}_j^{(m)})}=\frac{\pi({\bf x}_j^{(m)})}{\frac{1}{J}\sum_{k=1}^{J}q_k({\bf x}_j^{(m)})},
\label{f_dm_weights_static}
\end{equation}
for $j=1,...,J$  and $m=1,\ldots,M$, and
where $\psi({\bf x})=\frac{1}{J}\sum_{j=1}^{J}q_j({\bf x})$ is the mixture pdf, composed of all the proposal pdfs.
This approach is based on the considerations provided in Appendix \ref{BellissimaSect}.
\end{itemize}
In both  cases, the consistency of the estimators is ensured \cite{ElviraMIS15}.
The main advantage of the DM-MIS weights is that they yield more efficient estimators than using the standard importance weights  \citep{CORNUET12,Owen00,LetterVictor,APIS15}.
However, the DM-MIS estimator is computationally more expensive, as it requires $JM$ total evaluations for each proposal instead of just $M$, for computing all the weights. The number of evaluations of the target $\pi({\bf x})$ is the same regardless of whether the weights are calculated according to Eq. \eqref{is_weights_static} or \eqref{f_dm_weights_static}, so this increase in computational cost may not be relevant in many applications. However, in some other cases this additional computational load may
be excessive (especially for large values of $J$) and alternative efficient solutions are desirable. For instance, the use of partial mixtures has been proposed in \citep{LetterVictor}:
\begin{itemize}
\item[(c)] {\it Partial} DM-MIS (P-DM-MIS) \citep{LetterVictor}: divide the $J$ proposals in $L=\frac{J}{P}$ disjoint groups forming $L$ mixtures with $P$ components. Let us denote the set of $P$ indices corresponding to the $\ell$-th mixture ($\ell=1,\ldots, L$) as $\mathcal{S}_\ell=\{k_{\ell,1},\ldots,k_{\ell,P}\}$ (i.e., $|\mathcal{S}_\ell|=P$), where each $k_{\ell,p}\in\{1,\ldots,J\}$. Thus, we have
\begin{equation}
	\mathcal{S}_1 \cup \mathcal{S}_2\cup \ldots \cup \mathcal{S}_L= \{1,\ldots,J\},
\end{equation}
with $\mathcal{S}_r \cap \mathcal{S}_\ell = \emptyset$, for all $\ell=1,\ldots,L,$ and $r\neq \ell$. In this case, the importance weights are defined as
\begin{equation} 
	w_{j}^{(m)}=\frac{\pi({\bf x}_j^{(m)})}{\frac{1}{P}\sum_{k\in \mathcal{S}_\ell}q_{k}({\bf x}_j^{(m)})},
\label{p_dm_weights_static}
\end{equation}
with $j\in\mathcal{S}_{\ell}$, $\ell=1,\ldots, L$, and $m=1,\ldots,M$.
\end{itemize}
All the previous cases can be captured by a generic mixture-proposal $\Phi_j({\bf x})$, under which the MIS weights can be defined as
\begin{equation} 
 w_j^{(m)}= \frac{\pi({\bf x}_j^{(m)})}{{\Phi_j({\bf x}_j^{(m)})}}, 
  \label{GEN_weights_static}
\end{equation}
with $m=1,\ldots,M$, where $\Phi_j({\bf x}_j^{(m)})=q_j({\bf x}_j^{(m)})$  in Eq. \eqref{is_weights_static}, $\Phi_j({\bf x}_j^{(m)})=\frac{1}{J}\sum_{k=1}^{J}q_k({\bf x}_j^{(m)})$ in Eq. \eqref{f_dm_weights_static}, and 
\begin{equation}
\Phi_j({\bf x}_j^{(m)})=\frac{1}{P}\sum_{k\in \mathcal{S}_\ell}q_{k}({\bf x}_j^{(m)}), \quad \mbox{with} \quad  j\in \mathcal{S}_{\ell},
\end{equation}
  in Eq. \eqref{p_dm_weights_static}. In any case, the weights are always normalized as 
\begin{equation}
\bar{\rho}_j^{(m)}=\frac{w_j^{(m)}}{\sum_{i=1}^J\sum_{r=1}^M w_i^{(r)}}.
\end{equation}
Table \ref{FantasticTable0} shows these three choices of $\Phi_j({\bf x}_j^{(m)})$, whereas Table \ref{MISgenTable} summarizes a generalized static MIS procedure.
\begin{table}[!htb]
\caption{Three possible functions $\Phi_{j}({\bf x})$ for MIS.}
%($J=$ total number of proposals; $L=$  total number of mixtures; $P=$ number of components per mixture).}
\begin{center}
\begin{tabular}{|c||c|c|c|}
\hline
 \multirow{2}{*}{{\bf MIS approach}} & {\bf Function $\Phi_{j}({\bf x})$},     & ${\bf L}$ & ${\bf P}$    \\
\cline{3-4}
 & $(j=1,\ldots,J)$ & \multicolumn{2}{c|}{$LP=J$}   \\
\hline
\hline
Standard MIS & $q_{j}({\bf x})$  & $J$ & $1$  \\
\hline
 DM-MIS & $\psi({\bf x})= \frac{1}{J}\sum_{j=1}^J q_{j}({\bf x})$    & $1$ & $J$   \\
\hline
Partial DM-MIS & $\frac{1}{P}\sum_{k\in \mathcal{S}_\ell}q_{k}({\bf x})$    & $L$ & $P$\\ 
\hline
\end{tabular}
\end{center}
\label{FantasticTable0}
\end{table}%
%%%%%%%%%%%%%%%%
%%%%%%%%%%%%%%%%
\begin{table}[!t]
%	\centering
%\small
\caption{Generalized static MIS scheme.}
%\vspace{-1cm}
	\begin{tabular}{|p{0.95\columnwidth}|}
    \hline
\footnotesize
\begin{enumerate}
\item {\bf Generation:} Draw $M$ samples  from each $q_{j}$, i.e., 
$${\bf x}_{j}^{(m)} \sim q_{j}({\bf x}),$$
for  $j=1,\ldots,J$, and with $m=1,\ldots,M$.
\item {\bf Weighting:} Assign to each sample ${\bf x}_{j}^{(m)}$ the weight
\begin{equation}
w_{j}^{(m)}=\frac{\pi({\bf x}_{j}^{(m)})}{\Phi_{j}( {\bf x}_{j}^{(m)})}, 
\end{equation} 
where $\Phi_{j}$ is a mixture of $q_{j}$'s, as shown in Table \ref{FantasticTable0}.
\item {\bf Normalization:} Set 
$$
\bar{\rho}_j^{(m)}=\frac{w_j^{(m)}}{\sum_{i=1}^J\sum_{r=1}^M w_i^{(r)}}.
$$
\item {\bf Output:} Return all the pairs $\{{\bf x}_{j}^{(m)},\bar{\rho}_j^{(m)}\}$, for  $j=1,\ldots,J$ and  $m=1,\ldots,M$.
%	The final locations of the Gaussians (i.e., their means, ${\bm \mu}_{i}^{(M)}$ for $i = 1, \ldots, N$) could also be used to estimate the locations of the modes of $\pi({\bf x})$. %and the global maximum of $\pi({\bf x})	
%	(i.e., the MAP estimator of ${\bf x}$), as the proposal converges towards the areas of high probability (i.e., the maxima) of the target.
 \end{enumerate} \\
\hline 
\end{tabular}
\label{MISgenTable}
\end{table}
%%%%%%%%%%%%%%%%
%%%%%%%%%%%%%%%%

Note that the IS estimator $\hat{I}$ of a specific moment of $\bar{\pi}$, i.e., the integral $I$ given in Eq. \eqref{eq:integral}, and the approximation $\hat{Z}$ of the normalizing constant in Eq. \eqref{eq:normConst}, can now  be approximated as
\begin{gather}
\begin{split}
{\hat I}&=\sum_{j=1}^J \sum_{m=1}^M  \bar{\rho}_j^{(m)}  f({\bf x}_j^{(m)}), \\
 {\hat Z}&=\frac{1}{JM}\sum_{j=1}^J \sum_{m=1}^M w_j^{(m)}.
 \end{split}
\end{gather}
Then, the particle approximation of the measure of $\bar{\pi}$ is given by
\begin{equation}
\hat{\pi}^{(JM)}({\bf x})=\frac{1}{JM\hat{Z}}\sum_{j=1}^J \sum_{m=1}^M w_{j}^{(m)} \delta({\bf x}-{\bf x}_{j}^{(m)}).
\end{equation}
In Section \ref{AISschemes}, we describe a framework where a partial grouping of the proposal pdfs arises naturally from the sampler's definition.

 %%%%%%%%%%%%%%%%%%%%%%%%%%%%%%%%%%%%%%%%%%
\subsection{Generalized Adaptive Multiple Importance Sampling}
\begin{figure}[!htb]
\centering 
\centerline{
 \includegraphics[width=8cm]{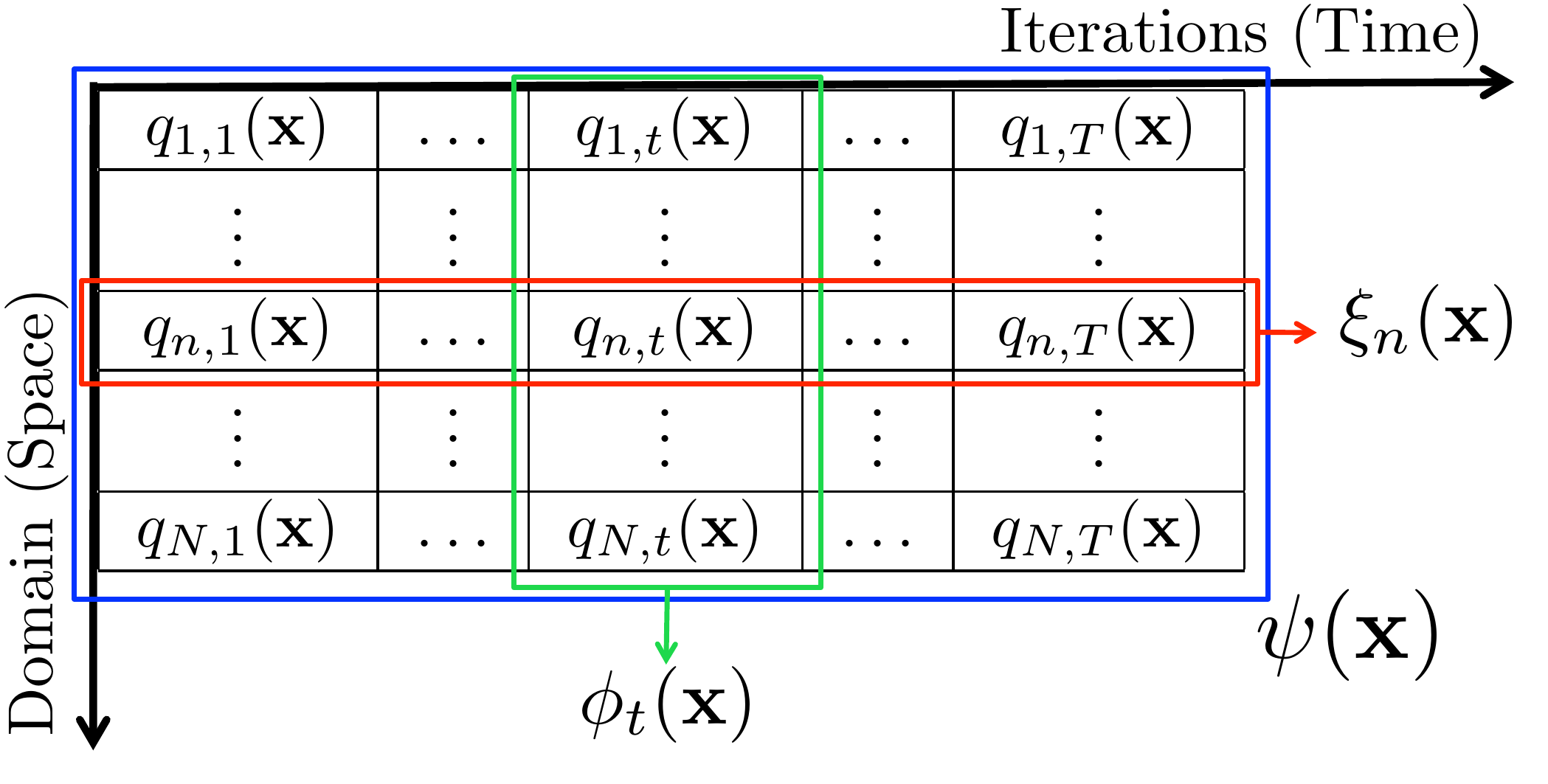}
  }
  \caption{Graphical representation of the $J=NT$ proposal pdfs used in the generalized adaptive multiple IS scheme, spread through the state space $\mathcal{X}$ ($n=1,\ldots,N$) and adapted over time ($t=1,\ldots,T$). Three different mixtures are displayed: $\psi({\bf x})$ involving all the proposals, $\phi_t({\bf x})$  involving only the proposals at the $t$-th iteration, and $\xi_n({\bf x})$ considering the temporal evolution of the $n$-th proposal pdf.
   %The DM-MIS approach can be applied considering the $J=NT$  proposals (providing only one mixture $\psi$), or the $T$ mixtures    
  }
\label{figSpaceTime}
\end{figure}

%%%%%%%%%%%%%%%%%%%%%%%%%%%%%%%%%%%%%%%%%%
\label{AISschemes}
In order to decrease the mismatch between the proposal and the target, several Monte Carlo methods adapt the parameters of the proposal iteratively using the information of the past samples \citep{Cappe04,CORNUET12,APIS15}.
In this adaptive scenario, we have a set of proposal pdfs $\{q_{n,t}({{\bf x}})\}$, with $n=1,\ldots,N$ and $t=1,\ldots,T$, where the subscript $t$ indicates the iteration index, $T$ is the total number of adaptation steps, and $J=NT$ is the total number of proposal pdfs. In the following, we present a unified framework, called generalized adaptive multiple importance sampling (GAMIS), which includes several methodologies proposed independently in the literature, as particular cases. In GAMIS, each proposal pdf in the population  $\{q_{n,t}\}$ is updated at every iteration $t=1,\ldots,T$, forming the sequence 
$$
q_{n,1}({\bf x}),q_{n,2}({\bf x}),\ldots, q_{n,T}({\bf x}),
$$
for the $n$-th proposal (see Figure \ref{figSpaceTime}). At the $t$-th iteration, the adaptation procedure takes into account statistical information about the target distribution gathered in the previous iterations, $1,\ldots,t-1$, using one of the many procedures that have been proposed in the literature \citep{pmc-cappe08,Cappe04,CORNUET12,APIS15}. Furthermore, at the $t$-th iteration, $M$ samples are drawn from each proposal $q_{n,t}$, 
$$
{\bf x}_{n,t}^{(m)} \sim q_{n,t}({{\bf x}}),  \quad \mbox{ with } \quad m=1,\ldots,M,
$$
$n=1,\ldots,N$ and $t=1,\ldots,T$. An importance weight $w_{n,t}^{(m)}$ is then assigned to each sample ${\bf x}_{n,t}^{(m)}$. Several strategies can be applied to build $w_{n,t}^{(m)}$ considering the different MIS approaches, as discussed in the previous section. Figure \ref{figSpaceTime} provides a graphical representation of this scenario, by showing both the spatial and temporal evolution of the $J=NT$ proposal pdfs. 

In an AIS algorithm, one weight 
\begin{equation}
\label{Gen_W_Eq}
w_{n,t}^{(m)}=\frac{\pi({\bf x}_{n,t}^{(m)})}{\Phi_{n,t}({\bf x}_{n,t}^{(m)})},
\end{equation}
is associated  to each sample ${\bf x}_{n,t}^{(m)}$. In a standard MIS approach, the function employed in the denominator is  
\begin{equation}
  \label{Case1}
\Phi_{n,t}({\bf x})=q_{n,t}({\bf x}).
\end{equation}
In the complete DM-MIS case, the function $\Phi_{n,t}$ is 
\begin{equation}
  \label{CompleteProposal}
\Phi_{n,t}({\bf x})= \psi({\bf x})= \frac{1}{NT}\sum_{k=1}^N \sum_{r=1}^T q_{k,r}({\bf x}).
\end{equation}
This case corresponds to the external blue rectangle in Fig. \ref{figSpaceTime}. Two natural alternatives of partial DM-MIS schemes appear in this scenario. The first one uses the following partial mixture 
 \begin{equation}
  \label{TemporalMixProposal}
 \Phi_{n,t}({\bf x})= \xi_{n}({\bf x})= \frac{1}{T}\sum_{r=1}^T q_{n,r}({\bf x}),
 \end{equation}
  with $n=1,\ldots,N$, in the denominator of the IS weight. Namely, we consider the temporal evolution of the $n$-th single proposal $q_{n,t}$. Hence, we have $L=N$ mixtures, each one formed by $P=T$ components (horizontal red rectangle in Fig. \ref{figSpaceTime}). The other possibility is considering the mixture of all the $q_{n,t}$'s at the $t$-th iteration, i.e., 
 \begin{equation}
  \label{SpatialMixProposal}
  \Phi_{n,t}({\bf x})=\phi_{t}({\bf x})= \frac{1}{N}\sum_{k=1}^N q_{k,t}({\bf x}),  
\end{equation}  
for $t=1,\ldots,T$, so that  we have $L=T$ mixtures, each one formed by $P=N$ components (vertical green rectangle in Fig. \ref{figSpaceTime}). The function $\Phi_{n,t}$ in Eq. \eqref{Case1} is used in the standard PMC scheme \citep{Cappe04}; Eq. \eqref{TemporalMixProposal} with $N=1$ has been considered in adaptive multiple importance sampling (AMIS) \citep{CORNUET12}. Eq. \eqref{SpatialMixProposal} has been applied in  the adaptive population importance sampling (APIS) algorithm \citep{APIS15}, whereas in other techniques, such as Mixture PMC \citep{pmc-cappe08,Douc07a,Douc07b},  %\citep{Douc07a,Douc07b} 
 a similar strategy is employed but using a standard sampling of the mixture $\phi_{t}({\bf x})$. 

%%%%%%%%%%%%%%%%
%%%%%%%%%%%%%%%%
\begin{table*}[!htb]
\caption{ Summary of possible MIS strategies in an adaptive framework.}
%($J=$ total number of proposals; $L=$  total number of mixtures; $P=$ number of components per mixture).}
\begin{center}
\begin{tabular}{|c||c|c|c|c||c|}
\hline
 \multirow{2}{*}{{\bf MIS approach}} &  \multirow{2}{*}{{\bf Function $\Phi_{n,t}({\bf x})$}}    &   \multirow{2}{*}{${\bf J}$}  & ${\bf L}$ & ${\bf P}$ & \multirow{2}{*}{{\bf Corresponding Algorithm} }    \\
\cline{4-5}
 && & \multicolumn{2}{c||}{$LP=J$} &  \\
\hline
\hline
Standard MIS & $q_{n,t}({\bf x})$ & & $NT$ & $1$ & PMC \citep{Cappe04} \\
\cline{1-2} \cline{4-6} 
 DM-MIS & $\psi({\bf x})= \frac{1}{NT}\sum_{n=1}^N \sum_{t=1}^T q_{n,t}({\bf x})$  &  & $1$ & $NT$ & suggested in \citep{LetterVictor}  \\
\cline{1-2} \cline{4-6} 
Partial DM-MIS & $\xi_{n}({\bf x})= \frac{1}{T}\sum_{t=1}^T q_{n,t}({\bf x})$ & $NT$ & $N$ & $T$ &   AMIS \citep{CORNUET12}, with $N=1$  \\
\cline{1-2} \cline{4-6} 
Partial DM-MIS & $\phi_{t}({\bf x})= \frac{1}{N}\sum_{n=1}^N q_{n,t}({\bf x})$&  & $T$ & $N$ &  APIS \citep{APIS15} and  \citep{pmc-cappe08,Douc07a,Douc07b} \\
\cline{1-2} \cline{4-6} 
Partial DM-MIS & generic $\Phi_{n,t}({\bf x})$ in Eq. \eqref{PartialMixProposal2}  &  & $L$ & $P$ & suggested in \citep{LetterVictor}\\ 
\hline
\end{tabular}
\end{center}
\label{FantasticTable}
\end{table*}% 

Table \ref{FantasticTable} summarizes all the possible cases discussed above.
The last row corresponds to a generic grouping strategy of the proposal pdfs $q_{n,t}$. As previously described, we can also divide the $J=NT$ proposals into $L=\frac{NT}{P}$ disjoint groups forming $L$ mixtures with $P$ components. We denote the set of $P$ pairs of indices corresponding to the $\ell$-th mixture ($\ell=1,\ldots, L$) as $\mathcal{S}_\ell=\{(k_{\ell,1},r_{\ell,1}), \dots,(k_{\ell,P},r_{\ell,P})\}$, where $k_{\ell,p}\in\{1,\ldots,N\}$,  $r_{\ell,p}\in\{1,\ldots,T\}$ (i.e., $|\mathcal{S}_\ell|=P$, with each element being a pair of indices), and $\mathcal{S}_r \cap \mathcal{S}_\ell = \emptyset$ for all $\ell=1,\ldots,L,$ and $r\neq \ell$. In this scenario, we have
   \begin{equation}
  \label{PartialMixProposal2}
  \Phi_{n,t}({\bf x})= \frac{1}{P}\sum_{(k,r)\in \mathcal{S}_\ell} q_{k,r}({\bf x}),  \quad \mbox{with} \quad  (n,t) \in \mathcal{S}_\ell.  
\end{equation}  
Note that, using $\psi({\bf x})$ and $\xi_n({\bf x})$, the computational cost per iteration increases as the total number of iterations $T$ grows.
Indeed, at the $t$-th iteration all the previous proposals $q_{n,1},\ldots,q_{n,t-1}$ (for all $n$) must be evaluated at all the new samples ${\bf x}_{n,t}^{(m)}$. 
Hence, algorithms based on these proposals quickly become unfeasible as the number of iterations grows.
On the other hand, using $\phi_t({\bf x})$ the computational cost per iteration is controlled by $N$, remaining invariant regardless of the number of adaptive steps performed.

Observe also that a suitable AIS scheme builds iteratively a global IS estimator which uses the normalized weights
 \begin{equation}
\label{RHOnorm2} 
	\bar{\rho}_{n,t}^{(m)} = \frac{w_{n,t}^{(m)} }{\sum_{\tau=1}^T\sum_{n=1}^N\sum_{m=1}^M  w_{n,\tau}^{(m)} }, 
\end{equation}
for $n=1,\ldots,N$,  $m=1,\ldots,M$, and $t=1,\ldots, T$. %It is important to remark that the estimators in a GAMIS scheme can be constructed in two different ways.  In a {\it batch mode}, the entire population of $J=NT$ proposal pdfs $\{q_{n,t}\}$'s is generated in advance performing the usual adaptation operations according to the chosen algorithm. After that, the algorithm is converted into a simpler static MIS technique and the importance weights are computed and normalized as in Table \ref{MISgenTable}. This version is simpler than the iterative mode described below, but no output of the algorithm is provided until adapting all the proposals have been updated, i.e., after $T$ iterations. In an {\it iterative mode},  an output is provided at each iteration $t$. 

Table \ref{GAMISgenTable} shows an iterative version of GAMIS. We remark that, at the $t$-th iteration, the weights of the samples previously generated need to be recalculated, as shown in step 2(c-3) of Table \ref{GAMISgenTable}. The choices $\Phi_{n,t}({\bf x})=q_{n,t}({\bf x})$ or $\Phi_{n,t}({\bf x})=\phi_t({\bf x})$ allow avoiding completely this re-computation step of the weights.   
 %However, the output after $T$ iterations is the same in both cases. 
 For simplicity, in Table \ref{GAMISgenTable} we have provided the output of the algorithms as weighted samples, i.e., all the pairs $\{{\bf x}_{n,t}^{(m)},\bar{\rho}_{n,t}^{(m)}\}$. However, the output can be equivalently expressed as an estimator of a specific moment of the target. In this case, the final IS estimators $\hat{I}_T$ and $\hat{Z}_T$ are 
\begin{gather}
\begin{split}
\label{equation_whatever_1}
{\hat I}_T&=\sum_{\tau=1}^T\sum_{n=1}^N \sum_{m=1}^M \bar{\rho}_{n,\tau}^{(m)}  f({\bf x}_{n,\tau}^{(m)}), \\ 
{\hat Z}_T&=\frac{1}{NMT}\sum_{\tau=1}^T\sum_{n=1}^N \sum_{m=1}^M w_{n,\tau}^{(m)},
\end{split}
\end{gather}
where $\bar{\rho}_{n,\tau}^{(m)}=\frac{w_{n,\tau}^{(m)}}{{NMT\hat Z}_T}$. Moreover, the final particle approximation is 
\begin{equation}
\hat{\pi}^{(NMT)}({\bf x})=\frac{1}{NMT\hat{Z}_T}\sum_{\tau=1}^T\sum_{n=1}^N \sum_{m=1}^M w_{n,\tau}^{(m)} \delta({\bf x}-{\bf x}_{n,\tau}^{(m)}).
\end{equation}
The estimators in Eq. \eqref{equation_whatever_1} can be expressed recursively, thus providing an estimate at each iteration $t$, as stated before. Starting with $H_0=0$, ${\hat I}_0=0$,  and setting $S_t=\sum_{n=1}^N \sum_{m=1}^M w_{n,t}^{(m)}$ and $H_t=H_{t-1}+S_t$, we have
\begin{eqnarray}
\label{equation_whatever_2}
{\hat I}_t&=&\frac{1}{H_t} \left[ H_{t-1}{\hat I}_{t-1}+\sum_{n=1}^N \sum_{m=1}^M w_{n,t}^{(m)} f({\bf x}_{n,t}^{(m)})\right], \nonumber \\
&=&\frac{H_{t-1}}{H_{t-1}+S_t}  {\hat I}_{t-1}+\frac{S_t}{H_{t-1}+S_t} {\hat A}_t, 
\end{eqnarray}
where ${\hat A}_t= \sum_{n=1}^N \sum_{m=1}^M \frac{w_{n,t}^{(m)}}{S_t} f({\bf x}_{n,t}^{(m)})$ is the partial IS estimator using only the samples drawn at the $t$-th iteration. Therefore, ${\hat I}_t$ can be seen as a convex combination of the two IS estimators ${\hat I}_{t-1}$ and ${\hat A}_t$ (for further explanations see Eqs. \eqref{General_standIS_0}-\eqref{General_standIS} in Appendix \ref{PobreDiabla3}). %Observe that, the final global estimator $\hat{I}_T$ in Eq. \eqref{equation_whatever_1}, obtained recursively in Eq. \eqref{equation_whatever_2}, is simply a standard IS estimator using all the samples ${\bf x}_{n,t}^{(m)}$ and considering the mixtures $\Phi_{n,t}({\bf x})$ as proposal pdfs in the IS weight ratio. 
Finally, note that 
\begin{equation}
{\hat Z}_t=\frac{1}{t}\frac{1}{NM} H_t.
\end{equation}
A brief discussion about the consistency of ${\hat I}_t$ and ${\hat Z}_t$ is provided in Appendix \ref{ConsApp}.

%%%%%%%%%%%%%%%%
%%%%%%%%%%%%%%%%
\begin{table}[!t]
%	\centering
%\small
\caption{GAMIS scheme: iterative version.}
%\vspace{-1cm}
\begin{tabular}{|p{0.95\columnwidth}|}
    \hline
\footnotesize
\begin{enumerate}
\item {\bf Initialization:} Set $t=1$, $H_0=0$ and choose $N$ initial proposal pdfs $q_{n,0}({\bf x})$.
\item For $t=1,\ldots,T$:
\begin{enumerate}
\item {\bf Adaptation:} update the proposal pdfs $\{q_{n,t-1}\}_{n=1}^N$ providing $\{q_{n,t}\}_{n=1}^N$, using a preestablished procedure (e.g., see \citep{Cappe04,pmc-cappe08,CORNUET12,APIS15} for some specific approaches).
\item {\bf Generation:} Draw $M$ samples  from each $q_{n,t}$, i.e., ${\bf x}_{n,t}^{(m)} \sim q_{n,t}({\bf x})$, with  $n=1,\ldots,N$ and  $m=1,\ldots,M$.
\item {\bf Weighting:}
 \begin{enumerate}
 \item[(c-1)] Update the function $\Phi_{n,t}({\bf x})$ given the current population $\{q_{1,t},\ldots, q_{N,t}\}$.
\item[(c-2)] Assign the weights to the new samples ${\bf x}_{n,t}^{(m)}$, 
\begin{equation}
w_{n,t}^{(m)}=\frac{\pi({\bf x}_{n,t}^{(m)})}{\Phi_{n,t}( {\bf x}_{n,t}^{(m)})}, 
\end{equation} 
  for $n=1,\ldots,N$, and $m=1,\ldots,M$.
 \item[(c-3)] Re-weight the previous samples ${\bf x}_{n,\tau}^{(m)}$ for $\tau=1,\ldots, t-1$ as
\begin{equation}
w_{n,\tau}^{(m)}=\frac{\pi({\bf x}_{n,\tau}^{(m)})}{\Phi_{n,t}( {\bf x}_{n,\tau}^{(m)})}, 
\end{equation}  
with  $\tau=1,\ldots, t-1$, $n=1,\ldots,N$, and $m=1,\ldots,M$.
\end{enumerate}
\item {\bf Normalization:} Set $S_t=\sum_{m=1}^M\sum_{n=1}^N w_{n,t}^{(m)}$, $H_{t}=H_{t-1}+S_t$ , and re-normalize all the weights,
\begin{equation}
\bar{\rho}_{n,\tau}^{(m)} = \bar{\rho}_{n,\tau-1}^{(m)} \frac{H_{t-1}}{H_t},
\end{equation}
for  $\tau=1,\ldots, t$, $n=1,\ldots,N$, and $m=1,\ldots,M$.
\item {\bf Output:}  Return all the pairs $\{{\bf x}_{n,\tau}^{(m)},\bar{\rho}_{n,\tau}^{(m)}\}$, for $\tau=1,\ldots,t$, $n=1,\ldots,N$,  and  $m=1,\ldots,M$.
 \end{enumerate}
%\item If $t<T$, set $t=t+1$, and repeat from step 2.  Otherwise, if $t=T$, then stop.
%	The final locations of the Gaussians (i.e., their means, ${\bm \mu}_{i}^{(M)}$ for $i = 1, \ldots, N$) could also be used to estimate the locations of the modes of $\pi({\bf x})$. %and the global maximum of $\pi({\bf x})	
%	(i.e., the MAP estimator of ${\bf x}$), as the proposal converges towards the areas of high probability (i.e., the maxima) of the target.
 \end{enumerate} \\
\hline 
\end{tabular}
\label{GAMISgenTable}
\end{table}
%%%%%%%%%%%%%%%%
%%%%%%%%%%%%%%%%

%%%%%%%%%%%%%%%%%%
%%%%%%%%%%%%%%%%%%
%\input{GAMIS_Batch_framework}
%%%%%%%%%%%%%%%%%%
%%%%%%%%%%%%%%%%%%

%%%%%%%%
%%%%%%%%

%%%%%%%%
%%%%%%%%

%%%%%%%%%%%%%%%%%%%%%%%%%%%%%%%%%%%%%%%%
\section{Markov adaptation for GAMIS} 
%%%%%%%%%%%%%%%%%%%%%%%%%%%%%%%%%%%%%%%%
\label{SuperSect}
In this section, we design efficient adaptive importance sampling (AIS) techniques by combining the main ideas discussed in the two previous sections. More specifically, we apply the hierarchical MC approach to adapt the proposal pdfs within a GAMIS scheme. Therefore, a Markov GAMIS technique, or simply {\it Markov Adaptive Importance Sampling} (MAIS) algorithm, consists of the following two layers:
\begin{itemize}
\item[1.] {\it Upper level (Adaptation):}  Given the set of mean vectors,
$$
\mathcal{P}_{t-1}=\{{\bm \mu}_{1,t-1},\ldots,{\bm \mu}_{N,t-1}\},
$$
obtain the new set $\mathcal{P}_{t}=\{{\bm \mu}_{1,t},\ldots,{\bm \mu}_{N,t}\}$ according to MCMC transitions with $\bar{\pi}$ as invariant density.  More specifically, a kernel $K({\bm \mu}_{1:N,t}|{\bm \mu}_{1:N,t-1})$ leaving invariant the distribution $\prod_{n=1}^{N} {\bar \pi}({\bm \mu}_n)$ is applied.

%we apply the hierarchical approach in Section \ref{HMCsect}, in order to update the location parameters of the proposal pdfs $q_{n,t}$) using $\bar{\pi}({\bm \mu})$ as prior. Namely, we desire to yield ${\bm \mu}_{n,t}\sim \bar{\pi}({\bm \mu})$.
\item[2.] {\it Lower level (MIS estimator):} Given the population of proposals,
$$
q_{1,t}({\bf x}|{\bm \mu}_{1,t}, {\bf C}_1), \ldots, q_{N,t}({\bf x}|{\bm \mu}_{N,t}, {\bf C}_N), 
$$
 choose a function $\Phi_{n,t}({\bf x})$ for the computation of the weights in Eq. \eqref{Gen_W_Eq}, and perform a MIS approximation of the target as described in Section \ref{AISschemes}.   
\end{itemize}

%%%%%%%%%%%%%%%%%%%%%%%%%%%%%%%%%%%%%%%%
\subsection{Theoretical support: adaptation and consistency} 
\label{TeoOtra}
%%%%%%%%%%%%%%%%%%%%%%%%%%%%%%%%%%%%%%%%

The motivation behind the MCMC adaptation has been described in Section \ref{Prior2} and \ref{Prior4}: the functions $q_{n,t}$, located at the ${\bm \mu}_{n,t}$'s, jointly provide a kernel estimate of the target ${\bar \pi}$. 

Furthermore, we recall that the generation of the means, ${\bm \mu}_{n,t}$, is {\it completely independent} from the samples ${\bf x}_{n,t}$ drawn in the lower level. This is a key point from a theoretical and practical point of view. Indeed, the generic MAIS algorithm can be divided in two steps: (a) first generate all the means $\{{\bm \mu}_{n,t}\}_{t=1}^T$ for $n=1,\ldots,N$, (b) then perform the MIS estimation considering all the proposals $q_{n,t}({\bf x}|{\bm \mu}_{n,t}, {\bf C}_n)$, $\forall n$ and $\forall t$. Namely, any MAIS technique can be converted into a generalized static MIS scheme (see Section \ref{LOCURASect2_1}). As a consequence, the unique conditions required for ensuring the consistency of the corresponding estimators are \cite{ElviraMIS15,Robert04}:
\begin{itemize}
\item All the proposal pdfs, $q_{n,t}$, must have heavier tails than the target ${\bar \pi}$. 
\item A suitable  function  $\Phi_{n,t}({\bf x})$ for the denominator of the importance weights must be chosen. Namely, the use of  $\Phi_{n,t}({\bf x})$ provides consistent estimators \cite{ElviraMIS15}, like the functions $\Phi_{n,t}({\bf x})$ described in Section \ref{AISschemes}. 
\end{itemize}
Moreover, the independence of the upper level from the lower level of the hierarchical approach, helps the parallelization of the algorithms as we discuss later. 

Table \ref{FantasticTable1negativo} compares different AIS schemes. In the standard AIS method \cite{Bugallo15}, the sequence of $\{{\bm \mu}_{n,t}\}$ converges to a  unknown fixed vector as $t\rightarrow \infty$. In the standard PMC algorithm \cite{Cappe04}, the limiting distribution of  $\{{\bm \mu}_{n,t}\}$ is unknown. Furthermore, in both cases, standard AIS and PMC, the adaptation depends on the previously generated samples ${\bf x}$'s. In MAIS techniques, the use of an ergodic chain (with invariant pdf ${\bar \pi}$) for generating the $n$-th mean vector ${\bm \mu}_{n,t}$ ensures that its asymptotic density is ${\bar \pi} ({\bm \mu})$.

%The theoretical motivation of this adaptation procedure is supported by the previously discussed kernel density estimation argument. Observe that the adaptation process is independent from the underlying IS steps. 

 \begin{table}[!htb]
\caption{Adaptation of the mean vectors $\{{\bm \mu}_{n,t}\}$ using different AIS techniques.}
%($J=$ total number of proposals; $L=$  total number of mixtures; $P=$ number of components per mixture).}
\begin{center}
\begin{tabular}{|c||c|c|c|}
\hline
%\backslashbox{{\bf Features}}{{\bf Method}}
\multirow{2}{*}{{\bf Features}}  & \multirow{2}{*}{{\bf Stand. AIS}}  & \multirow{2}{*}{{\bf PMC}} & \multirow{2}{*}{{\bf MAIS}}    \\
 &  &  &   \\
\hline
\hline
limiting&   (unknown)   & unknown &  \\
distribution of & fixed & (if/when & ${\bar \pi}({\bm \mu})$ \\
 $\{{\bm \mu}_{n,t}\}$ for $t\rightarrow \infty$ & vector  &  exists)& \\
\hline
 dependence of  &    & &   \\
 the adaptation  & yes  & yes & no \\
 w.r.t. the ${\bf x}$'s && &\\ 
\hline
\end{tabular}
\end{center}
\label{FantasticTable1negativo}
\end{table}%

%%%%%%%%%%%%%%%%%%%%%%%%%%%%%%%%%%%%%%%%
\subsection{The new class of algorithms} 
%%%%%%%%%%%%%%%%%%%%%%%%%%%%%%%%%%%%%%%%

Markov GAMIS framework can lead to many different algorithms, depending on the MCMC strategy used to update the mean vectors and the specific choice of the function $\Phi_{n,t}$. Table \ref{FantasticTableOtra} provides several examples of novel techniques determined by the value of $N$,  the choice of $\Phi_{n,t}$, and the type of MCMC adaptation. Some of them are variants of well-known techniques like PMC \citep{Cappe04} and AMIS \citep{CORNUET12}, where the Markov adaptation procedure is employed. Others, such as the {\it Random Walk Importance Sampling} (RWIS), the {\it Parallel Interacting Markov Adaptive Importance Sampling} (PI-MAIS) and {\it Doubly Interacting  Markov Adaptive Importance Sampling} (I$^2$-MAIS), are described below in detail. 
For these completely novel algorithms we have set $\Phi_{n,t}({\bf x})=\phi_t({\bf x})$, so that the computational cost is directly controlled by $N$ and the re-weighting step 2(c-3) in Table \ref{GAMISgenTable} is not required.

RWIS is the simplest possible Markov GAMIS algorithm. Specifically,  for the MCMC adaptation we consider a standard MH technique, setting $N=1$ and choosing $\Phi_{n,t}({\bf x})=\phi_{t}({\bf x})=q_{n,t}({\bf x})$ (since $N=1$, the two cases  coincide). Table \ref{RWISTable} shows the RWIS algorithm, which is a special case of the more general scheme described in Table \ref{PopMAIS} when $N=1$. Note that we have a proposal pdf used for the MH adaptation, $\varphi({\bm \mu}|{\bm \mu}_{t-1}, {\bf \Lambda})$,  which is different from the proposal pdf used for the IS estimation, $q({\bf x}|{\bm \mu}_{t}, {\bf C})$. 

%Namely, there are now two different proposal densities, one at each level of the hierarchical Monte Carlo. The MH technique is applied to obtain good mean vectors for the underlying IS, and the particle approximation of $\bar{\pi}$ is then obtained iteratively using $NT$ samples.  

%Some of these well-known AIS methods, are also based on MIS, are Population Monte Carlo (PMC) and its variants \citep{Cappe04,Douc07b,Cappe08,bugallo2009marginalized}, adaptive multiple importance sampling (AMIS) \citep{CORNUET12}, and more 
%recently, adaptive population importance sampling (APIS) \citep{APIS15}.

%%%%%%%%%%%%%
%%%%%%%%%%%%%
%The RWIS is 

%{\color{black} Table and discussion about the choice of $\Phi_{n,t}$}

%%%%%%%%%%%%%%%
%%%%%%%%%%%%%%%%
\begin{table*}[!htb]
\caption{Example of possible Markov GAMIS algorithms.}
%($J=$ total number of proposals; $L=$  total number of mixtures; $P=$ number of components per mixture).}
\begin{center}
\begin{tabular}{|c|c|c|c|}
\cline{3-4}
\multicolumn{2}{c|}{  } & {\bf Parallel adaptation}  & {\bf Interacting  adaptation}  \\
\hline
\Xhline{3\arrayrulewidth}
 \multirow{2}{*}{{\bf Function $\bm \Phi_{n,t}({\bf x})$}} &  \multirow{2}{*}{ ${\bf N=1}$ }    & \multirow{2}{*}{ ${\bf N>1}$}  &  \multirow{2}{*}{${\bf N>1}$}   \\
& & & \\
\Xhline{3\arrayrulewidth}
%\cline{4-5}
\hline
\hline
 \multirow{2}{*}{ $q_{n,t}({\bf x})$}& RWIS & \multicolumn{2}{c|}{Markov PMC (related to \citep{Cappe04})}    \\
 & (see Table \ref{RWISTable})  &   \multicolumn{2}{c|}{}   \\
\hline
\hline
 \multirow{2}{*}{$\xi_{n}({\bf x})= \frac{1}{T}\sum_{t=1}^T q_{n,t}({\bf x})$} & Markov AMIS  & $N$ parallel  & Population-based \\
% \cline{2-4}
 & (related to \citep{CORNUET12}) & Markov  AMIS (rel. to \citep{CORNUET12}) & Markov  AMIS (rel. to \citep{CORNUET12}) \\
\hline
\hline
 \multirow{2}{*}{$\phi_{t}({\bf x})= \frac{1}{N}\sum_{n=1}^N q_{n,t}({\bf x})$} & RWIS & PI-MAIS  & I$^2$-MAIS \\
 &  (see Table \ref{RWISTable}) & (see Section \ref{sec:MAMIS}) & (see Section \ref{sec:MAMIS}) \\
\hline
\hline
\multirow{2}{*}{$\psi({\bf x})= \frac{1}{NT}\sum_{n=1}^N \sum_{t=1}^T q_{n,t}({\bf x})$} & Markov AMIS &\multicolumn{2}{c|}{Full Markov GAMIS }   \\
 &  (related to \citep{CORNUET12}) &  \multicolumn{2}{c|}{}  \\
\hline
\hline
generic $ \Phi_{n,t}({\bf x})$ & \multicolumn{3}{c|}{Partial Markov GAMIS } \\
\hline
\end{tabular}
\end{center}
\label{FantasticTableOtra}
\end{table*}%

%%%%%%%
\begin{table}[!t]
%	\centering
%\small
\caption{Random Walk Importance Sampling (RWIS) algorithm. }
%\vspace{-1cm}
\begin{tabular}{|p{0.95\columnwidth}|}
    \hline
\footnotesize
\begin{enumerate}
\item {\bf Initialization:} start with $t=1$, $H_0=0$, choose the values $M$ and $T$, the initial location parameter ${\bm\mu}_0$, the scale parameters ${\bf C}$ and ${\bm \Lambda}$.
\item For $t=1,\ldots,T$:
\begin{enumerate}
\item {\bf MH step:} 
\begin{enumerate}
\item[(a-1)] Draw ${\bm \mu}' \sim \varphi({\bm \mu}|{\bm \mu}_{t-1},{\bm \Lambda})$.
\item[(a-2)] Set ${\bm \mu}_{t}={\bm \mu}'$ with probability 
$$
\alpha= \min\left[1,\frac{\pi({\bm \mu}')\varphi({\bm \mu}_{t}|{\bm \mu}',{\bm \Lambda}),}{\pi({\bm \mu}_{t})\varphi({\bm \mu}'|{\bm \mu}_{t-1},{\bm \Lambda})}\right],
$$
otherwise set ${\bm \mu}_{t}={\bm \mu}_{t-1}$ (with probability $1-\alpha$).
\end{enumerate}
\item {\bf IS steps:}
\begin{enumerate}
\item[(b-1)] Draw ${\bf x}_{t}^{(m)}\sim q_{t}({\bf x}|{\bm \mu}_{t},{\bf C}_n)$ for  $m=1,\ldots,M$.
\item[(b-2)] Weight the samples as 
$$
w_{t}^{(m)}=\frac{\pi({\bf x}_{t}^{(m)})}{q_{t}({\bf x}_{t}^{(m)}|{\bm \mu}_{t},{\bf C}_n)}.
$$
\item[(b-3)] Set  $S_t= \sum_{m=1}^M w_{t}^{(m)}$, $H_{t}=H_{t-1}+S_t$, and normalize the weights 
$$
\bar{\rho}_{t}^{(m)}= \frac{w_{t}^{(m)}}{\sum_{\tau=1}^t \sum_{r=1}^M w_{\tau}^{(r)}}=\bar{\rho}_{t-1}^{(m)} \frac{H_{t-1}}{H_t}.
$$
\end{enumerate}
\item {\bf Output:} Return all the pairs $\{{\bf x}_{\tau}^{(m)},\bar{\rho}_{\tau}^{(m)}\}$ for $m=1,\ldots,M$ and $\tau=1,\ldots, t$.
\end{enumerate}
%\item {\bf Stopping rule:}  if $t<T$ then $t=t+1$ and repeat from step 2. Otherwise, if $t\geq T$ then stop.
%\item If $t<T$, set $t=t+1$, and repeat from step 2.  Otherwise, if $t=T$, then stop.
%	The final locations of the Gaussians (i.e., their means, ${\bm \mu}_{i}^{(M)}$ for $i = 1, \ldots, N$) could also be used to estimate the locations of the modes of $\pi({\bf x})$. %and the global maximum of $\pi({\bf x})	
%	(i.e., the MAP estimator of ${\bf x}$), as the proposal converges towards the areas of high probability (i.e., the maxima) of the target.
 \end{enumerate} \\
\hline 
\end{tabular}
\label{RWISTable}
\end{table}
%%%%%%%

%%%%%%%%%%%%%%%%%%%%%%%%%%%%%%%%%%%%%%%%%%%%%%
\subsection{Population-based algorithms} 
%%%%%%%%%%%%%%%%%%%%%%%%%%%%%%%%%%%%%%%%%%%%%%
\label{sec:MAMIS}

The RWIS technique can be easily extended by using a population of $N$ proposal pdfs. In this case, we choose  
$$
\Phi_{n,t}({\bf x})=\phi_{t}({\bf x})=\frac{1}{N}\sum_{n=1}^N q_{n,t}({\bf x}),
$$
so that the computational cost of evaluating the mixture $\Phi_{n,t}({\bf x})=\phi_{t}({\bf x})$ depends only on $N$, regardless of the number $t$ of iterations. Moreover, step 2(c-3) in Table \ref{GAMISgenTable} is not required in this case. Table \ref{PopMAIS} describes the corresponding algorithm without specifying the MCMC approach used for generating the population of means, ${\bf \mathcal{P}}_{t}=\{{\bm \mu}_{1,t},...,{\bm \mu}_{N,t}\}$, given ${\bf \mathcal{P}}_{t-1}$. 

Two possible adaptation procedures via MCMC are discussed below.  In the first one, we consider $N$ independent parallel chains for updating the $N$ mean vectors. We refer to this method as Parallel Interacting Markov Adaptive Importance Sampling (PI-MAIS). Although PI-MAIS is parallelizable, in the iterative version of Table \ref{PopMAIS} the $N$ independent processes cooperate together in Eq. \eqref{eq:ISweights} to provide unique global IS estimate. In the second adaptation scheme, we introduce the interaction also in the upper level. Hence, we refer to this method  as {\it Doubly Interacting Markov Adaptive  Importance Sampling} (I$^2$-MAIS). In both cases, the corresponding technique provides an IS approximation of the target or, equivalently, the estimators $\hat{I}_T$ and $\hat{Z}_T$ in Eq. \eqref{equation_whatever_1}, using $NMT$ samples.  %Figure \ref{figMAIS} depicts a generic sketch of the IS estimation steps in  a PI-MAIS and I$^2$-MAIS.
%drawn from $NMT$ proposal pdfs: $N$ initial proposals chosen by the user, and $N(T-1)$ proposals updated by the algorithm.

%%%%%%%%%%%%%%
\begin{table}[!t]
%	\centering
%\small
\caption{Population-Based MAIS algorithms.}
%\vspace{-1cm}
\begin{tabular}{|p{0.95\columnwidth}|}
    \hline
\footnotesize

\begin{enumerate}
\item {\bf Initialization:} Set $t = 1$, $\hat{I}_0 = 0$ and $H_0 = 0$. Choose the initial population
$$
{\bf \mathcal{P}}_{0}=\{{\bm \mu}_{1,0},...,{\bm \mu}_{N,0}\},
$$ 
and $N$ covariance matrices ${\bf C}_n$ ($n=1,\ldots,N$). Choose  also the parametric form of the $N$ normalized proposals $q_{i,t}$ with parameters ${\bm \mu}_{n,t}$  and ${\bf C}_n$.  Let $T$ be the total number of iterations.   
\item For $t=1,\ldots,T$:
\begin{enumerate}
\item {\bf Update of the location parameters:} Perform one transition of one or more MCMC techniques over the current population,
$${\bf \mathcal{P}}_{t-1}=\{{\bm \mu}_{1,t-1},...,{\bm \mu}_{N,t-1}\},$$ 
obtaining a new population,
$$
{\bf \mathcal{P}}_{t}=\{{\bm \mu}_{1,t},...,{\bm \mu}_{N,t}\}.
$$ 
\item {\bf IS steps:}
\begin{enumerate}
	\item[(b-1)] Draw ${\bf x}_{n,t}^{(m)} \sim q_{n,t}({\bf x}|{\bm \mu}_{n,t},{\bf C}_n)$ for $m=1,\ldots, M$ and $n=1,\ldots,N$.
	\item[(b-2)] Compute the importance weights,
		\begin{equation}
		%\nonumber
			w_{n,t}^{(m)}  = \frac{\pi({\bf x}_{n,t}^{(m)})}{\frac{1}{N} \sum_{k=1}^N q_{k,t}({\bf x}_{n,t}^{(m)} |{\bm \mu}_{k,t},{\bf C}_k)}, 
		\label{eq:ISweights}
\end{equation}
with  $n=1,\ldots,N$, and $m=1,\ldots,M$.
\item[(b-3)] Set  $S_t=\sum_{n=1}^N \sum_{m=1}^M w_{n,t}^{(m)}$, $H_{t}=H_{t-1}+S_t$, and normalize the weights 
\begin{eqnarray}
\bar{\rho}_{n,t}^{(m)}&=& \frac{w_{n,t}^{(m)}}{\sum_{\tau=1}^t \sum_{i=1}^N\sum_{r=1}^M w_{i,\tau}^{(r)}} \nonumber \\
&=&\bar{\rho}_{n,t-1}^{(m)} \frac{H_{t-1}}{H_t}.  \nonumber
\end{eqnarray}
\end{enumerate}
\item {\bf Outputs:} Return all the pairs $\{{\bf x}_{\tau}^{(m)},\bar{\rho}_{\tau}^{(m)}\}$ for $m=1,\ldots,M$ and $\tau=1,\ldots, t$.
\end{enumerate}
\end{enumerate}  \\
\hline 
\end{tabular}
\label{PopMAIS}
\end{table}
%%%%%%%%%%%%%%
 
%\begin{figure}[!htb]
%\centering 
%  \subfigure[]{\includegraphics[width=7cm]{AlterMCMC_2.pdf}}
%  \caption{Sketch of the IS estimation steps (underlying layer in the hierarchical procedure....)  used in  a PI-MAIS and I$^2$-MAIS.  }
%\label{figMAIS}
%\end{figure}

%%%%%%%%%%%%%%%%%%%%%%%%%
\subsubsection{MCMC adaptation for PI-MAIS}
\begin{figure*}[!tb]
\centering 
\centerline{
   \subfigure[For PI-MAIS]{\includegraphics[width=5cm]{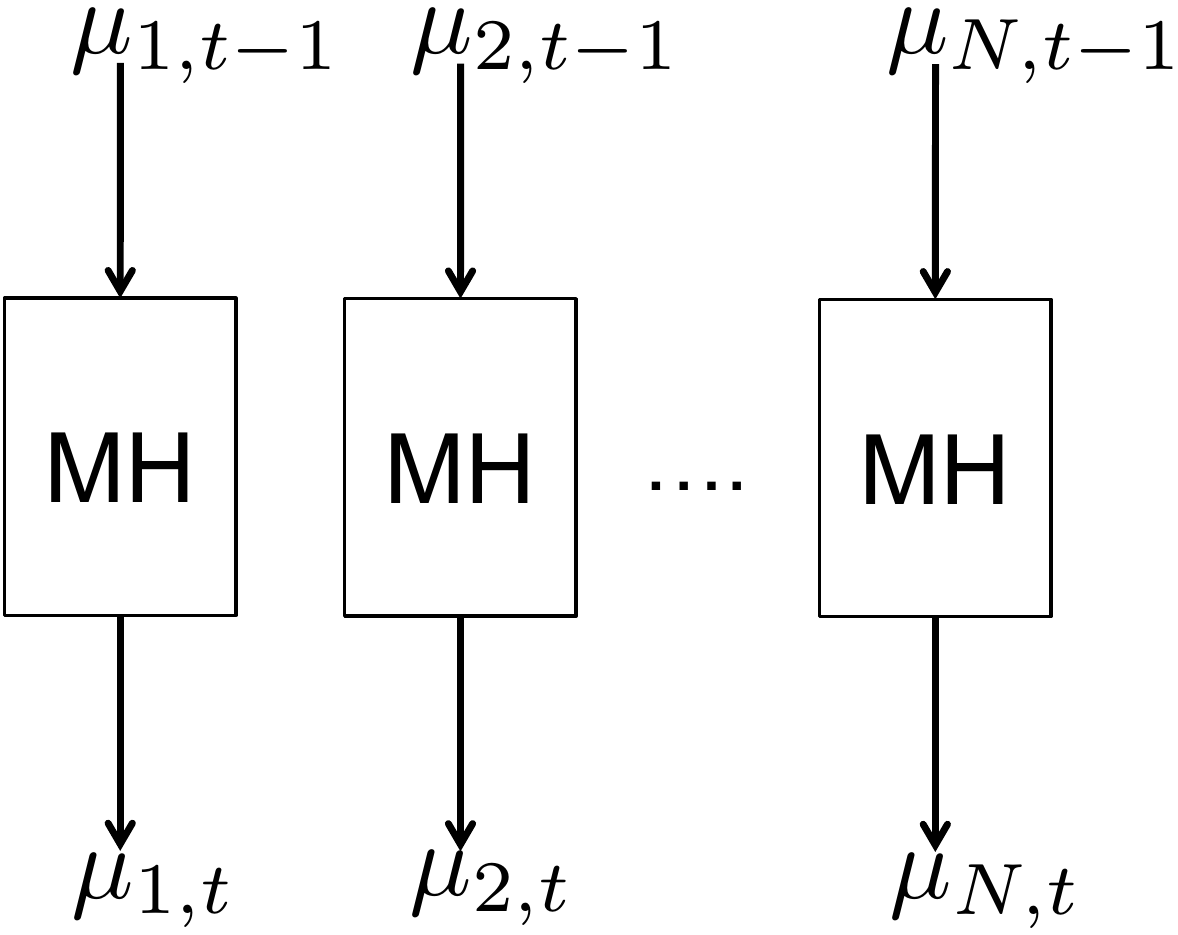}}
  % \subfigure[Possible adapt. in PI-MAIS]{\includegraphics[width=5cm]{AlterMCMC.png}}  
    \hspace{0.3cm}
    \subfigure[For I$^2$-MAIS]{\includegraphics[width=3.7cm]{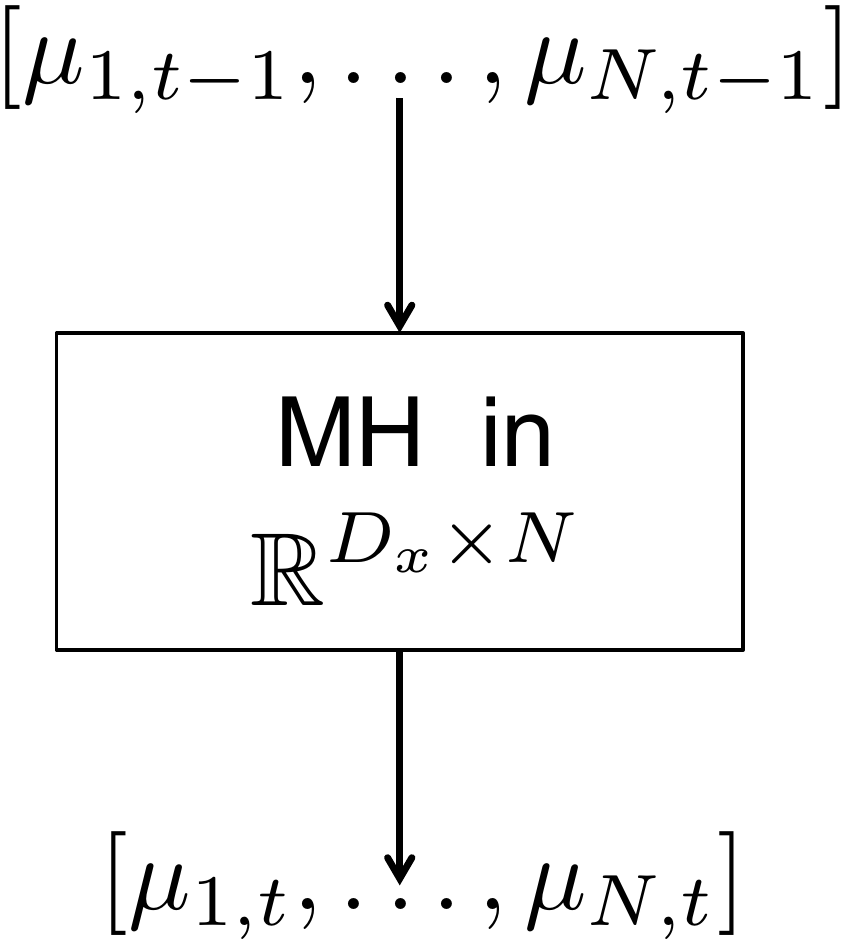}} 
  %  \subfigure[Possible adapt. in I$^2$-MAIS]{\includegraphics[width=3.7cm]{AlterMCMC3.png}}    
   \hspace{0.3cm}
    \subfigure[For I$^2$-MAIS]{\includegraphics[width=3.84cm]{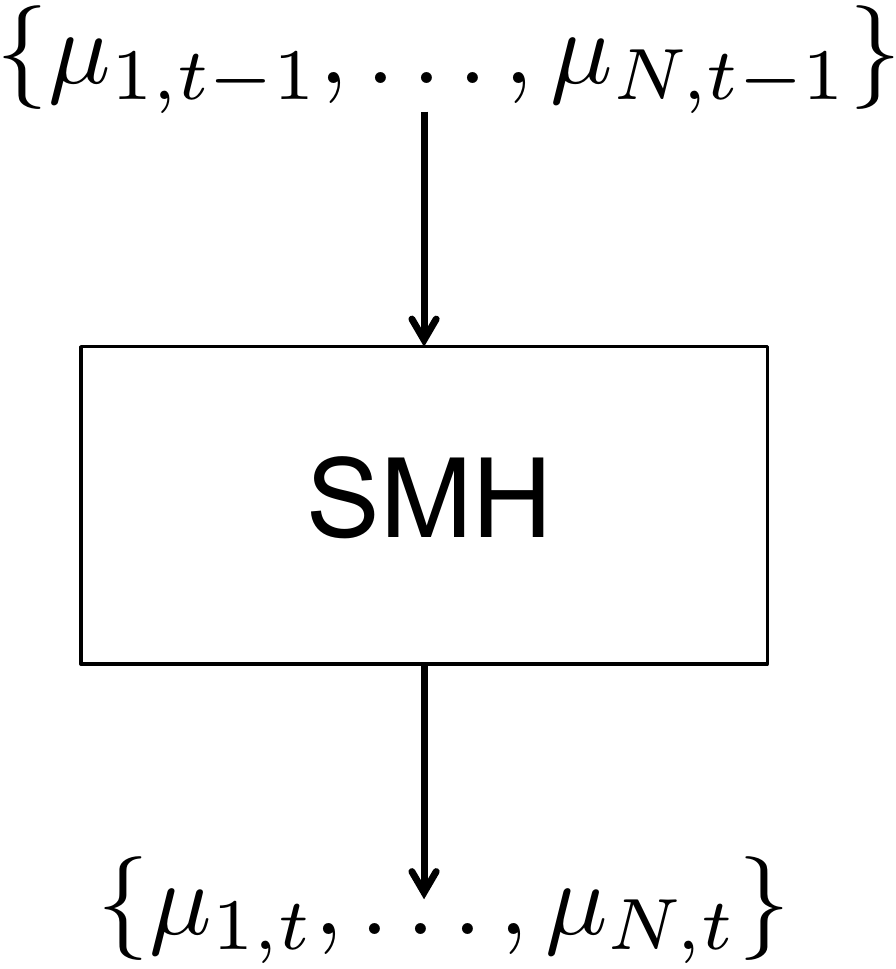}}
       \hspace{0.3cm}
    \subfigure[For I$^2$-MAIS]{\includegraphics[width=2.8cm]{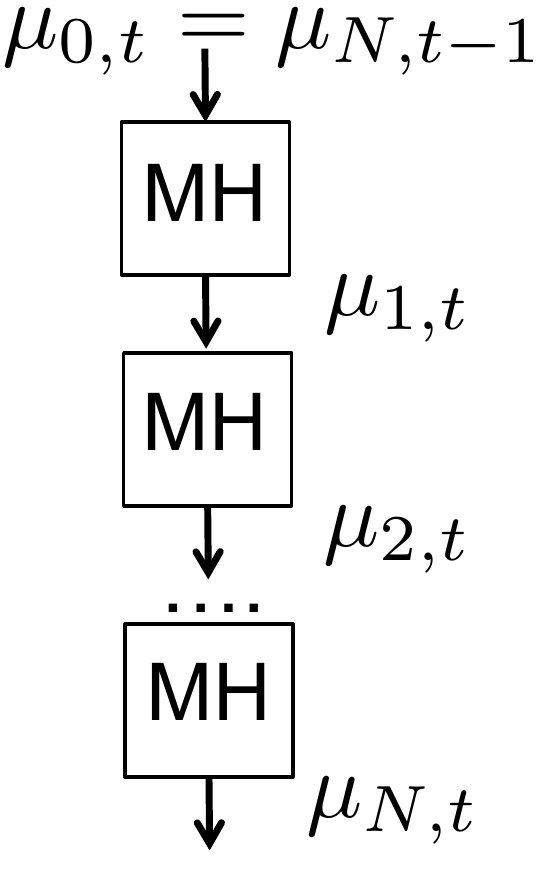}}
   % \subfigure[Possible adapt. in I$^2$-MAIS]{\includegraphics[width=3.84cm]{AlterMCMC4.png}}       
  }
  \caption{Different possible adaptation procedures for Population-based MAIS schemes. {\bf (a)} One transition of $N$ independent parallel MH chains (${\bm \mu}_{n,t}\in\mathbb{R}^{D_x}$) for PI-MAIS. {\bf (b)} One transition of an MH method working in the extended space $[{\bm \mu}_{1,t},\ldots,{\bm \mu}_{N,t}]\in \mathbb{R}^{D_x\times N}$. {\bf (c)} One  transition of SMH \citep[Chapter 5]{Liang10},  considering the population of mean vectors $\mathcal{P}_t=\{{\bm \mu}_{1,t},...,{\bm \mu}_{N,t}\}$. {\bf (d)} $N$ sequential transitions of (possibly) different MH kernels starting from ${\bm \mu}_{0,t}={\bm \mu}_{N,t-1}$. 
}
\label{figMHParallel}
\end{figure*}

%%%%%%%%%%%%%%%%%%%%%%%%%
\label{PI-MAIS}
The simplest option is applying one iteration of $N$ parallel MCMC chains, one for each ${\bm \mu}_{n,t-1}$ returning ${\bm \mu}_{n,t}$, for $n=1,\ldots, N$. For instance, given $N$ parallel MH transitions, each one employing (possibly) a different proposal pdf $\varphi_n$ with covariance matrix ${\bf \Lambda}_n$, we have: 
\newline
\newline
For $n=1,\ldots,N$: 
\begin{enumerate}
\item Draw ${\bm \mu}' \sim \varphi_n({\bm \mu}|{\bm \mu}_{n,t-1},{\bf \Lambda}_n)$.
\item Set ${\bm \mu}_{n,t}={\bm \mu}'$ with probability 
$$
\alpha= \min\left[1,\frac{\pi({\bm\mu}')\varphi_n({\bm \mu}_{n,t-1}|{\bm \mu}',{\bf \Lambda}_n)}{\pi({\bm \mu}_{n,t-1})\varphi_n({\bm \mu}'|{\bm \mu}_{n,t-1},{\bf \Lambda}_n)}\right],
$$
otherwise set ${\bm \mu}_{n,t}={\bm \mu}_{n,t-1}$ (with probability $1-\alpha$).
\end{enumerate}
Figure \ref{figMHParallel}(a) illustrates this scenario. Each mean vector ${\bm \mu}_{n,t}$ is updated independently from the rest.  Therefore, in PI-MAIS, the interaction among the different processes occurs only in the underlying IS layer of the hierarchical structure: the  importance weights in Eq. \eqref{eq:ISweights} are built using the partial DM-MIS strategy with  $\phi_t({\bf x})=\frac{1}{N} \sum_{n=1}^N q_{n,t}({\bf x} |{\bm \mu}_{n,t},{\bf C}_n)$. Considerations about the parallelization of PI-MAIS are given in Section \ref{ImplementSect}. 

%The proposal pdfs can be random walk proposals $\varphi_n({\bm \mu}_{n,t}|{\bm \mu}_{n,t-1},{\bf \Lambda}_n)=\varphi_n({\bm \mu}_{n,t}-{\bm \mu}_{n,t-1}|{\bf \Lambda}_n)$ or independent proposals $\varphi_n({\bm \mu}_{n,t}|{\bm \mu}_{n,t-1},{\bf \Lambda}_n)=\varphi_n({\bm \mu}_{n,t}|{\bf \Lambda}_n)$. Furthermore, $\varphi_{n}$ can easily incorporate gradient information as in the Metropolis adjusted Langevin algorithm (MALA) \citep{Robert04,Liang10}. 

%

 %{\color{black} chiarire della differente varianza....!!! exploration-approximation}

%The proposal pdfs $\varphi_n$ can coincide with $q_{n,t}$ using for the IS estimation, but clearly it is not necessary. 

%{\color{black} in the first layer...is completely parallel...so in a batch vision....
%(note that it could be adaptive and could coincide with $q_{n,t}$, i.e., $\varphi_{n}=q_{n,t}$)
%}

%%%%%%%%%%%%%%%%%%%%%%%%%%%
\subsubsection{MCMC adaptation for I$^2$-MAIS}
%%%%%%%%%%%%%%%%%%%%%%%%%%%
\label{I2-MAIS}

Let us consider an extended state space $\mathbb{R}^{D_x\times N}$ and an extended target pdf
\begin{equation}
\label{GenTarget_Pi_g}
\bar{\pi}_g({\bm \mu}_1,\ldots,{\bm \mu}_N)\propto \prod_{n=1}^{N} \pi({\bm \mu}_n),
\end{equation}
where each marginal $\pi({\bm \mu}_n)$, for $i=1,...,N$, coincides with the target in Eq. \eqref{eq:target}.
In this section, we describe three interacting adaptation procedures for the mean vectors, which consider the generalized pdf in Eq. \eqref{GenTarget_Pi_g} as invariant density.
They are represented graphically in Figs. \ref{figMHParallel}(b), (c) and (d). 
\newline
\newline
{\it  MH in the extended space $\mathbb{R}^{D_x\times N}$}
\newline
\newline
The simplest possibility is applying directly a block-MCMC technique, transitioning from the matrix 
$$
{\bf P}_{t-1}=[{\bm \mu}_{1,t-1},\ldots, {\bm \mu}_{N,t-1}],
$$ 
to the matrix  ${\bf P}_{t}=[{\bm \mu}_{1,t},\ldots, {\bm \mu}_{N,t}]$. Let us consider an MH method and a proposal pdf $\varphi({\bf P}_{t}|{\bf P}_{t-1}): \mathbb{R}^{D_x\times N} \rightarrow \mathbb{R}^{D_x\times N}$. For instance, one can consider a proposal of the type 
\begin{gather}
\begin{split}
\varphi({\bm \mu}_{1,t},\ldots, {\bm \mu}_{N,t}|{\bm \mu}_{1,t-1},\ldots, &{\bm \mu}_{N,t-1}) \nonumber\\
&=\prod_{n=1}^N \varphi_n({\bm \mu}_{n,t}|{\bm \mu}_{n,t-1},{\bf \Lambda}_n). \nonumber
\end{split}
\end{gather}
Thus, one transition is formed by the following steps:
\begin{enumerate}
\item Draw ${\bf P}' \sim \varphi({\bf P}|{\bf P}_{t-1})$, where ${\bf P}'=[{\bm \mu}_{1}',\ldots, {\bm \mu}_{N}']$.
\item Set ${\bf P}_{t}={\bf P}'$ with probability 
$$
\alpha= \min\left[1,\frac{\pi_g({\bf P}')\varphi({\bf P}_{t-1}|{\bf P}')}{\pi_g({\bf P}_{t-1})\varphi({\bf P}'|{\bf P}_{t-1})}\right],
$$
otherwise set ${\bf P}_{t}={\bf P}_{t-1}$ (with probability $1-\alpha$).
\end{enumerate}
At each iteration, $N$ new samples ${\bm \mu}_{n}'$ are drawn (as in PI-MAIS) and therefore $N$ new evaluations of $\pi$ are required (i.e., one evaluation of $\pi_g$).
When a new  ${\bf P}'$ is accepted, all the components of ${\bf P}_{t}$ differ from ${\bf P}_{t-1}$, unlike in the strategy described later. However, the probability of accepting a new population becomes very small for large values of $N$.
\newline
\newline
{\it  Sample Metropolis-Hastings (SMH) algorithm}
\newline
\newline
 SMH  is a population-based MCMC technique, suitable for our purposes  \citep[Chapter 5]{Liang10}. At each iteration $t$, given the previous set 
$$
{\bf \mathcal{P}}_{t-1}=\{{\bm \mu}_{1,t-1},...,{\bm \mu}_{N,t-1}\}, 
$$
 a new possible parameter ${\bm \mu}_{0,t-1}$, drawn from an independent proposal $\varphi({\bm \mu})$, is tested to be interchanged with another parameter in ${\bf \mathcal{P}}_{t-1}=\{{\bm \mu}_{1,t-1},...,{\bm \mu}_{N,t-1}\}$. The underlying idea of SMH is to replace one ``bad'' sample in the population ${\bf \mathcal{P}}_{t-1}$ with a potentially ``better'' one, according to a certain suitable probability $\alpha$.
The algorithm is designed so that, after a burn-in period, the elements in ${\bf \mathcal{P}}_{t}$ are distributed according to ${\bar \pi}_g({\bm \mu}_{1},\ldots,{\bm \mu}_{N})$. One iteration of SMH consists of the following steps:
\begin{enumerate}
	\item Draw a candidate ${\bm \mu}_{0,t-1}\sim \varphi({\bm \mu})$.
	\item Choose a ``bad'' sample, ${\bm \mu}_{k,t-1}$ with $k\in \{1,...,N\}$, from the population according to a probability
		proportional to $\frac{\varphi({\bm \mu}_{k,t-1})}{\pi({\bm \mu}_{k,t-1})}$, which corresponds to the inverse of the standard IS weights.
	\item Accept the new population, $\mathcal{P}_{t} = \{{\bm \mu}_{1,t},\ldots,{\bm \mu}_{N,t} \}$ with ${\bm \mu}_{n,t}= {\bm \mu}_{n,t-1}$ for all $n\neq k$ and ${\bm \mu}_{k,t}= {\bm \mu}_{0,t-1}$,		
	with probability
		\begin{equation}
		\nonumber
			\alpha(\mathcal{P}_{t-1},{\bm \mu}_{0,t-1})=\frac{\sum_{n=1}^{N} \frac{\varphi({\bm \mu}_{n,t-1})}
				{\pi({\bm \mu}_{n,t-1})}}{\sum_{i=0}^{N} \frac{\varphi({\bm \mu}_{i,t-1})}{\pi({\bm \mu}_{i,t-1})}
				-\min\limits_{0\leq i\leq N}\frac{\varphi({\bm \mu}_{i,t-1})}{\pi({\bm \mu}_{i,t-1})}}.
		%\label{AlfaSMH}
		\end{equation}
		Otherwise, set ${\bf \mathcal{P}}_{t}={\bf \mathcal{P}}_{t-1}$.
\end{enumerate}
Unlike in the previous strategy, the difference between ${\bf \mathcal{P}}_{t-1}$ and ${\bf \mathcal{P}}_{t}$ is at most one sample. Observe that $\alpha$ depends on $\mathcal{P}_{t-1}$ and the candidate ${\bm \mu}_{0,t-1}$. However, at each iteration, only one new evaluation of $\pi$ (and $\varphi$) is needed at ${\bm \mu}_{0,t-1}$, since the rest of the weights have already been computed in the previous steps (except for the initial iteration).
\newline
\newline
{\it  MH within Gibbs}
\newline
\newline
Another simple alternative, following an ``MH within Gibbs'' approach for sampling from ${\bar \pi}_g$, is to update sequentially each ${\bm \mu}_{n,t-1}$ using one MH step in $\mathbb{R}^{D_x}$. Hence, setting ${\bm \mu}_{0,t}={\bm \mu}_{N,t-1}$, we have:   
\newline
\newline
For $n=1,\ldots,N$: 
\begin{enumerate}
\item Draw ${\bm \mu}'$ from a proposal pdf $\varphi_n({\bm \mu}|{\bm \mu}_{n-1,t},{\bf \Lambda}_n)$.
\item Set ${\bm \mu}_{n,t}={\bm \mu}'$ with probability 
$$
\alpha= \min\left[1,\frac{\pi({\bm \mu}')\varphi_n({\bm \mu}_{n-1,t}|{\bm \mu}',{\bf \Lambda}_n)}{\pi({\bm \mu}_{n-1,t})\varphi_n({\bm \mu}'|{\bm \mu}_{n-1,t},{\bf \Lambda}_n)}\right],
$$
otherwise set ${\bm \mu}_{n,t}={\bm \mu}_{n-1,t}$.
\end{enumerate}
This scenario is illustrated in Fig. \ref{figMHParallel}(d). In this case, after $T$iterations of the I$^2$-MAIS scheme, we generate a unique MH chain with $NT$ total states, divided in $T$ parts of $N$ states. At each iteration of the I$^2$-MAIS scheme, each block of $N$ states is employed as mean vector of the $N$ proposal pdfs used in the lower level. %of the hierarchical procedure. %If the function $\Phi_{n,t}({\bf x})$ is chosen as $\phi_t({\bf x})$, in this case the resulting technique coincides an AMIS-type algorithm \cite{CORNUET12}  with an MCMC adaptation and applying a partial DM strategy.
%\newline
%\newline
%{\it  Final remarks}
%\newline
%\newline
%In all cases above, if $\varphi$'s are not properly chosen, the population of parameters hardly changes. 
%However, the diversity in the population is preserved, unlike in the resampling procedure, typically applied in several multiple AIS approaches \citep{Cappe04,Robert04}. %However, the parameters of  $\varphi$'s can be updated using an adaptive scheme \citep{Haario2001,Luengo13} in order to avoid this issue (see also Section \ref{AdaptSect}). 
%Techniques like the Normal Kernel Coupler \citep{Warnes01} are other possible alternatives to the use of SMH.

%%%%%%%%%%%%%%%%%%%%%%%%%%%%%%%%%%%%%%%%%%%%%%
\subsection{Computational cost: comparison between PI-MAIS and I$^2$-MAIS} 
%%%%%%%%%%%%%%%%%%%%%%%%%%%%%%%%%%%%%%%%%%%%%%
\label{CompCostComp}
%In the previously described algorithms, the samples generated by the Markov chain are not used for the estimation, but only for updating the mean vectors. 
In all cases, the total number of samples involved in the final estimation is $NMT$.  The total number of evaluations of the target, $E$, is larger due to the MCMC implementation, i.e., $E>NMT$. More precisely, the total number of evaluations of the target is:
\begin{itemize}
\item  $E=MNT+NT$, for PI-MAIS,
\item $E=MNT+NT$, for   I$^2$-MAIS with MH in the extended space $\mathbb{R}^{D_X\times N}$,
\item $E=MNT+T$, for I$^2$-MAIS with SMH,
\item $E=MNT+NT$, for   I$^2$-MAIS with the MH-within-Gibbs approach.
\end{itemize}
Note that we have taken into account that several evaluations of the target have been computed in the previous iterations. Moreover, the application of the  MCMC techniques  requires generation of $V$ additional uniform r.v.'s for performing the acceptance tests (and additional r.v.'s for choosing a ``bad'' candidate in SMH). Specifically, we need: $V=NT$ uniform r.v.'s in PI-MAIS and I$^2$-MAIS with MH-within-Gibbs, $V=T$ uniform r.v.'s for  I$^2$-MAIS with MH in the extended space, and$V=2T$, $T$ uniform r.v. and $T$ multinomial r.v., for I$^2$-MAIS with SMH. However, in practical applications, the main computational effort is usually required for the target evaluation. The computing time required in the multinomial sampling within SMH increases with $N$. 
Finally, we recall that  we have used a deterministic mixture weighting scheme with $\Phi_{n,t}({\bf x})=\phi_t({\bf x})$, which requires $MN^2T$ evaluations of the proposal pdfs, $q_{n,t}({\bf x})$, for $n=1,\ldots,N$ and $t=1,\ldots,T$.

\subsection{Non-iterative and parallel implementations} 
%%%%%%%%%%%%%%%%%%%%%%%%%%%%%%%%%%%%%%%%%%%%%%
\label{ImplementSect}

As remarked in Section \ref{TeoOtra}, the choice of the means ${\bm \mu}_{n,t}$'s is completely independent from the estimation steps. Thus, all the means can be selected in advance (also in parallel if the strategy in Section \ref{PI-MAIS} is used), and the MIS estimation steps can then be performed as in a completely static framework (i.e., as described in Section \ref{LOCURASect2_1}). This consideration is valid for any choice of $\Phi_{n,t}({\bf x})$.

Let us consider now the choice of $\Phi_{n,t}$'s as temporal mixtures, i.e.,  
   $\Phi_{n,t}=\frac{1}{T}\sum_{t=1}^T q_{n,t}({\bf x})$ or $\Phi_{n,t}({\bf x})=q_{n,t}({\bf x})$. Moreover, let us consider the use of $N$ parallel MCMC chains for adapting the means, i.e., one independent chain for each parameter ${\bm \mu}_{n,t}$, with $n=1,\ldots,N$. In this case, the corresponding algorithm is completely parallelizable. Indeed, it can be decomposed into $N$ parallel MAIS techniques, each one producing the  partial estimators ${\hat I}_{n,T}$ and ${\hat Z}_{n,T}$, after $T$ iterations. The global estimators are then given by   
 \begin{gather}
  \begin{split}
 {\hat I}_{T}&=\sum_{n=1}^N \frac{{\hat Z}_{n,T}}{\sum_{i=1}^N{\hat Z}_{i,T}} {\hat I}_{n,T}, \\
   {\hat Z}_{T}&=\frac{1}{N}\sum_{n=1}^N {\hat Z}_{n,T}.
  \end{split}
 \end{gather}
%Furthermore, if a unique independent proposal (i.e., $\varphi_n({\bm \mu})=\varphi({\bm \mu})$ for all $n$) is employed in the parallel adaptation of Section \ref{PI-MAIS}, the block independent Metropolis technique \cite{Jacob11} can be applied for reducing the total number of evaluations of the target.  
Furthermore, different strategies for sharing information among the parallel chains can also be applied \citep{Craiu09,O-MCMC,Smelly,Geyer91,Marinari92,Neal01}, or for reducing the total number of evaluations of the target \cite{Jacob11} (the scheme in \cite{Jacob11} can be applied if a unique independent proposal is employed, i.e., $\varphi_n({\bm \mu})=\varphi({\bm \mu})$ for all $n$).

% introducing interaction at the upper level, i.e., among the chains (hence, in this case PI-MAIS becomes an I$^2$-MAIS scheme). Moreover, a tempered version \cite{Geyer91,Marinari92,Neal01} of the generalized target ${\bar \pi}_g$ in Eq. \eqref{GenTarget_Pi_g} can be also considered:
%\begin{equation}
%\label{GenTarget_Pi_g2}
%\bar{\pi}_g({\bm \mu}_1,\ldots,{\bm \mu}_N)\propto \prod_{n=1}^{N} [\pi({\bm \mu}_n)]^{\beta_n},
%\end{equation} 
%where $0<\beta_n\leq 1$ for $n=1,\ldots,N$, so that each chain converges to a different ``modified'' target function. This is equivalent to consider different priors in the upper level hierarchical scheme.  

%%%%%%%%%%%%%%%%%%%%%
\section{Numerical simulations}
%%%%%%%%%%%%%%%%%%%%%
\label{SIMUsect}
In this section, we test the performance of the proposed scheme comparing them with other benchmark techniques. First of all,  we tackle two challenging issues for adaptive Monte Carlo methods: multimodality in Section \ref{sec:sims1} and nonlinearity in Section \ref{BananaExSect}. Furthermore, in Section \ref{LocSect} we consider an application of positioning and tuning model parameters in a wireless sensor network \citep{Ali07,Ihler05,MartinoStatCo10}.

%%%%%%%%%%%%%%%%%%%%%
\subsection{Multimodal target distribution}
%%%%%%%%%%%%%%%%%%%%%
\label{sec:sims1}

In this section, we test the novel proposed algorithms in a multimodal scenario, comparing with several other methods. Specifically, we consider a bivariate multimodal target pdf, which is itself a mixture of $5$ Gaussians, i.e., 
\begin{equation}
\label{Target1}
\bar{\pi}({\bf x})=\frac{1}{5}\sum_{i=1}^5 \mathcal{N}({\bf x};{\bf \nu}_i,{\bf \Sigma}_i), \quad {\bf x}\in \mathbb{R}^2,
\end{equation}
with means ${\bf \nu}_1=[-10, -10]^{\top}$, ${\bf \nu}_2=[0, 16]^{\top}$, ${\bf \nu}_3=[13, 8]^{\top}$, ${\bf \nu}_4=[-9, 7]^{\top}$, ${\bf \nu}_5=[14, -14]^{\top}$, and covariance matrices ${\bf \Sigma}_1=[2, \ 0.6; 0.6, \ 1]$, ${\bf \Sigma}_2=[2, \ -0.4; -0.4, \ 2]$, ${\bf \Sigma}_3=[2, \ 0.8; 0.8, \ 2]$, ${\bf \Sigma}_4=[3, \ 0; 0, \ 0.5]$ and ${\bf \Sigma}_5=[2, \ -0.1; -0.1, \ 2]$. The main challenge in this example is  the ability in discovering the $5$ different modes of $\bar{\pi}({\bf x})\propto \pi({\bf x})$. Since we know the moments of $\pi(\x)$, we can easily assess the performance of the different techniques.

Given a random variable (r.v.) ${\bf X}\sim {\bar \pi}({\bf x})$, we consider the problem of approximating via Monte Carlo  the expected value $E[{\bf X}]=[1.6, 1.4]^{\top}$ and the normalizing constant $Z=1$. Note that an adequate approximation of $Z$ requires the ability of  learning about all the $5$ modes. We compare the performances of different sampling algorithms in terms of  Mean Square Error (MSE):
{\bf (a)} the AMIS technique \citep{CORNUET12}, {\bf (b)} three different PMC schemes\footnote{The standard PMC method \citep{Cappe04} is described in Section \ref{PMClocura2}.}, two of them proposed in \citep{pmc-cappe08,Cappe04} and one PMC using a partial DM-MIS scheme with $\Phi_{n,t}({\bf x})=\phi_t({\bf x})$, {\bf (c)} $N$ parallel independent MCMC chains and {\bf (d)} the proposed PI-MAIS method. 
Moreover, we test two static MIS approaches, the standard MIS and a partial DM-MIS schemes with $\Phi_{n,t}({\bf x})=\phi_t({\bf x})$, computing iteratively the final estimator. %Thus, we are able to compare (a) these two static MIS schemes showing the advantages of the DM strategy, and (b) to see  the benefit of  Markov adaptation with respect to the static approach. 

\begin{figure*}[htb]
\centering 
\centerline{
\subfigure[PMC ($N=100$, $\sigma=3$)]{\includegraphics[width=4.4cm]{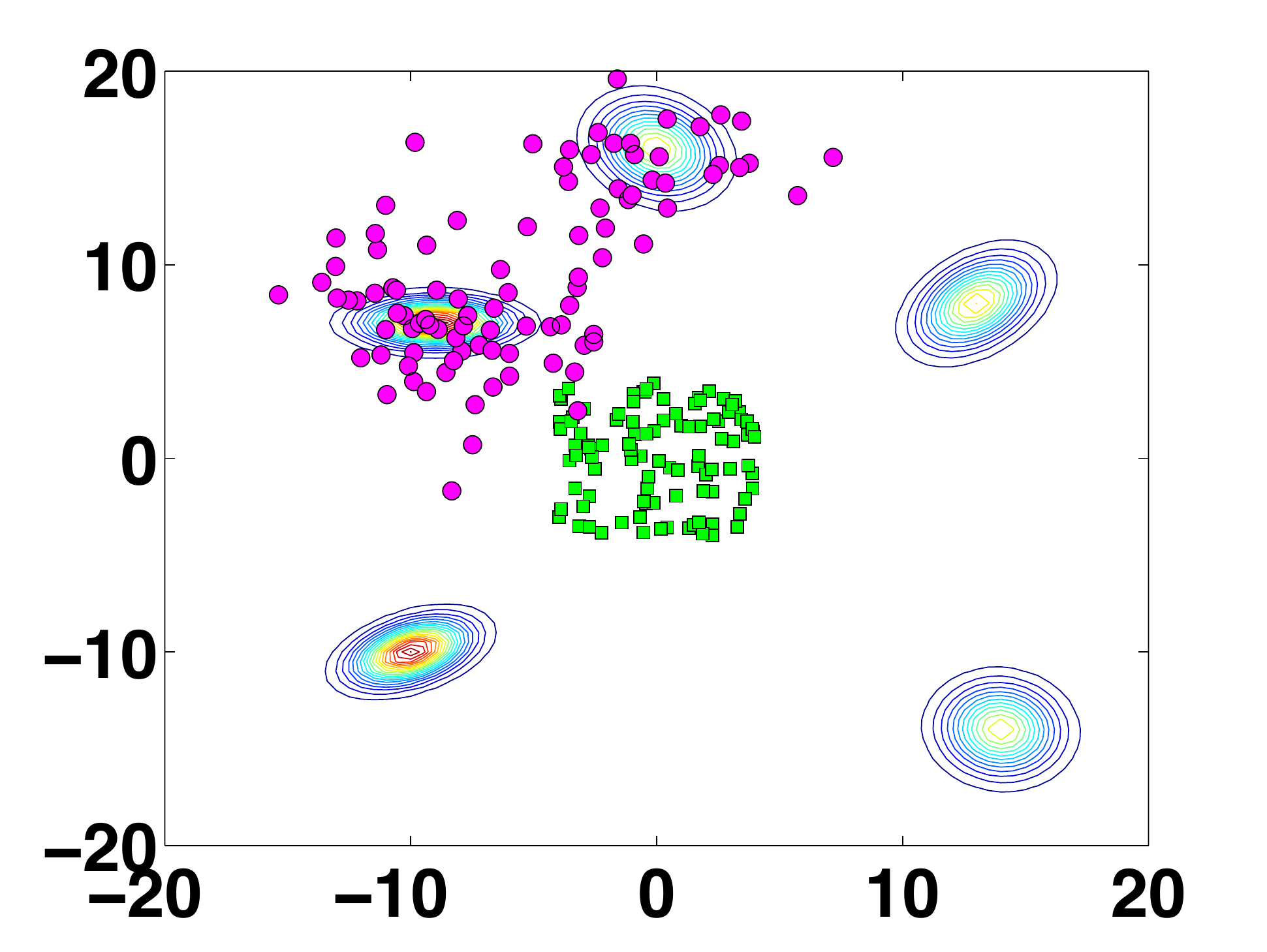}}
\subfigure[PMC ($N=100$, $\sigma=5$)]{\includegraphics[width=4.4cm]{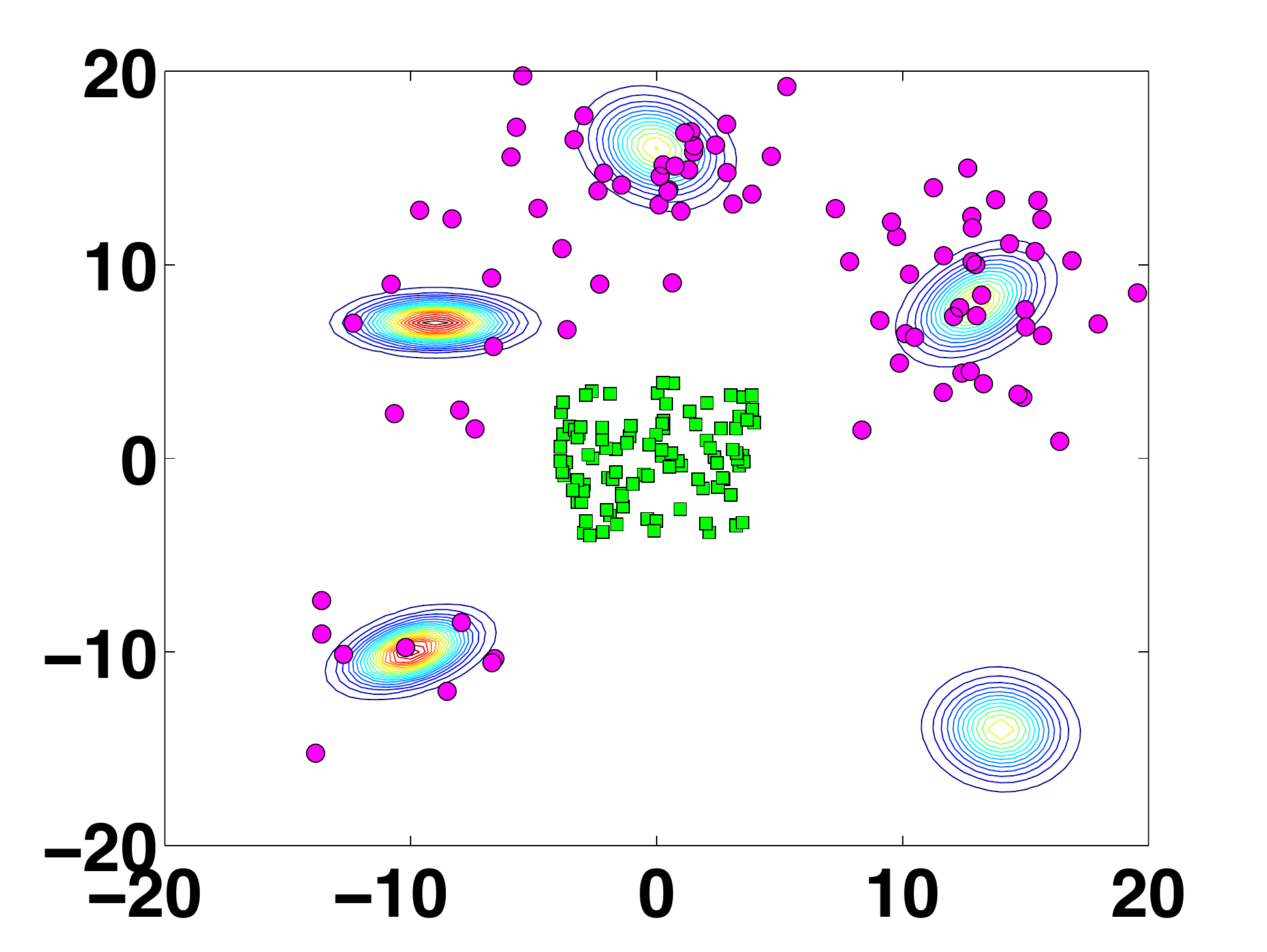}}
\subfigure[PI-MAIS ($N=100$, $\lambda=3$)]{\includegraphics[width=4.4cm]{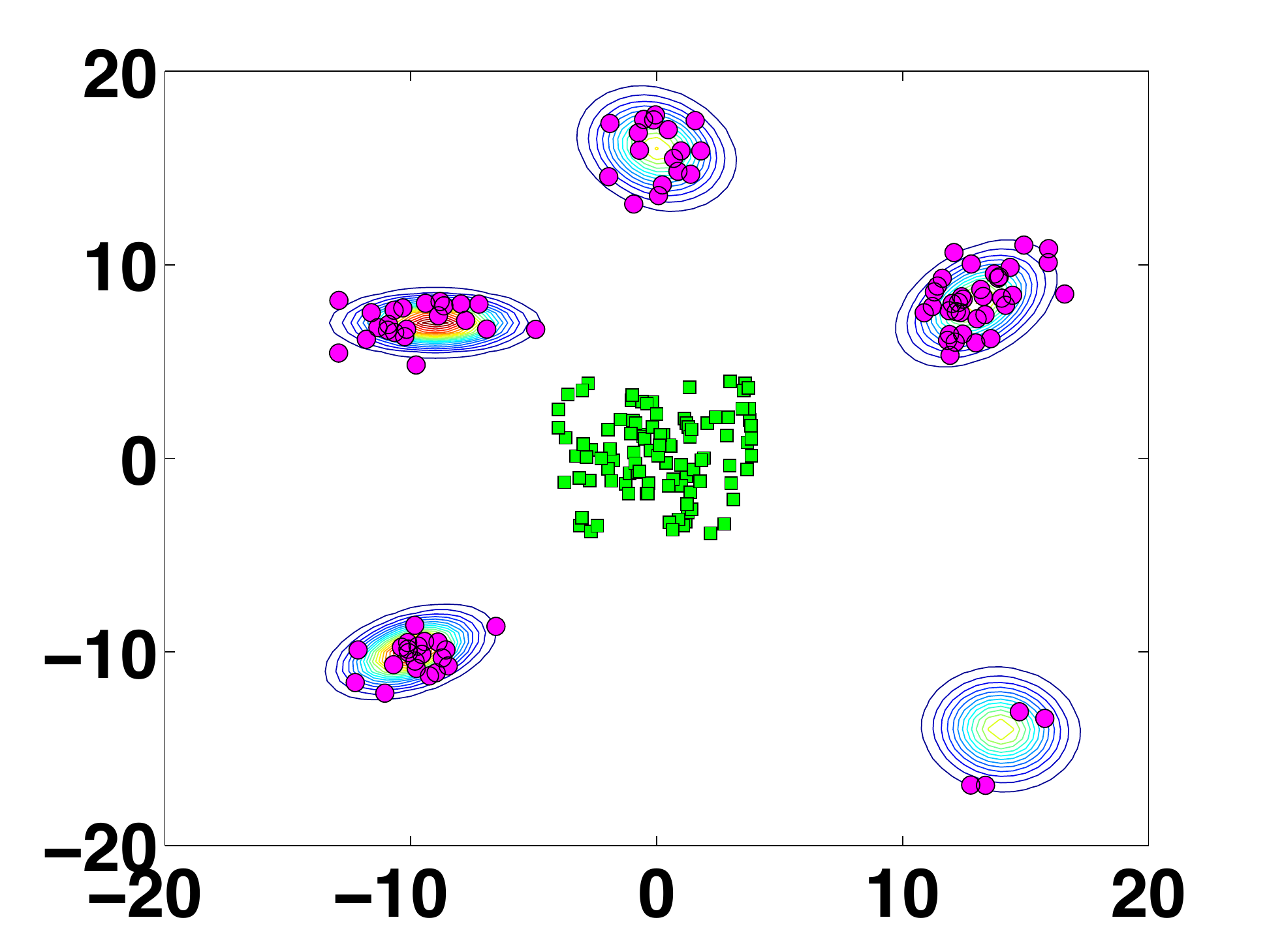}}
\subfigure[PI-MAIS ($N=100$, $\lambda=5$)]{\includegraphics[width=4.4cm]{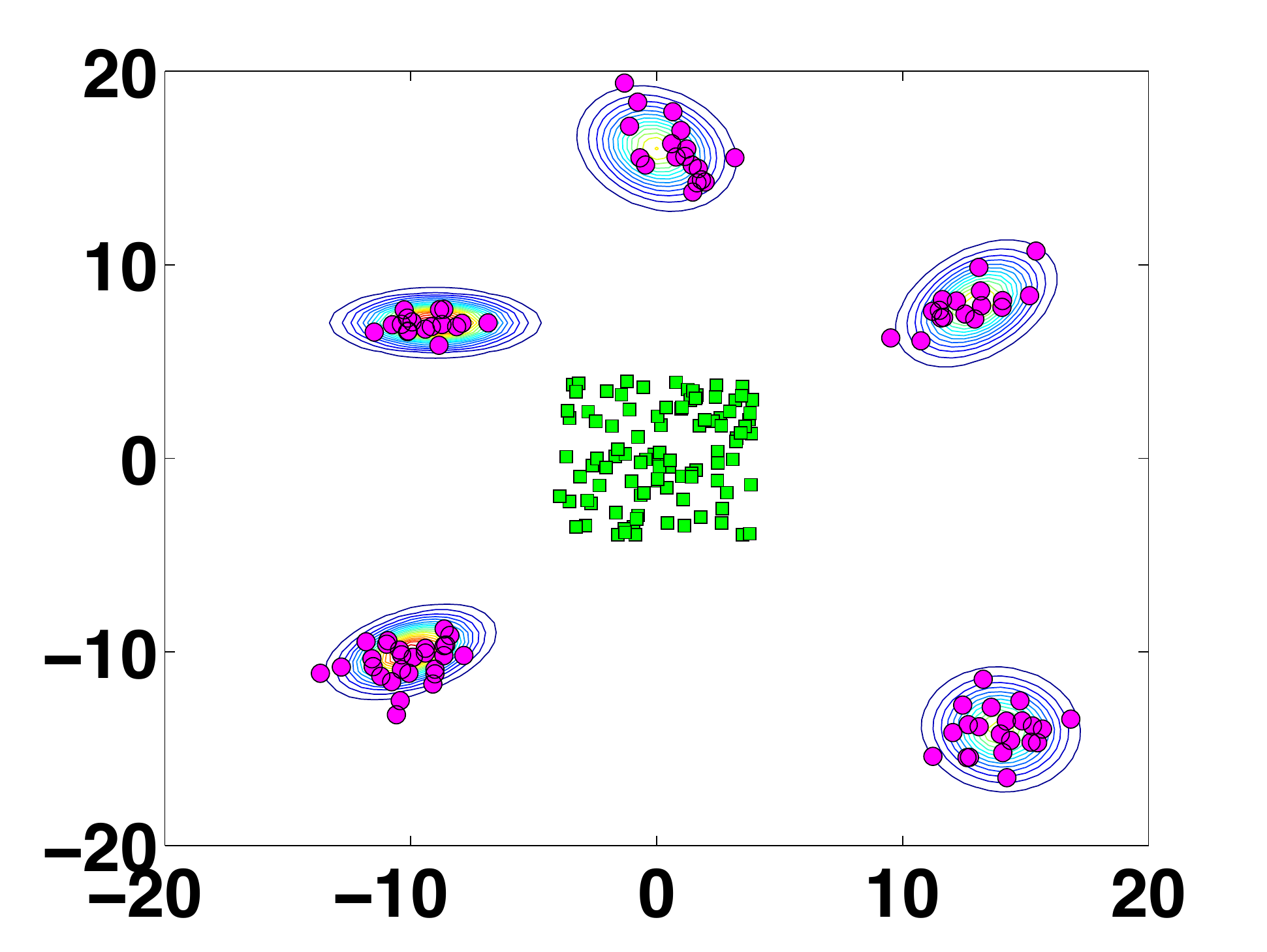}}
}
  \caption{%{\bf (Ex-Section-\ref{BananaExSect})}   
   Initial (squares) and final (circles) configurations of the mean vectors of the proposal densities for the standard PMC and the  PI-MAIS methods, in different specific runs. The initial configuration corresponds to {\bf In1}.  }
\label{figEx0}
\end{figure*}

For a fair comparison, all the mentioned algorithms have been implemented in such a way that the number of total evaluations of the target is $E=2\cdot 10^5$. %Namely, we test different combinations of the parameters in the different schemes maintaining fixed $E=2\cdot 10^5$. %\footnote{All the compared techniques require different other operations beyond the target evaluation. However, this is in general the more costly step.  } 
 All the involved proposal densities  are Gaussian pdfs.    
More specifically, in PI-MAIS,  we use the following parameters: $N=100$, $M\in\{1,19, 99\}$, $T\in\{20,100,1000\}$ in order to fulfill $E=MNT+NT=(M+1)NT=2\cdot 10^5$ (see Section \ref{CompCostComp}). The proposal densities of the upper level of the hierarchical approach, $\varphi_{n}(\x|{\bm \mu}_{n,t},{\bf \Lambda}_{n})$, are Gaussian pdfs with covariance matrices ${\bf \Lambda}_{n}=\lambda^2 {\bf I}_2$ and  $\lambda\in \{5,10,70\}$. The proposal densities used in the lower importance sampling level, $q_{n,t}(\x|{\bm \mu}_{n,t},{\bf C}_{n})$ are Gaussian pdfs with covariance matrices ${\bf C}_{n}=\sigma^2 {\bf I}_2$ and  $\sigma\in \{0.5,1,2,5,10,20,70\}$. We also try different non-isotropic diagonal covariance matrices in both levels, i.e, ${\bf \Lambda}_n = \textrm{diag}(\lambda_{n,1}^2, \lambda_{n,2}^2)$, where $\lambda_{i,j} \sim \mathcal{U}([1,10])$, and ${\bf C}_n = \textrm{diag}(\sigma_{n,1}^2, \sigma_{n,2}^2)$, where $\sigma_{n,j} \sim \mathcal{U}([1,10])$ for $j\in\{1,2\}$ and $n=1,\ldots N$.
We test all these techniques using two different initializations:  first, we choose deliberately a ``bad'' initialization of the initial mean vectors, denoted as {\bf In1}, in the sense that the  initialization region does not contain the modes of $\pi$. Thus, we can test the robustness of the algorithms and their ability to improve the corresponding {\it static} approaches. Specifically, the initial mean vectors are selected uniformly within the following square
$${\bm \mu}_{n,0}\sim \mathcal{U}([-4,4]\times[-4,4]),$$
for $n=1,\ldots,N$. Different examples of this configuration are shown in Fig. \ref{figEx0} with squares. Secondly, we also consider a better initialization, denoted as {\bf In2},  where the  initialization region contains all the modes. Specifically, the initial mean vectors are selected uniformly within the following square
$${\bm \mu}_{n,0}\sim \mathcal{U}([-20,20]\times[-20,20]),$$
for $n=1,\ldots,N$. 
All the results are averaged over $2\cdot 10^3$ independent experiments. Tables \ref{mean_gaussian_mixture_bad} and \ref{mean_gaussian_mixture_good} show the Mean Square Error (MSE) in the estimation of the first component of $E[{\bf X}]$, with the initialization {\bf In1} and  {\bf In2} respectively.  Table \ref{Z_gaussian_mixture_bad} provides the MSE in the estimation of $Z$ with {\bf In1}. The best results in each column are highlighted in bold-face.  In AMIS \citep{CORNUET12}, the mean vector and the covariance matrix of a single proposal (i.e., $N=1$)  are adapted, using  $\Phi_{1,t}({\bf x})=\xi_1({\bf x})$ in the computation of the IS weights. Hence, in AMIS, we have tested different values of samples per iterations $M\in \{500, 10^3 , 2 \cdot 10^3 , 5 \cdot 10^3 , 10^4\}$ and $T=\frac{E}{M}$.  For the sake of simplicity, we  directly show the worst and best results among the several simulations made with different parameters.
PI-MAIS outperforms the other algorithms virtually for all the choices of the parameters, with both initializations. In general, a greater value of $T$ is needed since the proposal pdfs are initially bad localized. Moreover, PI-MAIS always improves the performance of the static approaches. 
 These two consideration show the benefit of the Markov adaptation. Hence, PI-MAIS presents more robustness with respect to the initial values and the choice of the covariance matrices. Figure \ref{figSIMUafterR1_2}(a) providing a summary of the results in Table \ref{mean_gaussian_mixture_bad} showing the $\log(\mbox{MSE})$ as function of the $\log(\sigma)$, for the main compared methods.
Figure \ref{figEx0} depicts the initial (squares) and final (circles) configurations of the mean vectors of the proposal densities for the standard PMC and the  PI-MAIS methods, in a specific run and different values of $\sigma,\lambda\in\{3,5\}$.  In both cases,  PI-MAIS guarantees a better covering of the modes of $\pi({\bf x})$.

%%%%%%%%%%%%%%%%%%%%%%%%%%%%%%%
\subsection{Nonlinear banana-shaped  target distribution}
%%%%%%%%%%%%%%%%%%%%%%%%%%%%%%%%
\label{BananaExSect}

Here we consider a bi-dimensional ``banana-shaped'' target distribution \citep{Haario2001}, which is a benchmark function in the literature due to its nonlinear nature. Mathematically, it is  expressed  as 
{\footnotesize
$$
{\bar \pi}(x_1,x_2)\propto \exp\left(-\frac{1}{2\eta_1^2}\left(4-Bx_1-x_2^2\right)^2-\frac{x_1^2}{2\eta_2^2}-\frac{x_2^2}{2\eta_3^2}\right),
$$
}
where, we have set $B=10$, $\eta_1=4$, $\eta_2=5$, and $\eta_3=5$. The goal is to estimate the expected value $E[{X}]$, where ${X}=[X_1,X_2]\sim {\bar \pi}(x_1,x_2)$, by applying different Monte Carlo approximations. We approximately compute the true value $E[{X}]\approx[-0.4845,0]^{\top}$ using an exhaustive deterministic numerical method (with an extremely thin grid), in order to obtain the mean square error (MSE) of the following methods: standard PMC \citep{Cappe04}, the Mixture PMC \citep{pmc-cappe08}, the AMIS \citep{CORNUET12}, PI-MAIS and I$^2$-MAIS with SMH adaptation.

We consider Gaussian proposal distributions for all the algorithms. The initialization has been performed by randomly drawing the parameters of the Gaussians, with the mean of the $n$-th proposal given by 
${\bm \mu}_{n,0} \sim \mathcal{U}([-6,-3]\times[-4,4])$,
 and its covariance matrix given by ${\bf C}_{n} = [\sigma_{n,1}^2 \quad 0; \ 0  \quad \sigma_{n,2}^2 ]^{\top}$.
We have considered two cases: an {isotropic setting where $\sigma_{n,k} \in \{1,2,\ldots,10\}$ with $k=1,2$, and an anisotropic case with random selection of the parameters  where $\sigma_{n,k} \sim \mathcal{U}([1,20])$, with $k=1,2$. Recall that  in AMIS and Mixture PMC, the covariance matrices are also adapted. 

For each algorithm, we test several combinations of parameters, keeping fixed the total number of target evaluations, $E=2 \cdot 10^5$. In the standard PMC method, described in Section \ref{PMClocura2}), we consider  $N \in \{50, 100, 10^3, 5\cdot 10^3\}$ and $T=\frac{E}{N}$ (here $M=1$). In Mixture PMC, we consider different number of component in the mixture proposal pdf $N\in \{10, 50, 100\}$, and different samples per proposal $S\in\{100, 200, 10^3,  2 \cdot 10^3,  5 \cdot 10^3\}$ at each iteration (here $T=\frac{E}{S}$). In AMIS, we test $S\in \{500, 10^3, 2 \cdot 10^3, 5 \cdot 10^3, 10^4\}$ and $T=\frac{E}{S}$ (we recall $N=1$). The range of these values of parameters are chosen, after a preliminary study, in order to obtain the best performance from each technique. In PI-MAIS an I$^2$-MAIS,  we set $N\in \{50,100\}$. For the adaptation in PI-MAIS, we also consider Gaussian pdfs $\varphi_{n}(\x|{\bm \mu}_{n,t},{\bf \Lambda}_{n})$,  covariance matrices ${\bf \Lambda}_{n}=\lambda^2 {\bf I}_2$ with $\lambda\in \{3,5,10,20\}$. In  I$^2$-MAIS, for the SMH method we use a Gaussian pdf with mean $[0,0]^{\top}$ and covariance matrix  ${\bf \Lambda}=\lambda^2 {\bf I}_2$ and again $\lambda\in \{3,5,10,20\}$. We test $M\in\{1,9,19\}$ for both, so that $T=\frac{E}{N(M+1)}$ for PI-MAIS and $T=\lfloor\frac{E}{NM+1}\rfloor$ for I$^2$-MAIS (see Section \ref{CompCostComp}).

The results are averaged $500$ over independent simulations, for each combination of parameters. Table \ref{TableResults2} shows the smallest and highest MSE values obtained in the estimation of the expected value of the target, averaged between the two components of $E[{X}]$, achieved by the different methods. The smallest MSEs in each column (each $\sigma$) are highlighted in bold-face. PI-MAIS and I$^2$-MAIS outperform the other techniques virtually for all the values of $\sigma$.  In this example,  AMIS also provides good results. Figure \ref{figSIMUafterR1} show a graphical representation of the results in Table \ref{TableResults2}, with the exception of the last column.  

  Fig. \ref{fig2Ex1} displays the initial (squares) and final (circles) configurations of the mean vectors of the proposals for the different algorithms, in one specific run. Since in Mixture PMC and AMIS the covariance matrices are also adapted, we show the shape of some proposals as ellipses representing approximately $85\%$ of probability mass.  For, PMC we also depict a last resampling output with triangles, in order to show the loss in diversity. Unlike PMC, PI-MAIS ensures a better covering of the region of high probability. 

   %This is due to the additional degree of freedom in the c 
%However, due to this additional degree of freedom compared to the standard PMC, APIS is able to reach good performance.

\begin{figure*}[htb]
\centering 
  %\includegraphics[width=6cm]{FundTheor.pdf}
%\subfigure[]{\includegraphics[width=6cm,height=4cm]{ARMS_fig3v2.pdf}}
\centerline{
\subfigure[PMC ($N=100$, $\sigma=1$)]{\includegraphics[width=4.4cm]{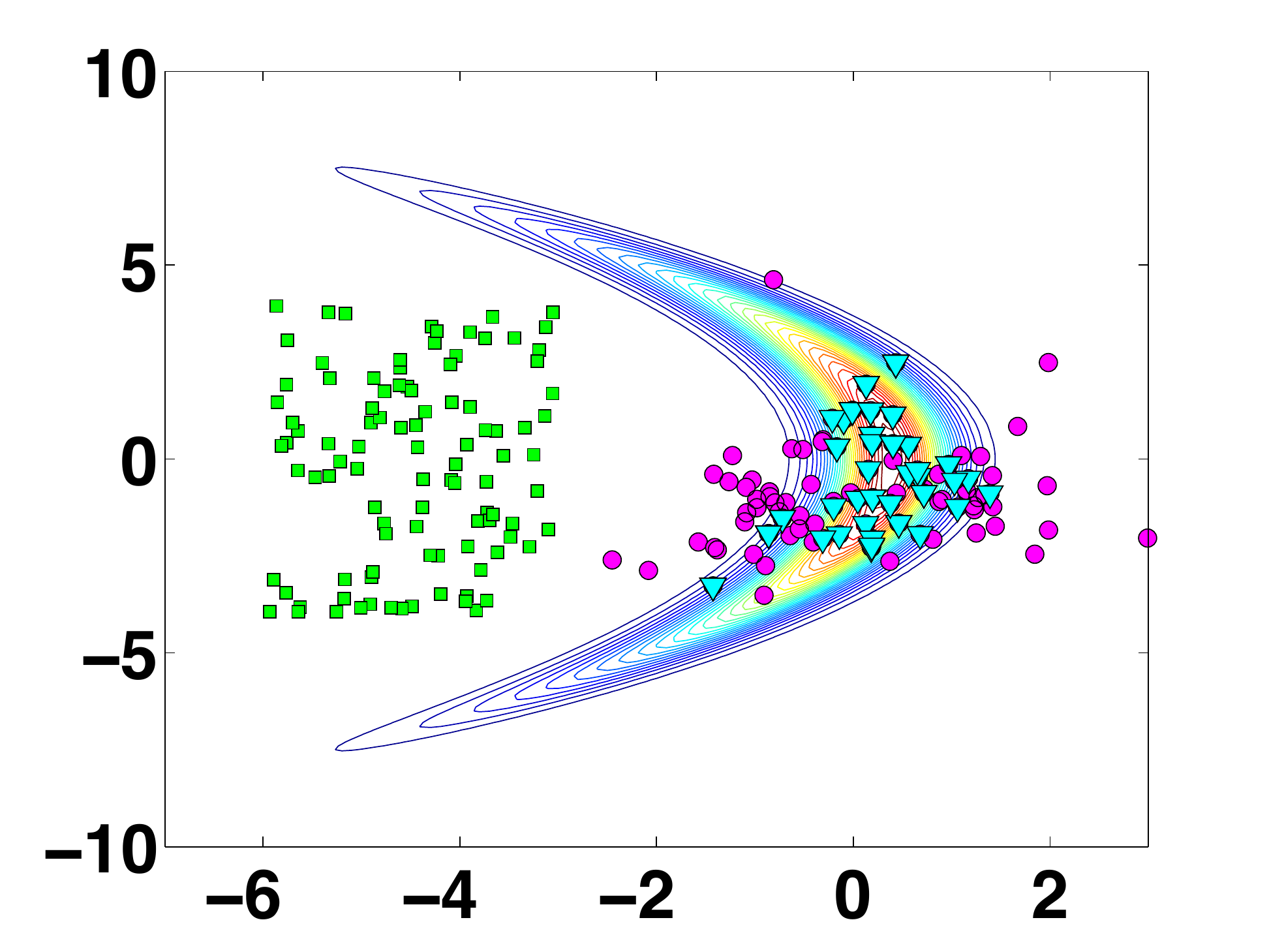}}
\subfigure[Mixture PMC with 10 mixands ($\sigma=5$)]{\includegraphics[width=4.4cm]{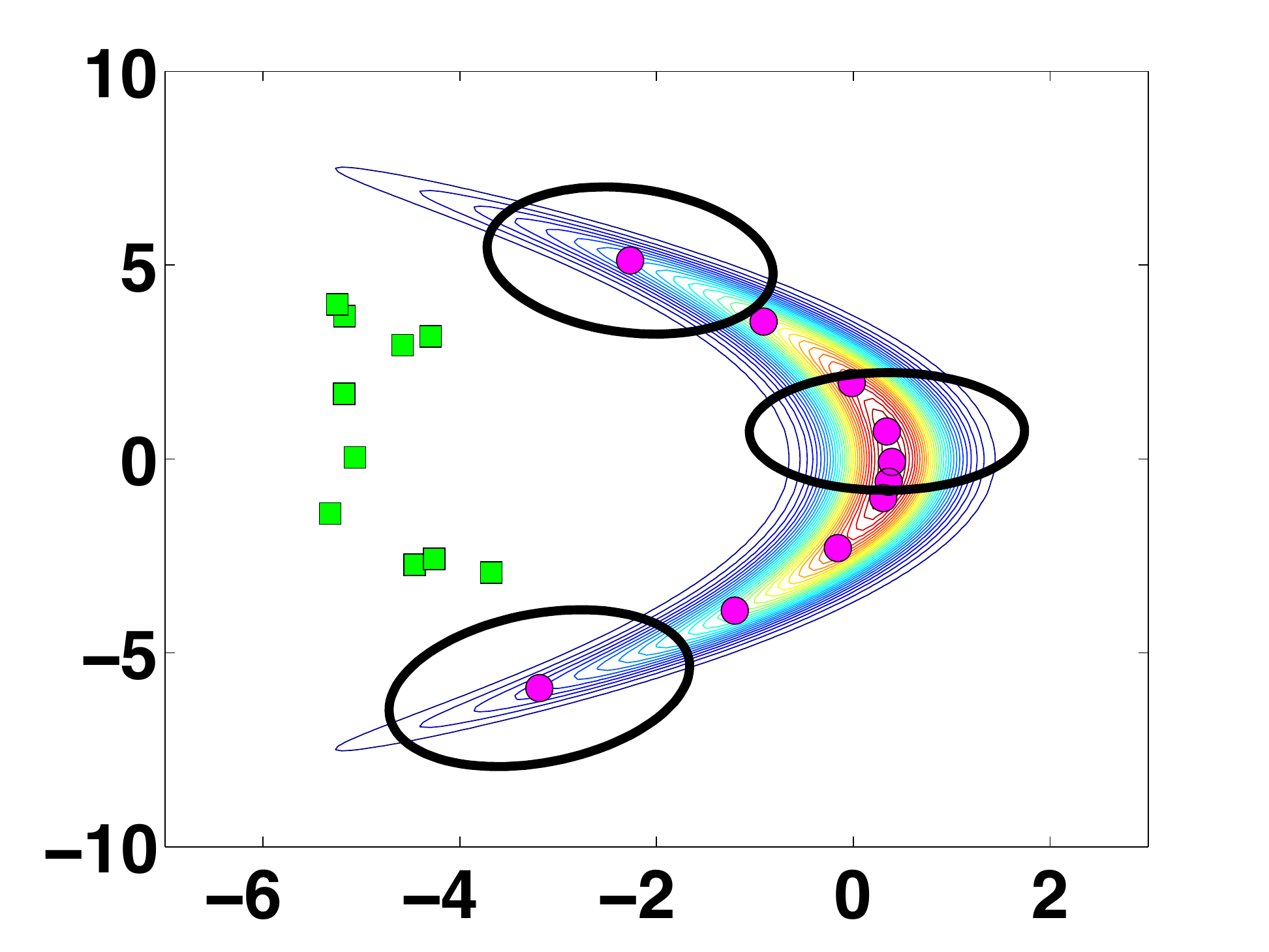}}
\subfigure[AMIS ($\sigma=5$)]{\includegraphics[width=4.4cm]{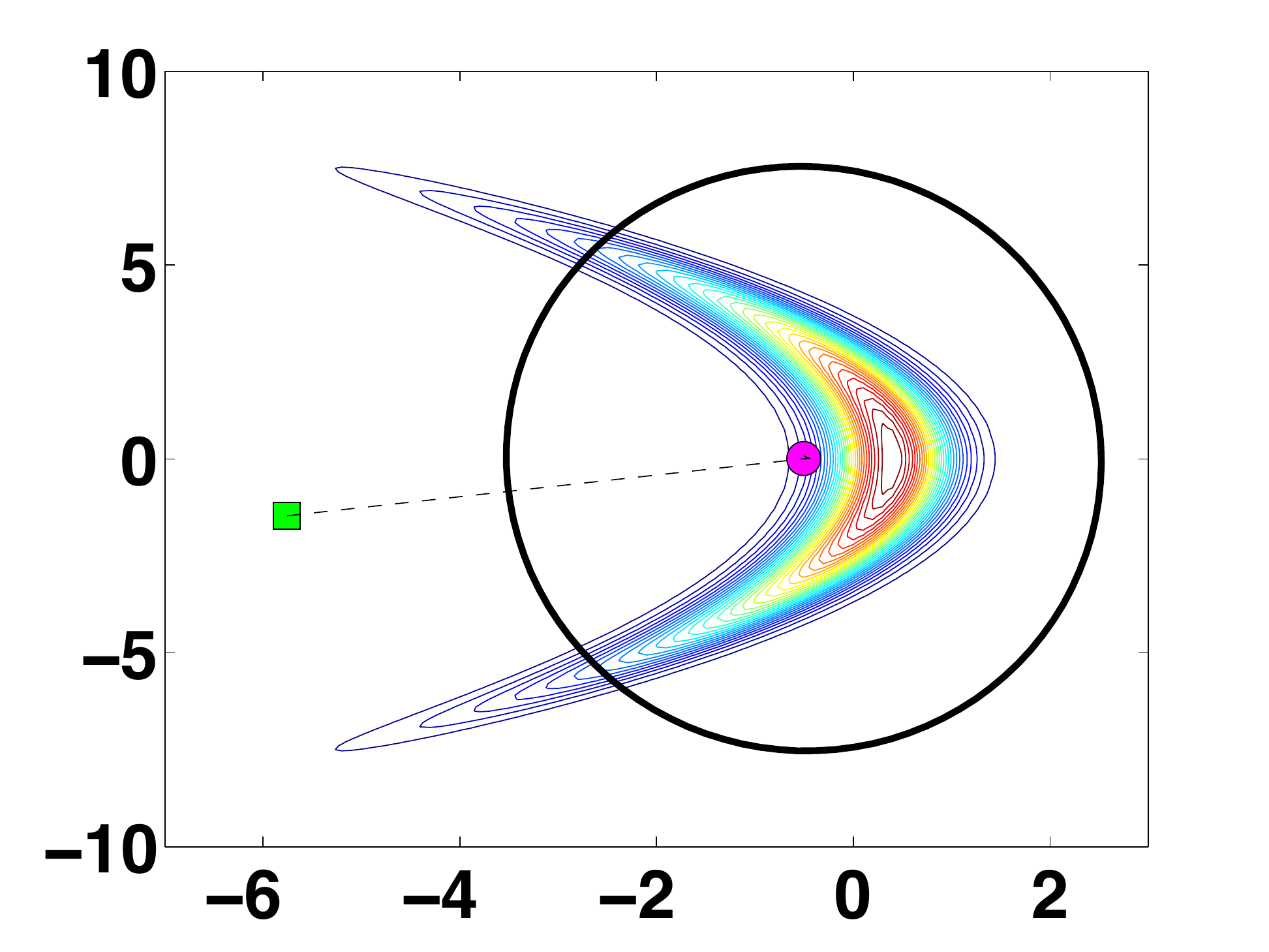}}
\subfigure[PI-MAIS ($N=100$, $\lambda=3$)]{\includegraphics[width=4.4cm]{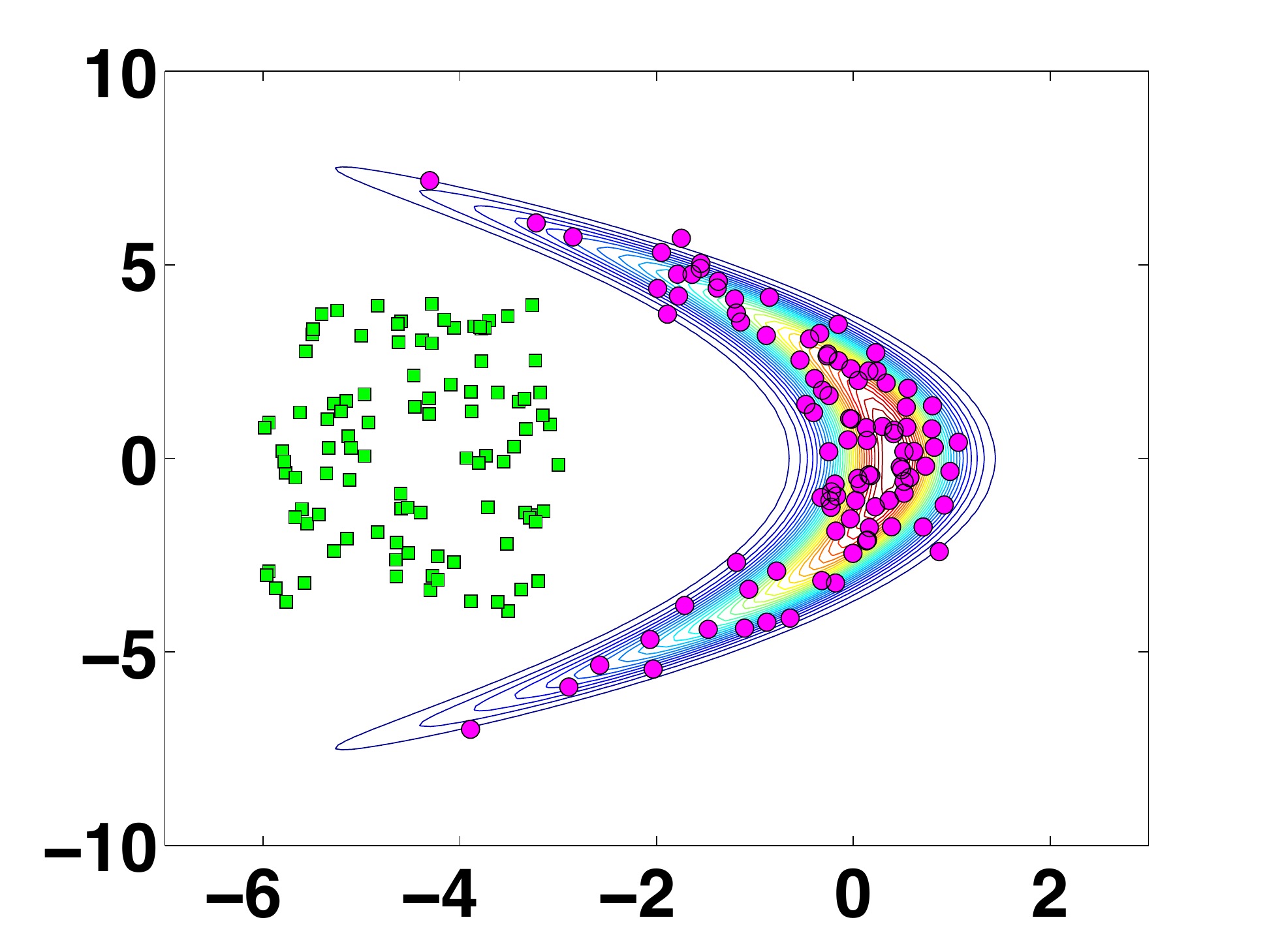}}
}
  \caption{%{\bf (Ex-Section-\ref{BananaExSect})}   
  Initial (squares) and final (circles) configurations of the mean vectors of the proposal densities for the  banana-shaped target distribution, in one specific run for the different methods. The Mixture PMC \citep{pmc-cappe08} and AMIS techniques \citep{CORNUET12} also adapt the covariance matrices (the ellipses show approximately $85\%$ of the probability mass). }
\label{fig2Ex1}
\end{figure*}

%%%%%%%%%%%%%%%%%%%%%%%%%%
\subsection{High dimensional target distribution}
%%%%%%%%%%%%%%%%%%%%%%%%%%
\label{sec:SIMU_HD}

Let us consider again a mixture of isotropic Gaussians as target pdf, i.e.,
\begin{equation}
\bar{\pi}({\bf x})  = \frac{1}{3}\sum_{k=1}^{3} \mathcal{N}({\bf x}; {\bm \nu}_k, {\bm \Sigma}_k), \quad {\bf x}\in \mathbb{R}^{D_x},
\label{eq:hdMixture}
\end{equation}  
where ${\bm \nu}_k=[\nu_{k,1}, \ldots,\nu_{k,D_x}]^{\top}$, and ${\bm \Sigma}_k=\chi_k^2 {\bf I}_{D_x}$ for $k \in \{1,2,3\}$, with $\bI_{D_x}$ being the $D_x \times D_x$ identity matrix.
We set $\nu_{1,j}=-5$, $\nu_{2,j}=6$, $\nu_{3,j}=3$ for all $j = 1,...,D_x$, and $\chi_k=8$ for all $k \in \{1,2,3\}$.
The expected value of the target ${\pi}({\bf x})$ is then $E[{X_j}]=\frac{4}{3}$ for $j=1,\ldots,D_x$. In order to study the performance of the proposed scheme as the dimension of the state space increases, we vary the dimension of the state space in Eq. \eqref{eq:hdMixture} testing different values of $D_x$ (with $2\leq D_x \leq 50$). 

We consider the problem of approximating via Monte Carlo the expected value of the target density, and we compare the performance of different methods:  {\bf (a)} the standard PMC scheme \citep{Cappe04}, {\bf (b)} $N$ parallel independent MH chains (Par-MH), {\bf (c)} a  standard Sequential Monte Carlo (SMC) scheme \citep{Moral06} and {\bf (d)} the proposed PI-MAIS method. 
We test the algorithms with $N\in\{100,500\}$. All the proposal pdfs involved in the experiments are Gaussians, with the same covariance matrices for all the techniques. The initial mean vectors in all techniques are selected randomly and independently as ${\bm \mu}_{n,0} \sim \mathcal{U}([-6\times 6]^{D_x})$ for $n=1,\ldots, N$.

Again, all the mentioned algorithms have been implemented in such a way that the number of total evaluations of the target is $E=2\cdot 10^5$. More specifically, in PI-MAIS, we use two sets of parameters: with $N=100$, $M=19$, $T=100$, and with $N=500$, $M=19$, $T=20$ in order to fulfill $E=(M+1)NT=2\cdot 10^5$ (see Section \ref{CompCostComp}). The proposal pdf of the upper level of the hierarchical approach, $\varphi_{n}(\x|{\bm \mu}_{n,t},{\bf \Lambda}_{n})$, are Gaussian pdfs with covariance matrices ${\bf \Lambda}_{n}=\lambda^2 {\bf I}_{D_x}$ and  $\lambda=10$. The proposal pdfs used in the lower importance sampling level, $q_{n,t}(\x|{\bm \mu}_{n,t},{\bf C}_{n})$ are Gaussian pdfs with covariance matrices ${\bf C}_{n}=\sigma^2 {\bf I}_{D_x}$ again with $\sigma=10$ (for a fair comparison with the other techniques). 
In PMC, Par-MH and SMC we use the same proposals with the same covariances and initial parameters. As described in App. \ref{PMClocura2}, in PMC the adaptation is carried out by resampling steps, in SMC an alternation of resampling and MH steps is performed whereas, in Par-MH, $N$ independent MH chains are carried out.

 The results are averaged over  $200$  independent simulations. Fig. \ref{figSIMU_HD} shows the log-MSE in the estimation of $E[{\bf X}]$ as a function of the dimension $D_x$ of the state-space.  Fig. \ref{figSIMU_HD}(a) compares the algorithms with $N=100$ proposal pdfs, whereas in Fig. \ref{figSIMU_HD}(b) we have $N=500$, keeping fixed the number of total evaluations of the target  $E=2\cdot 10^5$.
We observe, as expected, the performance of all the methods degenerate as the dimension of the problem, $D_x$ increases, since we maintain fixed the computational cost $E=2\cdot 10^5$.
PI-MAIS always provides the best results, with the exception for the cases corresponding to $N=100$ and $D_x=35,50$ where SMC obtains a lower MSE (for $N=100$ and $D_x=40$, they provide virtually the same MSE).

%%%%%%%%%%%%%%%%%%%%%%%%%%%%%%%%%%%%%
\subsection{Localization problem in a wireless sensor network}
%%%%%%%%%%%%%%%%%%%%%%%%%%%%%%%%%%%%%
\label{LocSect}
We consider the problem of positioning a target in a $2$-dimensional space using range measurements. This problem appears frequently in localization applications in wireless sensor networks \citep{Ali07,Ihler05,MartinoStatCo10}. %\citep{Pitt99} 
 Namely, we consider a random vector $\textbf{X}=[X_1,X_2]^{\top}$ to denote the target position in the plane $\mathbb{R}^{2}$. The position of the target is then a specific realization ${\bf X}={\bf x}$. 
The range measurements are obtained from $3$ sensors located at $\textbf{h}_1=[-10,2]^{\top}$, $\textbf{h}_2=[8,8]^{\top}$ and $\textbf{h}_3=[-20,-18]^{\top}$. The observation equations are given by
\begin{gather}
\label{IStemaejemplo}
\begin{split}
Y_{j}=a\log\left(\frac{||{\bf x}-{\bf h}_j ||}{0.3}\right)+\Theta_{j}, \quad j=1,\ldots, 3, \\
\end{split}   
\end{gather}   
where $\Theta_{j}$ are independent Gaussian variables with identical pdfs, $\mathcal{N}(\vartheta_j;0,\omega^2)$, $j=1,2$. We also consider a prior density over $\omega$, i.e., $\Omega \sim p(\omega)=\mathcal{N}(\omega;0,25) I(\omega>0),$
where $I(\omega>0)$ is $1$ if $\omega>0$ and $0$ otherwise. The parameter $A=a$ is also unknown and we again consider a Gaussian prior $A\sim p(a)=\mathcal{N}(a;0,25)$. 
Moreover, we also apply Gaussian priors over ${\bf X}$, i.e., $p(x_i)=\mathcal{N}(x_i;0,25)$ with $i=1,2$. Thus, the posterior pdf is
\begin{eqnarray}
\bar{\pi}(x_1,x_2,a,\omega)&=&p(x_1,x_2,a,\omega|\textbf{y}) \nonumber \\
&\propto& \ell(\textbf{y}|x_1,x_2,a,\omega)p(x_1)p(x_2)p(a)p(\omega), \nonumber 
\end{eqnarray}
where ${\bf y}\in \mathbb{R}^{D_y}$ is the vector of received measurements. We simulate $d=30$ observations from the model ($D_y/3=10$ from each of the three sensors) fixing $x_1=3$, $x_2=3$, $a=-20$ and $\omega=5$. With $D_y=30$, the expected value of the target ($E[X_1]\approx 2.8749$, $E[X_2]\approx 3.0266$, $E[A]\approx 5.2344$, $E[\Omega]\approx 20.1582$)\footnote{These values have been obtained with a deterministic, expensive and exhaustive numerical integration method, using a thin grid.} is quite close to the true values. 

Our goal is computing the expected value of 
$$(X_1,X_2,A,\Omega)\sim \bar{\pi}(x_1,x_2,a,\omega)$$ 
via Monte Carlo, in order to provide an estimate of the position of the target, the parameter $a$ and the standard deviation $\omega$ of the noise in the system. We apply PI-MAIS and three different PMC schemes (see example in Section \ref{sec:sims1}, for a description), all using $N$ Gaussian proposals.
We initialize the mean vectors so that they are randomly spread within the space of the variables of interest, i.e.,
 $$
 {\bm \mu}_{n,0}\sim \mathcal{N}({\bm \mu};{\bf 0}, 30^2 {\bf I}_4), \quad n=1,..., N,
 $$
 and the covariance matrices ${\bf C}_n=\mbox{diag}(\sigma_{n,1}^2,\ldots, \sigma_{n,4}^2) {\bf I}_4$ with $n=1,\ldots, N$. The values of the standard deviations $\sigma_{n,j}$ are chosen randomly for each Gaussian pdf. Specifically,  $\sigma_{n,j} \sim \mathcal{U}([1,Q])$, $j=1,\ldots,4$, where we have considered three possible values for $Q$, i.e., $Q\in \{5,10,30\}$.  For the adaptation process of PI-MAIS, we consider also Gaussian proposals with covariance matrices ${\bf \Lambda}_{n}=\lambda^2 {\bf I}_2$ and  $\lambda\in \{5,10,70\}$. We also try different non-isotropic diagonal covariance matrices, i.e, ${\bf \Lambda}_n = \textrm{diag}(\lambda_{n,1}^2, \lambda_{i,2}^2)$, where $\lambda_{n,j} \sim \mathcal{U}([1,30])$.

For a fair comparison, all the techniques have been simulated with sets of parameters that yield the same number of target evaluations, fixed to $E=2\cdot 10^5$. In PI-MAIS, we have chosen parameters $N=100$, $M=\{1,19, 99\}$, $T=\{20,100,1000\}$. The PMC algorithms has been simulated with $N=100$ and $T=2000$. The MSE of the different estimators (averaged over $3000$ independent runs) are provided in Table \ref{mean_sensors} and the $\log(\mbox{MSE})$ in Figure \ref{figSIMUafterR1_2}(b).
 PI-MAIS outperforms always PMC when $\sigma_{n,j}\sim \mathcal{U}([1,5])$ and $\sigma_{n,j}\sim \mathcal{U}([1,10])$ whereas PMC provides better results for $\sigma_{n,j}\sim \mathcal{U}([1,30])$. Therefore, the results show jointly the  robustness and flexibility of the proposed PI-MAIS technique.

%%%%%%%%%%%%%%%%%%%%%%%%%%%%%%
\vspace{-0.5cm}
\section{Conclusions}
\label{sec:conclusions}
\vspace{-0.3cm}
In this work, we have introduced a layered (i.e., hierarchical) framework for designing adaptive Monte Carlo methods. In general terms, we have shown that such a hierarchical interpretation lies behind the good performance of two well-known algorithms; a random walk proposal within an MH scheme and the standard PMC method. Furthermore, we have used this approach to introduce a novel class of adaptive importance sampling (AIS) schemes. The novel class of AIS algorithms employs the determinist mixture (DM) idea \citep{Owen00,Veach95} in order to reduce the variance of the resulting IS estimators. We have extended the use of the DM strategy with respect to other algorithms available in the literature,  providing a more general and flexible framework. From an estimation perspective, this framework includes different schemes proposed in literature \citep{CORNUET12,APIS15} as special cases, although they differ to an extent in terms of the employed adaptation procedure. Our framework also contains several other sampling schemes considering full or partial DM approaches. Finally, we have discussed several aspects of the trade-offs in terms of the computational cost and advantages due to improved accuracy of the resulting estimators. Numerical comparisons with different algorithms on benchmark models have confirmed the benefit of the layered adaptive sampling approaches.

\vspace{-0.7cm}
%%%%%%%%%%%%%%%%%%%
\section*{Acknowledgements}
%%%%%%%%%%%%%%%%%
\vspace{-0.2cm}
This work has been supported by the projects COMONSENS (CSD2008 00010), ALCIT (TEC2012 38800C03 01), DISSECT (TEC2012 38058 C03 01), OTOSiS (TEC 2013 41718 R), and COMPREHENSION (TEC 2012 38883 C02 01), by the BBVA Foundation with "I Convocatoria de Ayudas Fundaci—n BBVA a Investigadores, Innovadores y Creadores Culturales"- MG FIAR project, by the ERC grant 239784 and AoF grant 251170, and by the European Union 7th Framework Programme through the Marie Curie Initial Training Network ``Machine Learning for Personalized Medicine'' MLPM2012, Grant No. 316861.
\bibliographystyle{plain}
\bibliography{bibliografia}

\appendix
%%%%%%%%%%%%%%%%%%%%%%%%%%%%%%%%%%%%%%%%%%%%%
%\section{Estimation of normalizing constant with different procedures with MCMC or alternative ways.... }
%%%%%%%%%%%%%%%%%%%%%%%%%%%%%%%%%%%%%%%%%%%%

%{\color{black}
%%%%%%%%%%%%%%%%%%%%%%%%%%%
%\section{Optimal characteristic function}
%\label{sec:charFun}
%%%%%%%%%%%%%%%%%%%%%%%%%%

%According to Bochner's theorem \cite[Theorem 1.8.9]{ushakov1999selected}, a $D_x$-dimensional complex-valued function ${\mathbf \Omega}: \mathbb{R}^{D_x} \rightarrow \mathbb{C}$ is a characteristic function if and only if:
%\begin{enumerate}
%	\item ${\mathbf \Omega}({\bm \nu})$ is non-negative definite.
%	\item ${\mathbf \Omega}({\bf 0}) = 1$.
%\end{enumerate}
%It is straightforward to show that $H^*({\bm \nu}|{\bf C})$ fulfills these two requirements:
%\begin{enumerate}
%	\item $\bar{\Pi}({\bm \nu})$ and $Q({\bm \nu}|{\bf C})$ are non-negative definite (as they are the characteristic functions of $\bar{\pi}({\bf x})$ and
%		$q({\bf x}|{\bf C})$ respectively), and the ratio of two non-negative definite functions results in another non-negative definite function. Therefore,
%		$H^*({\bm \nu}|{\bf C}) = \frac{\bar{\Pi}({\bm \nu})}{Q({\bm \nu}|{\bf C})}$ is non-negative definite.
%	\item $\bar{\Pi}({\bf 0}) = 1$ and $Q({\bf 0}|{\bf C}) = 1$, implying that $H^*({\bf 0}|{\bf C}) = \frac{\bar{\Pi}({\bf 0})}{Q({\bf 0}|{\bf C})} = 1$.
%\end{enumerate}
%Hence, $H^*({\bm \nu}|{\bf C})$ is the optimal characteristic function associated to the optimal pdf $h^*({\bf x}|{\bf C})$.}

%%%%%%%%%%%%%%%%%%%%%%%%%%
\section{Consistency of GAMIS estimators}
%%%%%%%%%%%%%%%%%%%%%%%%%
\label{ConsApp}

First of all,  we remark that the complete analysis should take in account the chosen adaptive procedure since, in general, the adaptation uses the information of previous weighted samples. However, in this work we consider an adaption procedure completely independent of the estimation steps, as clarified in Sections \ref{RelationSect}-\ref{TeoOtra}.  This simplifies substantially the analysis as described in Section \ref{TeoOtra}.  
  
The consistency of the global estimators in Eq. \eqref{equation_whatever_1} provided by GAMIS can be considered when number of samples per time step ($M\times N$) and the number of iterations of the algorithm ($T$) grow to infinity. For some exhaustive studies of specific cases, see the analysis  in \citep{Robert04,Douc07a} and \citep{Marin12}. 
Here we provide some brief arguments for explaining why $\hat{I}_T$ and ${\hat Z}_T$ obtained by a GAMIS scheme are, in general, consistent.
Let us assume that $q_{n,t}$'s have heavier tails than $\bar{\pi}({\bf x}) \propto \pi({\bf x})$. Note that the global estimator $\hat{I}_T$ can be seen as a result of a static batch MIS estimator involving $L$ different mixture-proposals $\Phi_{n,t}({\bf x})$ and $J=NMT$ total number of samples. The  weights $w_{n,t}^{(m)}$ built using $\Phi_{n,t}({\bf x})$ in the denominator of the IS ratio are suitable importance weights yielding consistent estimators, as explained in detail in Appendix\ref{BellissimaSect}.
Hence, for a finite number of iterations $T < \infty$, when $M\rightarrow \infty$ (or $N \rightarrow \infty$), the consistency can be guaranteed by standard IS arguments, since it is well known that $\hat{Z}_{T} \to Z$ and $\hat{I}_{T} \to I$ as $M \to \infty$, or  $N\to \infty$  \citep{Robert04}.

Furthermore, for $T \rightarrow \infty$ and $N,M < \infty$, we have a convex combination, given in Eq. \eqref{equation_whatever_2}, of conditionally independent (consistent but biased) IS estimators \citep{Robert04}. Indeed, although in an adaptive scheme the proposals depend on the previous configurations of the population, the samples drawn at each iteration are conditionally independent of the previous ones, and independent of each other drawn at the same iteration. The bias is due to  unknown $Z$ (see Eq. \eqref{eq:normConst}), and hat $\hat{Z}_{T}$ is used to replace $Z$.
However, $\hat{Z}_{T} \to Z$ as $T \to \infty$, as discussed in \citep[Chapter 14]{Robert04}: hence, $\hat{I}_{T}$ is asymptotically unbiased  as $T \to \infty$.

\section{Importance sampling with multiple proposals}
\label{BellissimaSect}
%%%%%%%%%%%%%%%%%%%%%%%%%%%%%%%%%%%%%

Recall that our goal is computing efficiently the integral $I = \frac{1}{Z} \int_{\mathcal{X}} f({\bf x}) \pi({\bf x}) d{\bf x}$ where $f$ is any square-integrable function (w.r.t. $\bar{\pi}({\bf x})$) of ${\bf x}$, and $Z=\int_{ \mathcal{X}} \pi({\bf x}) d{\bf x}<\infty$ with $\pi({\bf x}) \geq 0$ for all ${\bf x}\in \mathcal{X}\subseteq \mathbb{R}^{D_x}$.
Let us assume that we have two proposal pdfs, $q_1({\bf x})$ and $q_2({\bf x})$, from which we intend to draw $M_1$ and $M_2$ samples respectively: 
$$
{\bf x}_1^{(1)},\ldots,{\bf x}_{1}^{(M_1)}\sim q_1({\bf x}) \quad\quad \mbox{ and } \quad \quad {\bf x}_2^{(1)},\ldots,{\bf x}_{2}^{(M_2)}\sim q_2({\bf x}).
$$
There are at least two procedures to build a joint IS estimator: the standard multiple importance sampling (MIS) approach and the full deterministic mixture (DM-MIS) scheme. %Both are briefly reviewed in the following but, first of all, we clarify and described two procedures for drawing from a mixture of pdfs. 

%%%%%%%%%%%%%%%%%%%%
\subsection{Standard IS approach}
\label{PobreDiabla}
%%%%%%%%%%%%%%%%%%%%

The simplest approach \citep[Chapter 14]{Robert04} is computing the classical IS weights:
\begin{align}
\label{eq:weightsIS}
	w_1^{(i)} = \frac{\pi({\bf x}_1^{(i)})}{q_1({\bf x}_1^{(i)})}, \quad w_2^{(k)} = \frac{\pi({\bf x}_2^{(k)})}{q_2({\bf x}_2^{(k)})},
\end{align}
with $ i=1,\ldots, M_1$ and $k=1,\ldots, M_2$.
The IS estimator is then built by normalizing them jointly, i.e., computing
\begin{equation}
	\hat{I}_{IS}
		= \frac{1}{S_{tot}}\left(\sum_{i=1}^{M_1} w_1^{(i)} f({\bf x}_1^{(i)})+\sum_{k=1}^{M_2} w_2^{(k)} f({\bf x}_2^{(k)}) \right),
\label{Total_IS_standard}
\end{equation}
where $S_{tot}=\sum_{i=1}^{M_1} w_1^{(i)}+\sum_{k=1}^{M_2} w_2^{(k)}$. For $J>2$ proposal pdfs and ${\bf x}_j^{(1)},\ldots,{\bf x}_{j}^{(M_j)}\sim q_j({\bf x})$,  for $j=1,\ldots,J$, we have
\begin{gather}
\nonumber
\left\{
\begin{split}
w_j^{(m_j)} &= \frac{\pi({\bf x}_j^{(m_j)})}{q_j({\bf x}_j^{(m_j)})}, \quad \mbox{ and }  \\
	\hat{I}_{IS} &= \frac{1}{\sum_{n=1}^{J} \sum_{m_j=1}^{M_j}w_{j}^{(m_j)}} \sum_{j=1}^{J} \sum_{m_j=1}^{M_j}w_{j}^{(m_j)} f({\bf x}_{j}^{(m_j)}).
\end{split}
\right.
\end{gather}
In this case, $S_{tot}=\sum_{n=1}^{J} \sum_{m_j=1}^{M_j}w_{j}^{(m_j)}$.

%%%%%%%%%%%%%%%%%%%%%%%%
\subsection{Deterministic mixture approach}
\label{PobreDiabla2}
%%%%%%%%%%%%%%%%%%%%%%%%

An alternative approach is based on the deterministic mixture sampling idea \citep{Owen00,Veach95,ElviraMIS15}. %described previously in Appendix \ref{LucaApp}.
Considering $N=2$ proposals $q_1$, $q_2$, and setting 
\begin{equation*}
	\mathcal{Z}= \left\{{\bf x}_{1}^{(1)},\ldots,{\bf x}_{1}^{(M_1)},{\bf x}_{2}^{(1)},\ldots,{\bf x}_{2}^{(M_2)}\right\},
\end{equation*}
with ${\bf x}_{j}^{(m_j)} \in \mathbb{R}^{D_x}$ ($n \in \{1,2\}$ and $1 \le m_j \le M_j$), the weights are now defined as 
\begin{equation}
	w_{j}^{(m_j)} = \frac{\pi({\bf x}_{j}^{(m_j)})}{\frac{M_1}{M_1+M_2} q_1({\bf x}_{j}^{(m_j)})+\frac{M_2}{M_1+M_2} q_2({\bf x}_{j}^{(m_j)})}.
\label{eq:weightsDM}
\end{equation}
In this case, the {\it complete} proposal is considered to be a mixture of $q_1$ and $q_2$, weighted according to the number of samples drawn from each one.
Note that, unlike in the standard procedure for sampling from a mixture, a deterministic and fixed number of samples are drawn from each proposal in the DM approach  \cite{ElviraMIS15}. It can be shown that the set $\mathcal{Z}$ of samples drawn in this deterministic way is distributed according to the mixture $q({\bf z})=\frac{M_1}{M_1+M_2} q_1({\bf z})+\frac{M_2}{M_1+M_2} q_2({\bf z})$ \cite[Chapter 9, Section 11]{mcbookOwen}.
The DM estimator is finally given by
\begin{equation}
	\hat{I}_{DM} = \frac{1}{S_{tot}}\sum_{j=1}^{2} \sum_{m_j=1}^{M_j}{w_{j}^{(m_j)} f({\bf x}_{j}^{(m_j)})},
\end{equation}
where $S_{tot} =\sum_{j=1}^{2} \sum_{m_j=1}^{M_j}w_{j}^{(m_j)}$ and the $w_{j}^{(m_j)}$ are given by \eqref{eq:weightsDM}.
For $J>2$ proposal pdfs, the DM estimator can also be easily generalized:
%$$
%\begin{cases} 
%w_{i}^{(m_i)} = \frac{\pi({\bf x}_{i}^{(m_i)})}{\sum_{j=1}^{J}{\frac{M_j}{M_{tot}} q_j({\bf x}_{j}^{(m_j)})}},\quad \mbox{ and }  \\
%\hat{I}_{DM} = \frac{1}{\sum_{n=1}^{J} \sum_{m_j=1}^{M_j}w_{j}^{(m_j)}} \sum_{j=1}^{J} \sum_{m_j=1}^{M_j}w_{j}^{(m_j)} f({\bf x}_{j}^{(m_j)}), 
%\end{cases}
%$$
%\begin{eqnarray}
%\left\{
%w_{i}^{(m_i)} &=& \frac{\pi({\bf x}_{i}^{(m_i)})}{\sum_{j=1}^{J}{\frac{M_j}{M_{tot}} q_j({\bf x}_{j}^{(m_j)})}},\quad \mbox{ and }  \\
%\hat{I}_{DM} &=& \frac{1}{\sum_{n=1}^{J} \sum_{m_j=1}^{M_j}w_{j}^{(m_j)}} \sum_{j=1}^{J} \sum_{m_j=1}^{M_j}w_{j}^{(m_j)} f({\bf x}_{j}^{(m_j)}), \nonumber 
%\right.
%\end{eqnarray}
\begin{gather}
\nonumber
\left\{
\begin{split}
w_{i}^{(m_i)} &= \frac{\pi({\bf x}_{i}^{(m_i)})}{\sum_{j=1}^{J}{\frac{M_j}{M_{tot}} q_j({\bf x}_{j}^{(m_j)})}},\quad \mbox{ and }  \\
\hat{I}_{DM} &= \frac{1}{\sum_{n=1}^{J} \sum_{m_j=1}^{M_j}w_{j}^{(m_j)}} \sum_{j=1}^{J} \sum_{m_j=1}^{M_j}w_{j}^{(m_j)} f({\bf x}_{j}^{(m_j)}), 
\end{split}
\right.
\end{gather}
with $i=1,\ldots,J$, $M_{tot}=M_1+M_2+\ldots+M_J$ and $S_{tot} =\sum_{j=1}^{J} \sum_{m_j=1}^{M_j}w_{j}^{(m_j)}$.
On the one hand, the DM approach is more efficient than the IS method, thus providing a better performance in terms of a reduced variance of the corresponding estimator, as shown in the following section. On the other hand, it needs to evaluate every proposal $M_{tot}$ times instead of only  $M_j$ times (in the standard MIS procedure), and therefore is more costly from a computational point of view. However, this increased computational cost is negligible when the proposal is much cheaper to evaluate than the target, as it often happens in practical applications.

%%%%%%%%%%%%%%%%%%%%%%%%%%%%%
\subsection{Convex combination of partial IS estimators}
\label{PobreDiabla3}
%%%%%%%%%%%%%%%%%%%%%%%%%%%%%%

Regardless the type of weights employed in the IS scheme (either as in Eq. \eqref{eq:weightsIS} or as in Eq. \eqref{eq:weightsDM}), the resulting estimators can be written as convex combination of simpler ones. First of all, let us consider again the use of  $J=2$ proposals, $q_1$ and  $q_2$. We draw $M_j$ samples from each one, ${\bf x}_j^{(1)},\ldots,{\bf x}_{j}^{(M_j)}\sim q_j({\bf x})$, with  $j\in\{1,2\}$. The two partial sums of the weights corresponding only to the samples drawn from $q_1$ and $q_2$, are given by $S_1=\sum_{i=1}^{M_1} w_1^{(i)}$ and $S_2=\sum_{k=1}^{M_2} w_2^{(k)}$. The partial IS estimators, obtained by considering only one proposal pdf, are $\hat{I}_1=\sum_{i=1}^{M_1} \bar{w}_1^{(i)} f({\bf x}_1^{(i)})$ and $\hat{I}_2=\sum_{k=1}^{M_2} \bar{w}_2^{(k)} f({\bf x}_2^{(k)})$  where the normalized weights are $\bar{w}_1^{(i)}=\frac{w_1^{(i)}}{S_1}$  and $\bar{w}_2^{(k)}=\frac{w_2^{(k)}}{S_2}$, respectively.
%Note that, by defining $S_{tot}=S_1+S_2$, where. 
% Eq. \eqref{Total_IS_standard} can be rewritten as
The complete IS estimator, taking into account the $M_1+M_2$ samples jointly, is
\begin{eqnarray}
\label{General_standIS_0}
	\hat{I}_{tot} &=& \frac{1}{S_1+S_2}\left(S_1 \hat{I}_1+S_2 \hat{I}_2 \right) \nonumber \\
		&=& \frac{S_1}{S_1+S_2}\hat{I}_1+\frac{S_2}{S_1+S_2}\hat{I}_2.
\end{eqnarray}
This procedure can be easily extended for $J>2$ different proposal pdfs, obtaining the complete estimator as the convex combination of the $N$ partial estimators:
\begin{gather}
\begin{split}
	\hat{I}_{tot} &= \frac{\sum_{j=1}^{J}{S_j \hat{I}_j}}{\sum_{j=1}^{J}{S_j}}, \\
	 \hat{Z}_{tot} &=\frac{1}{\sum_{j=1}^{J} M_j}\sum_{j=1}^{J}{S_j}=\frac{1}{\sum_{j=1}^{J} M_j} \sum_{j=1}^{J}M_j {\hat{Z}_j}  , 
\label{General_standIS}
\end{split}
\end{gather}
where ${\bf x}_j^{(1)},\ldots,{\bf x}_{j}^{(M_j)} \sim q_j({\bf x})$,
 %$w_j^{(i)} = \pi({\bf x}_j^{(i)})/q_j({\bf x}_j^{(i)})$,
   $\hat{I}_j = \sum_{k=1}^{M_j}{w_j^{(k)} f({\bf x}_j^{(k)})}$, $S_j = \sum_{k=1}^{M_j}{w_j^{(k)}}$ and $\hat{Z}_j=\frac{1}{M_j} \sum_{k=1}^{M_j}w_j^{(k)}$.

\section{Hierarchical interpretation of PMC}
%%%%%%%%%%%%%%%%%%%%%%%%%%%%%%%%%%%%
\label{PMClocura2}

The standard Population Monte Carlo (PMC) \citep{Cappe04} method can be interpreted as  using a hierarchical procedure.     
%with a prior pdf $h({\bm \mu}) \approx \bar{\pi}({\bm \mu})$.
 Although  it is possible to recognize the two different layers,  there are some  differences w.r.t.  the hierarchical procedure in Section \ref{HMCsect}. The first one is that in PMC  the generation of ${\bm \mu}$'s is not independent of the previously generated ${\bf x}$'s. The second one is that the prior is instead $h({\bm \mu})=\hat{\pi}_t^{(N)}({\bm \mu})$,  where $\hat{\pi}_t^{(N)}$ is an approximation of the measure of $\bar{\pi}({\bm \mu})$ obtained using the previously generated samples ${\bf x}$'s (in the second level of the hierarchical approach). %The quality of this approximation increases with $N$, as discussed below and in Appendix \ref{SIR_appendix}
More specifically, a standard PMC method \citep{Cappe04} is an adaptive importance sampler using a population of proposals $q_1$, $\ldots$, $q_N$. PMC consists of the following steps, given an initial set, ${\bm \mu}_{1,0}$, $\ldots$, ${\bm \mu}_{N,0}$, of mean vectors: 
\begin{enumerate}
\item For $t=0,\ldots,T-1:$
\begin{enumerate}
\item Draw ${\bf x}_{n,t}\sim q_{n,t}({\bf x}|{\bm \mu}_{n,t},{\bf C}_n)$, for $n=1,\ldots,N$.
\item  Assign to each sample ${\bf x}_{n,t}$ the weights, 
\begin{equation}
w_{n,t}=\frac{\pi({\bf x}_{n,t})}{q_{n,t}({\bf x}_{n,t}|{\bm \mu}_{n,t},{\bf C}_n)}.
\end{equation}
\item {\it Resampling:} draw $N$ independent samples ${\bm \mu}_{n,t+1}$, $n=1,\ldots, N$, 
according to the particle approximation
\begin{equation}
\label{ApproxPMC}
\hat{\pi}_t^{(N)}({\bm \mu}|{\bf x}_{1:N,t})=\frac{1}{\sum_{n=1}^N w_{n,t}} \sum_{n=1}^N w_{n,t} \delta({\bm \mu}-{\bf x}_{n,t}),
\end{equation}
where we have denoted ${\bf x}_{1:N,t}=[{\bf x}_{1,t},\ldots,{\bf x}_{N,t}]^{\top}$.
Note that each ${\bm \mu}_{n,t+1} \in \{{\bf x}_{1,t},\ldots, {\bf x}_{N,t}\}$, for all $n$.
\end{enumerate}
\item Return all the pairs $\{{\bf x}_{n,t}, w_{n,t}\}$, 
%$$
%\bar{\rho}_{n,t}=\frac{w_{n,t}}{\sum_{t=1}^T\sum_{n=1}^N w_{n,t}},
%$$ 
$n=1,\ldots,N$ and  $t=0,\ldots,T-1$.
\end{enumerate}
Fixing an iteration $t$, the generating procedure used in one iteration of the standard PMC method can be cast in the hierarchical formulation:
\begin{enumerate}
\item Draw $N$ samples ${\bm \mu}_{1,t},\ldots,{\bm \mu}_{N,t}$ from $\hat{\pi}_{t-1}^{(N)}({\bm \mu}| {\bf x}_{1:N,t-1})$.
\item Draw ${\bf x}_{n,t}\sim q_{n,t}({\bf x}|{\bm \mu}_{n,t},{\bf C}_n)$, for $n=1,\ldots,N$.
\end{enumerate}
Note that $\hat{\pi}_{t-1}^{(N)}$ plays the role of the prior $h$ in the hierarchical scheme above. Differently from the novel proposed scheme, the two levels of hierarchical procedure are not independent since the pdf $\hat{\pi}_{t}^{(N)}({\bm \mu}|{\bf x}_{1:N,t})$ depends on the samples drawn in the lower level. Furthermore, $\hat{\pi}_{t}^{(N)}$ also varies with $t$ and $N$, whereas in our procedure we consider a fixed prior $h$. 
However, note that $\hat{\pi}_{t}^{(N)}$ is an empirical measure approximation of ${\bar \pi}$ that improves when $N$ grows. An equivalent formulation of the hierarchical scheme for PMC is given below, involving a probability of generating a new mean ${\bm \mu}$ given the previous ones 
${\bm  \mu}_{1:N,t-1}=[{\bm  \mu}_{1,t-1},\ldots,{\bm  \mu}_{N,t-1}]^{\top},$
  denoted as $K_t^{(N)}({\bm \mu}|{\bm \mu}_{1:N,t-1})$.

\subsection{Distribution after one resampling step}
\label{SIR_appendix}
%%%%%%%%%%%%%%%%%%%%%%%%%%%%%%%%%%

 Consider the $t$-th iteration of PMC. Let us define as
\begin{equation*}
	{\bf m}_{\neg n}=[{\bf x}_{1,t},\ldots, {\bf x}_{n-1,t},{\bf x}_{n+1,t},\ldots, {\bf x}_{N,t}]^{\top},
\end{equation*}
the vector containing all the generated samples except for the $n$-th.
Let us also denote as ${\bm \mu}_{i,t+1}\in\{{\bf x}_{1,t}\ldots,{\bf x}_{N,t}\}$, a generic mean vector, i.e. $i\in\{1,\ldots,N\}$ at the iteration $t+1$, after applying one resampling step (i.e., a multinomial sampling according to the normalized weights).
Hence, the distribution of ${\bm  \mu}$ given the previous means ${\bm  \mu}_{1:N,t-1}$ is 
\begin{gather}
\begin{split}
	&K_{t+1}^{(N)}({\bm \mu}_{i,t+1}|{\bm \mu}_{1,t},\dots,{\bm \mu}_{N,t}) =\\
	&=\int_{\mathcal{X}^{N}} \hat{\pi}_t^{(N)}({\bm \mu}_{i,t+1}|{\bf x}_{1:N,t})  \left[\prod_{n=1}^{N}{q_{n,t}({\bf x}_{n,t}|{\bm \mu}_{n,t},{\bf C}_n)}\right]
		d{\bf x}_{1:N,t},
\label{eq:phiIS}
\end{split}
\end{gather}
where $\hat{\pi}_t^{(N)}({\bm \mu}|{\bf x}_{1:N,t})$ is given in Eq. \eqref{ApproxPMC}. For simplicity, below we denote 
$$q_{n}({\bf x})=q_{n,t}({\bf x}|{\bm \mu}_{n,t},{\bf C}_n), \quad \mbox{ and } {\bm \mu}={\bm \mu}_{i,t}.$$
Then, after some straightforward rearrangements,  Eq. \eqref{eq:phiIS} can be rewritten as
{\scriptsize
\begin{eqnarray}
&&	K_{t+1}^{(N)}({\bm \mu}|{\bm \mu}_{1,t},\dots,{\bm \mu}_{N,t}) = \nonumber \\
&=&\sum_{j=1}^{N}\left(\int_{\mathcal{X}^{N-1}} \frac{\pi({\bf x}_{j,t})}{\sum_{n=1}^{N}{\frac{\pi({\bf x}_{n,t})}{q_n({\bf x}_{n,t})}}}\left[\prod_{\substack{n=1 \\ n \neq j}}^{N}{q_n({\bf x}_{n,t})}\right]
		 d{\bf m}_{\neg j}\right)\delta({\bm \mu}-{\bf x}_{j,t}). \nonumber 
%%%\label{eq:phiISxxxx}
\end{eqnarray}
}
Finally, we can write
\begin{eqnarray}
&&K_{t+1}^{(N)}({\bm \mu}|{\bm \mu}_{1,t},\dots,{\bm \mu}_{N,t}) =  \nonumber \\
&&	 \pi({\bm \mu}) \sum_{j=1}^{N} \left({\int_{\mathcal{X}^{N-1}} \frac{1}{N\hat{Z}} 
		\left[\prod_{\substack{n=1 \\ n \neq j}}^N q_n({\bf x}_{n,t})\right]  d{\bf m}_{\neg j}} \right),
\label{eq:phiIS2}
\end{eqnarray}
where $\hat{Z}= \frac{1}{N}\sum_{n=1}^N\frac{\pi({\bf x}_n)}{q_n({\bf x}_n)}$ is the estimate of the normalizing constant of the target obtained using the classical IS weights. The hierarchical formulation of PMC can be rewritten as:
\begin{enumerate}
\item Draw $N$ samples ${\bm \mu}_{1,t},\ldots,{\bm \mu}_{N,t}$ from $K_{t}^{(N)}({\bm \mu}|{\bm \mu}_{1:N,t-1})$ in Eq. \eqref{eq:phiIS} or  \eqref{eq:phiIS2}.
\item Draw ${\bf x}_{n,t}\sim q_{n,t}({\bf x}|{\bm \mu}_{n,t},{\bf C}_n)$, for $n=1,\ldots,N$.
\end{enumerate}
%This formulation is closer to scheme in Section \ref{} since the pdf $K_t^{(N)}({\bm \mu}|{\bm \mu}_{1:N,t-1})$, playing the role of prior $h({\bf \mu})$,   is not depending to the samples ${\bf x}$'s generated in the lower level. %Moreover, this formulation shows that there is also a dynamic 
When $N \to \infty$, then $\hat{Z} \to Z$ \citep{Robert04}, and thus $K_t^{(N)}({\bm \mu}|{\bm \mu}_{1:N,t-1}) \to \frac{1}{Z} \pi({\bm \mu}) = \bar{\pi}({\bm \mu})$, for all $t=1\ldots,T$.   Namely, when $N$ grows, 
the hierarchical scheme above tends to have $h({\bm \mu})={\bar \pi}({\bm \mu})$ as prior in the upper level. 
Figures \ref{figEquPMC} show three different examples of the conditional pdf $K_t^{(N)}$ (obtained via numerical approximation) for a fixed $t$ and different $N\in\{2,20,1000\}$. We can observe that $K_t^{(N)}$ becomes closer to the target $\bar{\pi}$ (depicted in solid line) as $N$ grows.

\begin{figure*}[!hbt]
\centering 
\centerline{
%\subfigure[]{\includegraphics[width=3.97cm]{FigEqPMC1.pdf}}
\subfigure[]{\includegraphics[width=5cm]{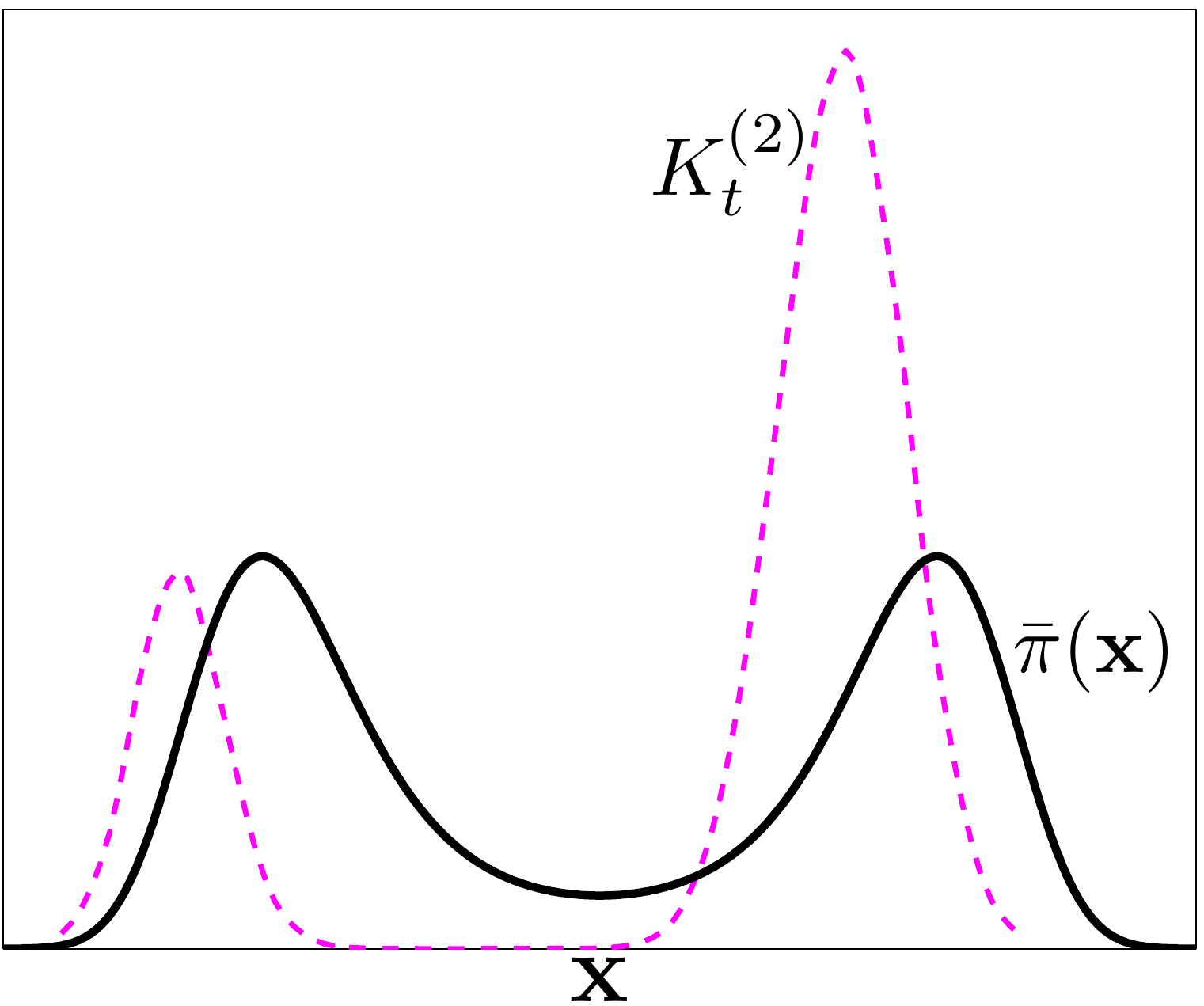}}
\subfigure[]{\includegraphics[width=5cm]{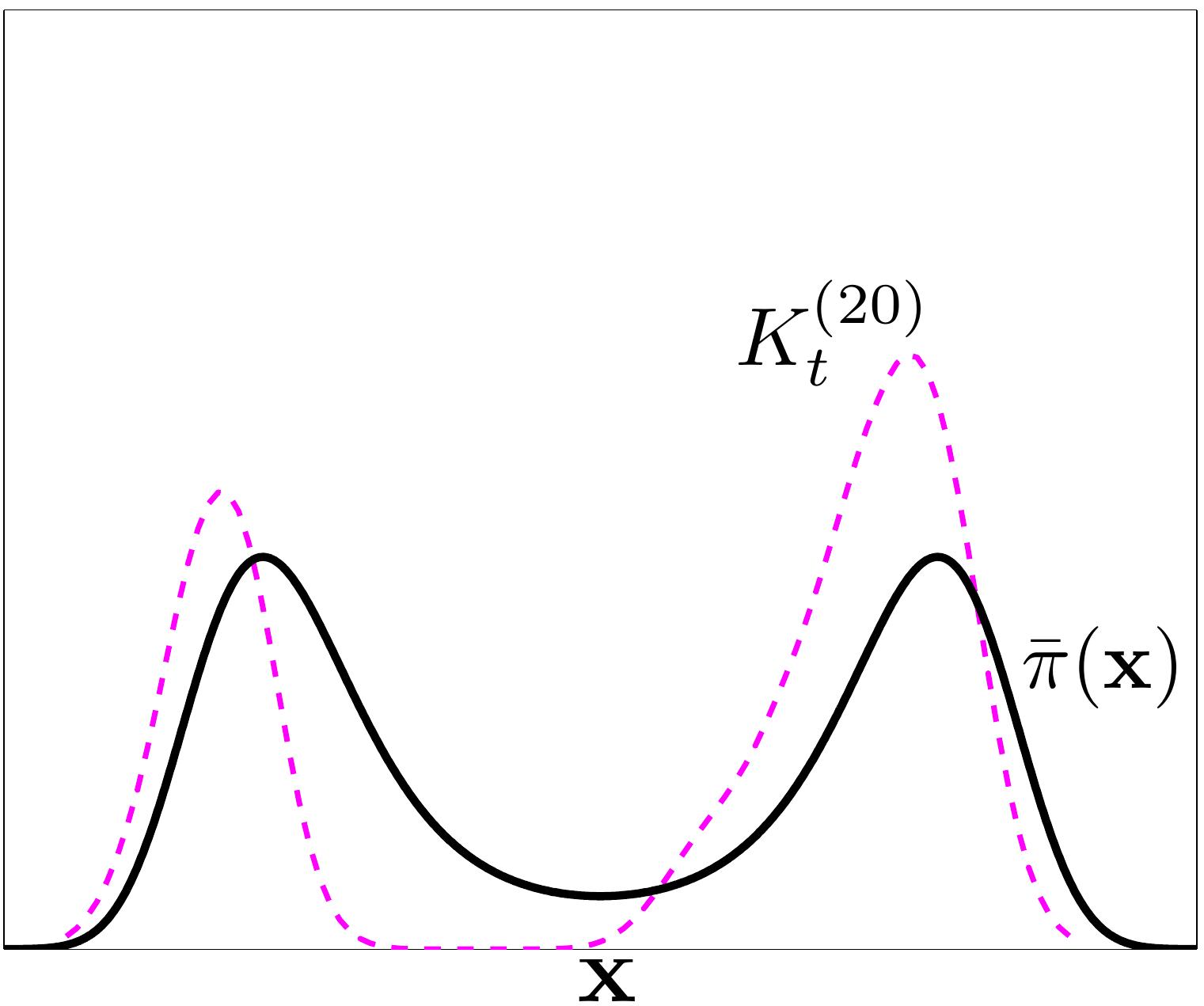}}
\subfigure[]{\includegraphics[width=5cm]{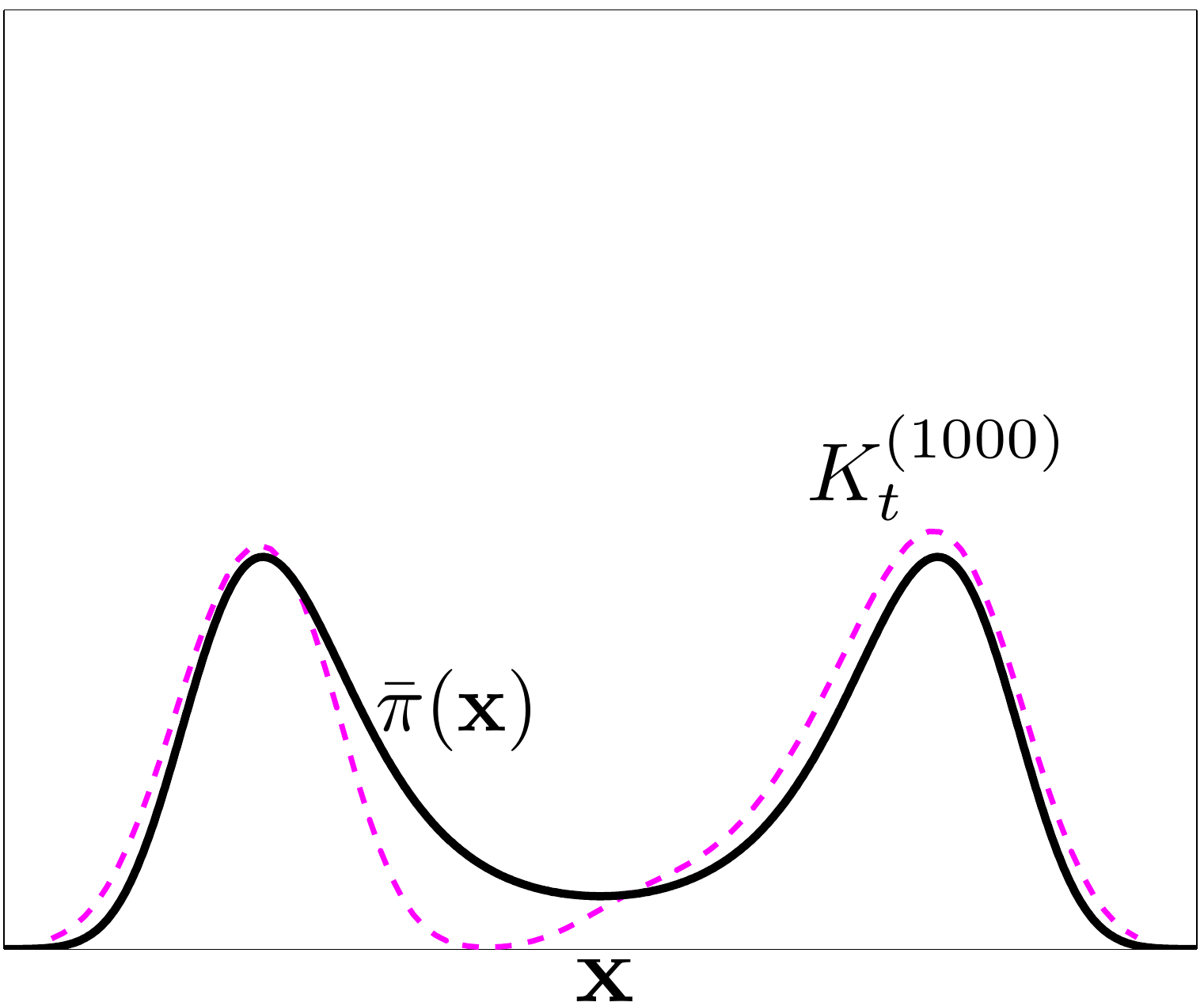}}
  }
  \caption{Examples of $K_t^{(N)}({\bm \mu}|{\bm \mu}_{1:N,t-1})$ (approximated numerically and shown with dashed line) and a bimodal target pdf ${\bar\pi}({\bf x})$ (solid line), fixing an iteration $t$ within a PMC method and for different $N$: {\bf (a)} $N=2$, {\bf (b)} $N=20$ and {\bf (c)} $N=1000$.  }
\label{figEquPMC}
\end{figure*}

% when the target $\bar{\pi}({\bf x})\propto \pi({\bf x})$ is univariate and bimodal (depicted in solid line). 

%%%%%%%%%%%%%%%%%%%%%%%%%%%%%%%%%%%%%%%%%%%
\subsubsection{Differences between PMC and MAIS algorithms}
%%%%%%%%%%%%%%%%%%%%%%%%%%%%%%%%%%%%%%%%%%%
In the Markov adaptive importance sampling (MAIS) schemes described in Section \ref{SuperSect}, since we are using MCMC methods for drawing from $h({\bm \mu})={\bar \pi}({\bm \mu})$, actually we have also a current prior 
$K_t^{(N)}({\bm \mu}_{1:N,t}|{\bm \mu}_{1:N,t-1})$,
determined for the kernels of the considered MCMC algorithms. For instance, in PI-MAIS we have 
$$
K_t^{(N)}({\bm \mu}_{1:N,t}|{\bm \mu}_{1:N,t-1})=\prod_{n=1}^N A_n({\bm \mu}_{n,t}|{\bm \mu}_{n,t-1}),
$$
where $A_n({\bm \mu}_{n,t}|{\bm \mu}_{n,t-1})$ is the kernel of the $n$-th chain. Unlike in PMC, since we are using ergodic chains with invariant pdf ${\bar \pi}$, we know that 
$K_t^{(N)}({\bm \mu}_{1:N,t}|{\bm \mu}_{1:N,t-1})\rightarrow \prod_{n=1}^N{\bar \pi}({\bm \mu}_{n})$ for $t\rightarrow \infty$, with a fixed $N$. Whereas PMC requires to increase $N$ for obtaining the same result. %This is one reason why MAIS schemes outperform PMC.
\begin{table*}[ht]
\setlength{\tabcolsep}{2pt}
\def\marginwidth{1.5mm}
\scriptsize
\begin{center}
\begin{tabular}{|l|c|c|c|c|c|c|c|c|c|}
%\begin{tabu}{|[1pt]l@{\hspace{\marginwidth}}|c@{\hspace{\marginwidth}}|c@{\hspace{\marginwidth}}|[1pt]c@{\hspace{\marginwidth}}|
%c@{\hspace{\marginwidth}}|c@{\hspace{\marginwidth}}|c@{\hspace{\marginwidth}}|c@{\hspace{\marginwidth}}|c@{\hspace{\marginwidth}}|c@{\hspace{\marginwidth}}|c@{\hspace{\marginwidth}}|c@{\hspace{\marginwidth}}|[1pt]}
%\multicolumn{10}{c}{{\bf Table 3.1}} \\
%\Xhline{2\arrayrulewidth}
\hline
\rowcolor[rgb]{.85,.85,.85}
\multicolumn{3}{|c|}{{\sc Algorithm}}    &  $ \sigma=0.5$ &  $\sigma=1$ & $\sigma=2$ & $\sigma=5$ & $\sigma=10$    & $\sigma=70$ & $\sigma_{n,j}\sim \mathcal{U}([1,10])$    \\
%\Xhline{2\arrayrulewidth}
\hline
\hline
%\Xhline{2\arrayrulewidth}  %%%% alternativa valida \tabucline[2pt]{-}
& \multirow{3}{*}{$\lambda=5$} & $M=99,T=20$  &1.2760 &0.5219 &0.5930 &0.0214 &0.0139 &0.1815 &0.0107\\
\cline{3-3} 
& & $M=19,T=100$  &0.2361 &0.1205 &0.0422 &0.0087 &0.0140 &0.1868 &0.0052\\
\cline{3-3} 
& & $M=1,T=1000$  &0.1719 &0.0019 &0.0155 &0.0103 &0.0273 &0.3737 &0.0070\\
\cline{2-10} 
& \multirow{3}{*}{$\lambda=10$} & $M=99,T=20$  &1.0195 &0.1546 &0.2876 &0.0178 &0.0133 &0.1789 &0.0098\\
\cline{3-3} 
& & $M=19,T=100$  &0.1750 &0.0120 &0.0528 &0.0086 &0.0136 &0.1856 &0.0050\\
\cline{3-3} 
{\sc PI-MAIS} ($N=100$)& & $M=1,T=1000$  &0.1550 &{\color{blue}\bf 0.0021} &{\color{blue}\bf 0.0020} &0.0095 &0.0252 &0.3648 &0.0066\\
\cline{2-10} 
 & \multirow{3}{*}{$\lambda=70$} & $M=99,T=20$  &16.9913 &5.5790 &1.4925 &0.0382 &0.0128 &0.1834 &0.0252\\
\cline{3-3} 
& &$M=19,T=100$  &2.6693 &0.9182 &0.1312 &0.0147 &0.0143 &0.1844 &0.0120\\
\cline{3-3} 
& & $M=1,T=1000$  &0.3014 &0.1042 &0.0136 &0.0115 &0.0267 &0.3697 &0.0093\\
\cline{2-10} 
& \multirow{3}{*}{$\lambda_{n,j} \sim \mathcal{U}([1,10])$ } & $M=99,T=20$  &1.0707 &0.5364 &0.3523 &0.0199 &{\color{blue} {\bf 0.0121}} &0.1919 &0.0094\\
\cline{3-3} 
& & $M=19,T=100$  &0.2481 &0.0595 &0.1376 &{\color{blue}\bf 0.0075} &0.0144 &0.1899 & {\color{blue}\bf 0.0049}\\
\cline{3-3} 
& & $M=1,T=1000$  &{\color{blue}\bf 0.1046} &0.0037 &0.0045 &0.0099 &0.0274 &0.3563 &0.0065\\
%\Xhline{2\arrayrulewidth}
\hline
\hline
%\Xhline{2\arrayrulewidth}
{\sc Static standard MIS} & \multicolumn{2}{c|}{$\Phi_{n,t}({\bf x})=q_{n,t}({\bf x})$}  & 29.56 & 41.95 & 64.51 & 2.17  & 0.0147 & 0.1914 & 4.55 \\
\cline{1-3}
{\sc Static partial DM-MIS}& \multicolumn{2}{c|}{$\Phi_{n,t}({\bf x})=\phi_t({\bf x})$} & 29.28 & 47.74 & 75.22 & 0.2424 &0.0124 &0.1789 & 0.0651\\
%\Xhline{2\arrayrulewidth}
\hline
\hline
%\Xhline{2\arrayrulewidth}
\multirow{2}{*}{{\sc AMIS \citep{CORNUET12}}}  &\multicolumn{2}{c|}{(best results)} &124.22&121.21 &100.23  &0.8640 & {\color{blue}\bf 0.0121} &  {\color{blue}\bf 0.0136} & 0.7328\\
\cline{2-3} 
& \multicolumn{2}{c|}{(worst results)} &125.43 &123.38 &114.82 &16.92  &0.0128 &18.66 & 13.49 \\
%\Xhline{2\arrayrulewidth}
\hline
\hline
%\Xhline{2\arrayrulewidth}
{\sc PMC \citep{Cappe04}} &     \multicolumn{2}{c|}{}  &112.99 &114.11 &47.97 &2.34 &0.0559 &2.41 &0.3017\\
\cline{1-1} %\cline{4-10}  
{\sc PMC with partial DM-MIS} &  \multicolumn{2}{c|}{$N=100$, $T=2000$}  & 111.92 &107.58 & 26.86 & 0.6731 & 0.0744 & 2.42 & 0.0700\\
\cline{1-1} 
{\sc Mixture PMC \citep{pmc-cappe08}} &   \multicolumn{2}{c|}{}  & 110.17 & 113.11  & 50.23 & 2.75 & 0.0521 & 2.57 & 0.6194  \\
\hline 
\hline
%\Xhline{2\arrayrulewidth} 
 {\sc Parallel Indep. MH chains} &  \multicolumn{2}{c|}{$N=100$,$T=2000$}   & 1.6910 & 1.7640 & 1.8832 &1.4133  & 0.2969  & 0.5475 & 7.3446\\
\hline
%\Xhline{2\arrayrulewidth} 
%\end{tabu}
\end{tabular}
\end{center}
\vspace{-0.3cm}
\caption{ {\bf (Ex-Sect \ref{sec:sims1})} MSE of the estimator of the $E[{\bf X}]$ (first component) with the initialization {\bf In1}. For all the techniques, the total number of evaluations of the target is $ E=2\cdot 10^5$. We recall that, in AMIS \citep{CORNUET12}, $N=1$ and $\Phi_{1,t}({\bf x})=\xi_1({\bf x})$. The last row corresponds to the application of $N=100$ (as in PI-MAIS) parallel MH chains where the random walk proposals have covariance matrices ${\bf C}=\sigma^2 {\bf I}_2$. The lengths of the chains, as well as of the PMC runs, is $T=2000$ for keeping $E=2\cdot 10^5$. For the techniques which adapt the covariance matrices of the proposal pdfs, the values of $\sigma$ have been employed as initial scale values for the covariance matrices. For AMIS, we show the best and worst results obtained testing different combinations of  $M$  %$M\in\{500, 10^3, 2\cdot 10^3,  5\cdot 10^3, 10^4\}$ 
and $T=\frac{E}{M}$. The best results, in each column, are highlighted with bold-faces.
}
\label{mean_gaussian_mixture_bad}
%\end{sidewaystable} 
\end{table*}

%%%%%%%%%%%%%%%%%%%%%%%%%%%%%%%%%
%%%%%%%%%%%%%%%%%%%%%%%%%%%%%%%%%
%%%%%%%%%%%%%%%%%%%%%%%%%%%%%%%%%
%%%%%%%%%%%%%%%%%%%%%%%%%%%%%%%%%

%%%%%%%%%%%%%%%%%%%%%%%%%
%%%%%%%%%%%%%%%%%%%%%%%%%
\begin{figure*}[!tb]
\centering 
\centerline{
 \subfigure[{\bf Ex-Sect \ref{sec:sims1}}]{\includegraphics[width=7.7cm]{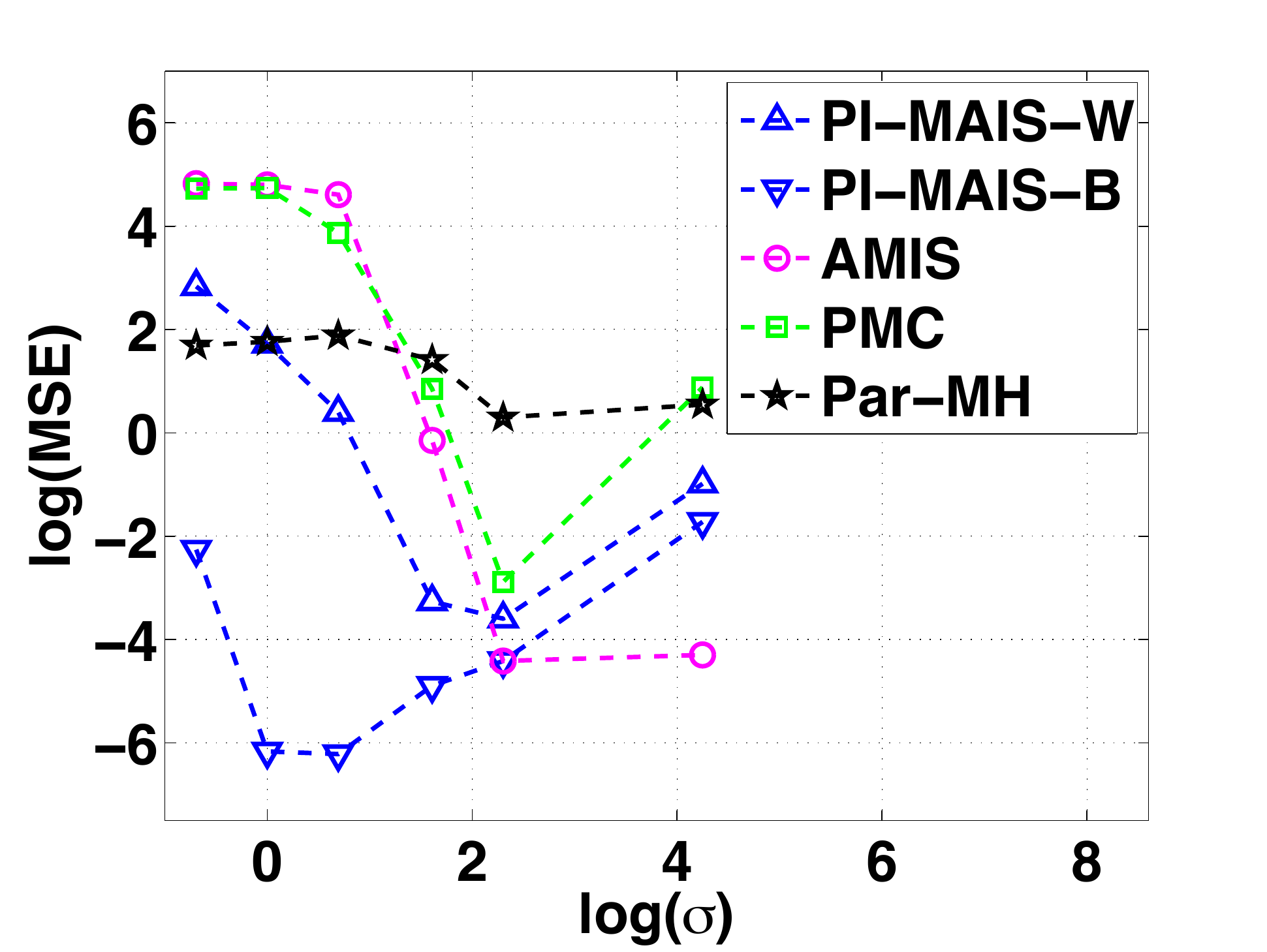}}
  \subfigure[{\bf Ex-Sect \ref{LocSect}}]{\includegraphics[width=7.7cm]{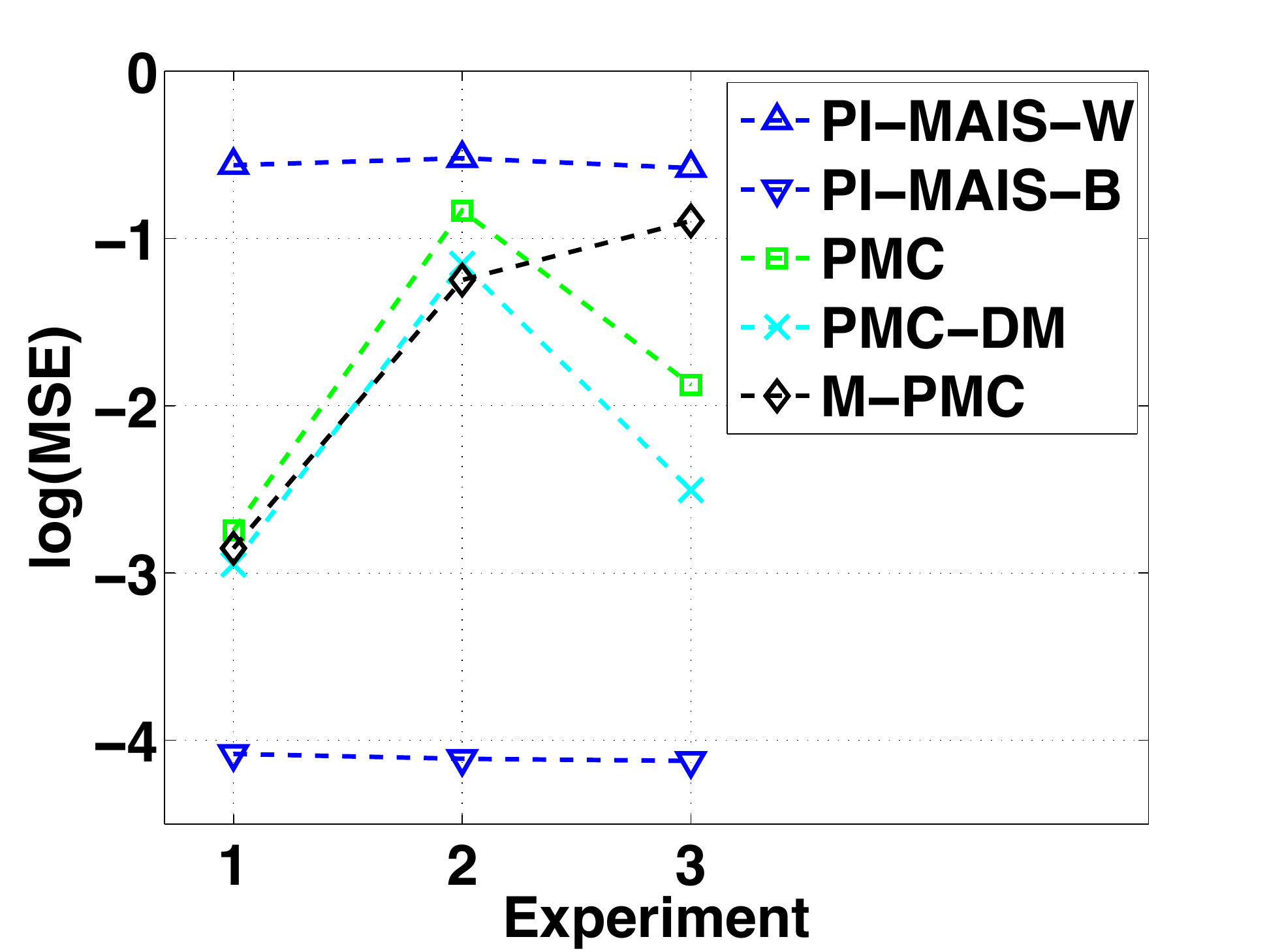}}
  }
  \caption{ {\bf (Ex-Sect \ref{sec:sims1}-\ref{LocSect})} Summary of the results in Table \ref{mean_gaussian_mixture_bad} in Fig. {\bf (a)}, and Table \ref{mean_sensors} in Fig. {(\bf b)}: the curve $\log(\mbox{MSE})$ of the different methods as function of $\log(\sigma)$ in Fig. {\bf (a)} ($\sigma\in\{0.5,1, 2, 5, 10, 70\}$), and as function of  the different experiments in Fig. {\bf (b)}. The worst and best results of PI-MAIS are depicted with triangles up and down, respectively. }
\label{figSIMUafterR1_2}
\end{figure*}
%%%%%%%%%%%%%%%%%%%%%%%%%
%%%%%%%%%%%%%%%%%%%%%%%%%

%%%%%%%%%%%%%%%%%%%%%%%%%
%%%%%%%%%%%%%%%%%%%%%%%%%
%%%%%%%%%%%%%%%%%%%%%%%%%

%%%%%%%%%%%%%%%%%%%%%%%%%%%%%%%%%
%%%%%%%%%%%%%%%%%%%%%%%%%%%%%%%%%
%%%%%%%%%%%%%%%%%%%%%%%%%%%%%%%%%
%%%%%%%%%%%%%%%%%%%%%%%%%%%%%%%%%

%\section{Tables}
%\newpage

\begin{table*}[ht]
%\begin{sidewaystable}[p]
%\begin{table}[p]
\setlength{\tabcolsep}{2pt}
\def\marginwidth{1.5mm}
%\footnotesize
\scriptsize
\begin{center}
\begin{tabular}{|l|c|c|c|c|c|c|c|c|c|}
%\begin{tabular}{|[2pt]l@{\hspace{\marginwidth}}|c@{\hspace{\marginwidth}}|c@{\hspace{\marginwidth}}|c@{\hspace{\marginwidth}}|
%\begin{tabu}{|[1pt]l@{\hspace{\marginwidth}}|c@{\hspace{\marginwidth}}|c@{\hspace{\marginwidth}}|[1pt]c@{\hspace{\marginwidth}}|
%c@{\hspace{\marginwidth}}|c@{\hspace{\marginwidth}}|c@{\hspace{\marginwidth}}|c@{\hspace{\marginwidth}}|c@{\hspace{\marginwidth}}|c@{\hspace{\marginwidth}}|c@{\hspace{\marginwidth}}|c@{\hspace{\marginwidth}}|[1pt]}
%\multicolumn{10}{c}{{\bf Table 3.1}} \\
%\Xhline{2\arrayrulewidth}
\hline
\rowcolor[rgb]{.85,.85,.85}
\multicolumn{3}{|l|}{{\sc Algorithm}}    &  $ \sigma=0.5$ &  $\sigma=1$ & $\sigma=2$ & $\sigma=5$ & $\sigma=10$    & $\sigma=70$ & $\sigma_{n,j}\sim \mathcal{U}([1,10])$    \\
\Xhline{2\arrayrulewidth}
\hline
\hline
%\Xhline{2\arrayrulewidth}  %%%% alternativa valida \tabucline[2pt]{-}
& \multirow{3}{*}{$\lambda=5$} & $M=99,T=20$  &0.6096 &0.0657 &0.0023 &0.0056 &{\color{blue}\bf 0.0124} &0.1768 &0.0051\\
\cline{3-3} 
& & $M=19,T=100$  &0.2878 &0.0358 &{\color{blue}\bf 0.0010} &0.0050 &0.0127 &0.1802 &0.0038\\
\cline{3-3} 
& & $M=1,T=1000$  &0.1244 &{\color{blue}\bf 0.0011} &0.0014 &0.0091 &0.0242 &0.3510 &0.0064\\
\cline{2-10} 
& \multirow{3}{*}{$\lambda=10$} & $M=99,T=20$  &0.9236 &0.0543 &0.0021 &0.0062 &0.0137 &0.1815 &0.0054\\
\cline{3-3} 
& & $M=19,T=100$  &0.2294 &0.0077 &0.0012 &0.0054 &0.0132 &0.1890 &0.0044\\
\cline{3-3} 
{\sc PI-MAIS} ($N=100$) & & $M=1,T=1000$  &0.0786 &0.0042 &0.0014 &0.0086 &0.0256 &0.3503 &0.0066\\
\cline{2-10} 
& \multirow{3}{*}{$\lambda=70$} & $M=99,T=20$  &5.9889 &0.3662 &0.0082 &0.0089 &0.0140 &0.1841 &0.0093\\
\cline{3-3} 
& &  $M=19,T=100$  &1.6670 &0.0871 &0.0045 &0.0080 &0.0139 &0.1971 &0.0074\\
\cline{3-3} 
& & $M=1,T=1000$  &0.2579 &0.0134 &0.0024 &0.0097 &0.0258 &0.3543 &0.0082\\
\cline{2-10} 
& \multirow{3}{*}{$\lambda_{n,j} \sim \mathcal{U}([1,10])$} & $M=99,T=20$  &0.5623 &0.0417 &0.0025 &0.0059 &{\color{blue}\bf 0.0124} &0.1848 &0.0056\\
\cline{3-3} 
& & $M=19,T=100$  &0.2704 &0.0204 &0.0011 &{\color{blue}\bf 0.0048} &0.0136 &0.1726 &{\color{blue}\bf 0.0037}\\
\cline{3-3} 
& & $M=1,T=1000$  &{\color{blue}\bf 0.0750} &0.0014 &0.0013 &0.0089 &0.0247 &0.3540 &0.0066\\
%\Xhline{2\arrayrulewidth}
\hline
\hline
%\Xhline{2\arrayrulewidth}
{\sc Static standard MIS} & \multicolumn{2}{c|}{$\Phi_{n,t}({\bf x})=q_{n,t}({\bf x})$} &  12.00 & 9.40 & 10.26 &  7.67  & 0.5443&  0.1764 & 4.37\\
\cline{1-3}
{\sc Static partial DM-MIS}& \multicolumn{2}{c|}{$\Phi_{n,t}({\bf x})=\phi_t({\bf x})$} &  10.14 & 0.9469 & 0.0139  &0.0100 & 0.0146
& 0.1756  & 0.0106 \\
%\Xhline{2\arrayrulewidth}
\hline
\hline
%\Xhline{2\arrayrulewidth}
%\multicolumn{2}{|[1pt]l|[1pt]}{{\bf AMIS} (best)}
\multirow{2}{*}{{\sc AMIS \citep{CORNUET12}}} & \multicolumn{2}{c|}{(best results)} &113.97 &112.70 &107.85  &44.93 &0.7404 & {\color{blue}\bf 0.0141} & 31.02 \\
\cline{2-3} 
&   \multicolumn{2}{c|}{(worst results)} &116.66 &115.62 &111.83 &70.62 &9.43  &18.62 &  58.63  \\
%\Xhline{2\arrayrulewidth}
\hline
\hline
%\Xhline{2\arrayrulewidth}
{\sc PMC \citep{Cappe04}} &  \multicolumn{2}{c|}{} & 111.54 & 110.78 & 90.21 & 2.29  & 0.0631 & 2.42 & 0.3082 \\
\cline{1-1} %\cline{4-10}  
{\sc PMC with partial DM-MIS} &  \multicolumn{2}{c|}{$N=100$, $T=2000$}  & 23.16 &7.43 & 7.56 & 0.6420 & 0.0720 & 2.37 & 0.0695\\
\cline{1-1}   
{\sc Mixture PMC \citep{pmc-cappe08}} & \multicolumn{2}{c|}{}   & 25.43 & 10.68 & 6.29 & 0.6142 & 0.0727  &2.55  &  0.1681 \\
%\Xhline{2\arrayrulewidth}
\hline 
\hline 
%\Xhline{2\arrayrulewidth}
{\sc Parallel Indep. MH chains} &  \multicolumn{2}{c|}{$N=100$,$T=2000$} &  1.3813 & 1.3657 & 1.2942 & 1.0178 & 0.3644 &  1.0405 & 5.3211  \\
\hline
%\Xhline{2\arrayrulewidth} 
%\end{tabu}
\end{tabular}
\end{center}
%\vspace{-0.4cm}
\caption{ {\bf (Ex-Sect \ref{sec:sims1})} MSE of the estimator of the expected value (first component). For all the techniques, the total number of evaluations of the target is again $ E=2\cdot 10^5$. In this case, we have applied the initialization {\bf In2}, differently from Table \ref{mean_gaussian_mixture_bad}. The best results, in each column, are highlighted with bold-faces. }
\label{mean_gaussian_mixture_good}
%\end{sidewaystable} 
\end{table*}

%%%%%%%%%%%%%%%%%%%%%%%%%
%%%%%%%%%%%%%%%%%%%%%%%%%

%%%%%%%%%%%%%%%%%%%%%%%%%
%%%%%%%%%%%%%%%%%%%%%%%%%

%%%%%%%%%%%%%%%%%%%%%%%%%
\begin{table*}[ht]
\setlength{\tabcolsep}{2pt}
\def\marginwidth{1.5mm}
\scriptsize
\begin{center}
\begin{tabular}{|l|c|c|c|c|c|c|c|c|c|}
%\begin{tabu}{|[1pt]l@{\hspace{\marginwidth}}|c@{\hspace{\marginwidth}}|c@{\hspace{\marginwidth}}|[1pt]c@{\hspace{\marginwidth}}|
%c@{\hspace{\marginwidth}}|c@{\hspace{\marginwidth}}|c@{\hspace{\marginwidth}}|c@{\hspace{\marginwidth}}|c@{\hspace{\marginwidth}}|c@{\hspace{\marginwidth}}|c@{\hspace{\marginwidth}}|c@{\hspace{\marginwidth}}|[1pt]}
%\multicolumn{10}{c}{{\bf Table 3.1}} \\
%\Xhline{2\arrayrulewidth}
\hline
\rowcolor[rgb]{.85,.85,.85}
\multicolumn{3}{|c|}{{\sc Algorithm}}    &  $ \sigma=0.5$ &  $\sigma=1$ & $\sigma=2$ & $\sigma=5$ & $\sigma=10$    & $\sigma=70$ & $\sigma_{n,j}\sim \mathcal{U}([1,10])$    \\
%\Xhline{2\arrayrulewidth}
\hline
\hline
%\Xhline{2\arrayrulewidth}  %%%% alternativa valida \tabucline[2pt]{-}
& \multirow{3}{*}{$\lambda=5$} & $M=99,T=20$  &0.0388 &0.0120 &0.0070 &0.0002 &0.0001 &0.0016 &0.0001\\
\cline{3-3} 
& & $M=19,T=100$  &0.0031 &0.0013 &0.0004 &0.0001 &0.0001 &0.0017 &0.0001\\
\cline{3-3} 
& & $M=1,T=1000$  &0.0016 &0.0001 &0.0001 &0.0001 &0.0002 &0.0031 &0.0001\\
\cline{2-10} 
& \multirow{3}{*}{$\lambda=10$} & $M=99,T=20$  &0.0217 &0.0046 &0.0040 &0.0001 &0.0001 &0.0016 &0.0002\\
\cline{3-3} 
& & $M=19,T=100$  &0.0019 &0.0002 &0.0005 &0.0001 &0.0001 &0.0017 &0.0001\\
\cline{3-3} 
{\sc PI-MAIS} ($N=100$)& & $M=1,T=1000$  &0.0016 &0.0001 &0.0001  & {\color{blue} {\bf 8} ${\bf \cdot 10^{-5}}$} &0.0002 &0.0031 &0.0001\\
\cline{2-10} 
 & \multirow{3}{*}{$\lambda=70$} & $M=99,T=20$  &6.3732 &0.2713 &0.0226 &0.0003 &0.0001 &0.0016 &0.0002\\
\cline{3-3} 
& &$M=19,T=100$  &0.1082 &0.0114 &0.0019 &0.0001 &0.0001 &0.0017 &0.0001\\
\cline{3-3} 
& & $M=1,T=1000$  &0.0038 &0.0009 &0.0001 &0.0001 &0.0002 &0.0033 &0.0001\\
\cline{2-10} 
& \multirow{3}{*}{$\lambda_{n,j} \sim \mathcal{U}([1,10])$ } & $M=99,T=20$  &0.0350 &0.0101 &0.0043 &0.0001 &0.0001 & 0.0015 &0.0001\\
\cline{3-3} 
& & $M=19,T=100$  &0.0029 &0.0007 &0.0010 &{\color{blue} {\bf 8} ${\bf \cdot 10^{-5}}$}  &  9 $\cdot 10^{-5}$  &0.0017 &{\color{blue} {\bf 9} ${\bf \cdot 10^{-5}}$} \\
\cline{3-3} 
& & $M=1,T=1000$ &0.0014 &0.0001 & {\color{blue} ${\bf 9 \cdot 10^{-5}}$} &0.0001 &0.0002 &0.0036 &0.0001\\
%\Xhline{2\arrayrulewidth}
\hline
\hline
%\Xhline{2\arrayrulewidth}
{\sc Static standard MIS} & \multicolumn{2}{c|}{$\Phi_{n,t}({\bf x})=q_{n,t}({\bf x})$}  & 3.94 $\cdot 10^{4}$ & 7.12 $\cdot 10^{7}$ & 1.07 $\cdot 10^{3}$ & 0.0113  & 0.0001 &  0.0016 & 0.2190 \\
\cline{1-3}
{\sc Static partial DM-MIS}& \multicolumn{2}{c|}{$\Phi_{n,t}({\bf x})=\phi_t({\bf x})$} & 9.51$\cdot 10^8$ &  4.60 $\cdot 10^5$ & 15.34 & 0.0016& 0.0001 & 0.0016 & 0.0005 \\
%\Xhline{2\arrayrulewidth}
\hline
\hline
%\Xhline{2\arrayrulewidth}
\multirow{2}{*}{{\sc AMIS \citep{CORNUET12}}}  &\multicolumn{2}{c|}{(best results)} & 15.92&15.66 & 12.81 & 0.0069 &  {\color{blue} {\bf 8} ${\bf \cdot 10^{-5}}$ } & {\color{blue}  \bf 0.0001} & 0.0002  \\
\cline{2-3} 
& \multicolumn{2}{c|}{(worst results)}    &15.97&  15.92& 14.87& 0.4559&0.0001 & 1.62 & 0.0084 \\
%\Xhline{2\arrayrulewidth}
\hline
\hline
%\Xhline{2\arrayrulewidth}
{\sc PMC \citep{Cappe04}} &     \multicolumn{2}{c|}{}  &  33.53 & 17.10 & 14.42 & 0.4249  &0.0015 & 0.0016 & 0.3542 \\
\cline{1-1} %\cline{4-10}  
{\sc PMC with partial DM-MIS} &  \multicolumn{2}{c|}{$N=100$, $T=2000$}   & 15.85& 14.31& 1.81 & 0.0402&  0.0002 & 0.0016&  0.0004 \\
\cline{1-1} 
{\sc Mixture PMC \citep{pmc-cappe08}} &   \multicolumn{2}{c|}{}    & 14.51  & 12.09 &  3.56 & 0.0287  &   0.0002 & 0.0015 & 0.0010 \\
\hline 
%\Xhline{2\arrayrulewidth} 
%\end{tabu}
\end{tabular}
\end{center}
\vspace{-0.3cm}
\caption{ {\bf (Ex-Sect \ref{sec:sims1})} MSE of the estimator of the normalizing constant $Z$ with the initialization {\bf In1}. For all the techniques, the total number of evaluations of the target is $ E=2\cdot 10^5$. The smallest MSE for each $\sigma$ is  bold-faced. 
}

\label{Z_gaussian_mixture_bad}
%\end{sidewaystable} 
\end{table*}

%%%%%%%%%%%%%%%%%%%%%%%%%
%%%%%%%%%%%%%%%%%%%%%%%%%

\begin{table*}
%\footnotesize
\begin{center}
\begin{tabular}{|l|l||c|c|c|c|c|c|c|c|}\hline
\rowcolor[rgb]{.85,.85,.85}\multicolumn{2}{|l||}{{\sc  Algorithm}}& $\sigma=0.5$ & $\sigma=1$ & $\sigma=2$ & $\sigma=3$ & $\sigma=5$ & $\sigma=10$  & $\sigma=70$ & $\sigma_{i,j}\sim \mathcal{U}([1,20])$  \\\hline\hline
\multirow{2}{*}{{\sc PI-MAIS} } & Worst   & 0.0083 & 0.0081 & 0.0012 & 0.0005 & 0.0050 & 0.0126 &  0.1126 & 0.0218    \\ 
\cline{2-10}
 & Best    & 0.0025 &{\color{blue}\bf 0.0001} &  {\color{blue}\bf0.0002} &  {\color{blue}\bf 0.0001} & 0.0002 &  0.0003  & 0.0361 &   0.0004  \\ 
 \hline\hline
 \multirow{2}{*}{{\sc I$^2$-MAIS} } & Worst   & 0.0335 & 0.0227 & 0.0053 & 0.0044 & 0.0041 & 0.0096 &  0.2130 & 0.0181    \\ 
\cline{2-10}
 & Best    & 0.0082 &  0.0025 & 0.0013 & 0.0008 &{\color{blue}\bf 0.0001} & {\color{blue}{\bf 0.0002}}  & 0.0265 &  {\color{blue}\bf 0.0003}  \\ 
  \hline\hline
\multirow{2}{*}{{\sc PMC} \citep{Cappe04}} & Worst & 0.0670 &0.0461 & 0.0209 & 0.0093 & 0.0055 &0.0072  &  9.4749 &  0.1065\\
\cline{2-10}
  & Best  & 0.0210 & 0.0164 &0.0069 & 0.0016 & 0.0015 & 0.0011 & 0.0262 & 0.0026\\\hline\hline 
\multirow{2}{*}{{\sc Mixture PMC} \citep{pmc-cappe08}}&  Worst  & 3.5772 & 0.0113  & 0.0044 & 0.0066 & 0.0174 & 0.0267 & 0.0913 &  0.0103   \\
\cline{2-10}
&   Best   & 0.0092 & 0.0020  & 0.0018 & 0.0035 & 0.0034 & 0.0055 & 0.0138  & 0.0025 \\  \hline\hline
\multirow{2}{*}{{\sc AMIS} \citep{CORNUET12}}   & Worst   & 0.0040 & 0.0039 & 0.0040 & 0.0016 & 0.0011 &  0.0012 &  0.0035 & 0.0013  \\
  \cline{2-10}
& Best  & {\color{blue}\bf 0.0023} & 0.0028 & 0.0023 & 0.0009 & 0.0003 & 0.0004 & {\color{blue}\bf 0.0023} & 0.0007 \\	
\hline
 \end{tabular}
\end{center}
%\vspace{-0.5cm}
\caption{{\bf (Ex-Section-\ref{BananaExSect})} 
Bi-dimensional banana-shaped distribution example: Best and worst results in terms of MSE, obtained with the different techniques for different values of $\sigma$.  The smallest MSE for each $\sigma$ is  bold-faced. }
\label{TableResults2}
\end{table*}

%%%%%%%%%%%%%%%%%%%%%%%%%
%%%%%%%%%%%%%%%%%%%%%%%%%
%%%%%%%%%%%%%%%%%%%%%%%%%
%%%%%%%%%%%%%%%%%%%%%%%%%

\begin{table*}[ht]
%\begin{sidewaystable}[p]
%\begin{table}[p]
%\setlength{\tabcolsep}{2pt}
%\def\marginwidth{1.5mm}
\footnotesize
\begin{center}
\begin{tabular}{|l|c|c|c|c|c|}
%\begin{tabu}{|[1pt]l@{\hspace{\marginwidth}}|c@{\hspace{\marginwidth}}|c@{\hspace{\marginwidth}}|[1pt]c@{\hspace{\marginwidth}}|
%c@{\hspace{\marginwidth}}|c@{\hspace{\marginwidth}}|c@{\hspace{\marginwidth}}|c@{\hspace{\marginwidth}}|c@{\hspace{\marginwidth}}|c@{\hspace{\marginwidth}}|c@{\hspace{\marginwidth}}|c@{\hspace{\marginwidth}}|[1pt]}
%\Xhline{2\arrayrulewidth}
\hline
\rowcolor[rgb]{.85,.85,.85}
 \multicolumn{3}{|l|}{{\sc Algorithm}}     &  $\sigma_{i,j}\sim \mathcal{U}([1,5])$ &  $\sigma_{i,j}\sim \mathcal{U}([1,10])$ & $\sigma_{i,j}\sim \mathcal{U}([1,30])$  \\
%\Xhline{2\arrayrulewidth}
\hline
\hline
%\Xhline{2\arrayrulewidth}  %%%% alternativa valida \tabucline[2pt]{-}
\multirow{9}{*}{{\sc PI-MAIS}}& \multirow{3}{*}{$\lambda=5$} & $M=99,T=20$  &0.3819 &0.3508 &0.3626\\
\cline{3-3} 
& & $M=19,T=100$  &0.0728 &0.0738 &0.0710\\
\cline{3-3} 
& & $M=1,T=1000$ &0.0173 &{\color{blue}\bf  0.0164} &0.0171\\
\cline{2-6} 
& \multirow{3}{*}{$\lambda=10$} & $M=99,T=20$  &0.5701 &0.5943 &0.5605\\
\cline{3-3} 
& & $M=19,T=100$  &0.1389 &0.1429 &0.1425\\
\cline{3-3} 
& & $M=1,T=1000$  &0.0401 &0.0408 &0.0393\\
%& \multirow{3}{*}{$\lambda=70$} & $M=99,T=20$  &4.3653 &4.0099 &4.1902\\
%\cline{3-6} 
%& &  $M=19,T=100$  &16.2317 &16.7786 &16.4043\\
%\cline{3-6} 
%& & $M=1,T=1000$  &35.5136 &215.1906 &64.2133\\
\cline{2-6} 
& \multirow{3}{*}{$\lambda_{i,j} \sim \mathcal{U}([1,30])$} & $M=99,T=20$  &0.3758 &0.3795 &0.4028\\
\cline{3-3} 
& & $M=19,T=100$  &0.0741 &0.0793 &0.0771\\
\cline{3-3} 
& & $M=1,T=1000$  & {\color{blue}\bf  0.0169} & 0.0167 & {\color{blue}\bf  0.0162}\\
%\Xhline{2\arrayrulewidth}
\hline
\hline
%\Xhline{2\arrayrulewidth}
{\sc PMC} \citep{Cappe04} &   \multicolumn{2}{c|}{}  &0.0642 &0.4345 & 0.1533 \\
\cline{1-1}
{\sc PMC with partial DM-MIS}  &  \multicolumn{2}{c|}{$N=100$, $T=2000$}  & 0.0524 & 0.3163 &  0.0817 \\
\cline{1-1} 
{\sc Mixture PMC} \citep{pmc-cappe08} &   \multicolumn{2}{c|}{} & 0.0577 & 0.2870 &  0.4083 \\
%\Xhline{2\arrayrulewidth} 
\hline
%\end{tabu}
\end{tabular}
\end{center}
%\vspace{-0.5cm}
\caption{{\bf (Ex-Sect \ref{LocSect})} MSE of the estimator of $E[(X_1,X_2,A,\Omega)]$ using different techniques, keeping constant the total number of target evaluation, $E=2 \ 10^5$. The best results, in each column,  are highlighted with bold-faces.
 }
\label{mean_sensors}
%\end{sidewaystable} 
\end{table*}

\begin{figure*}[!tb]
\centering 
\centerline{
   \subfigure[Worst results.]{\includegraphics[width=8cm]{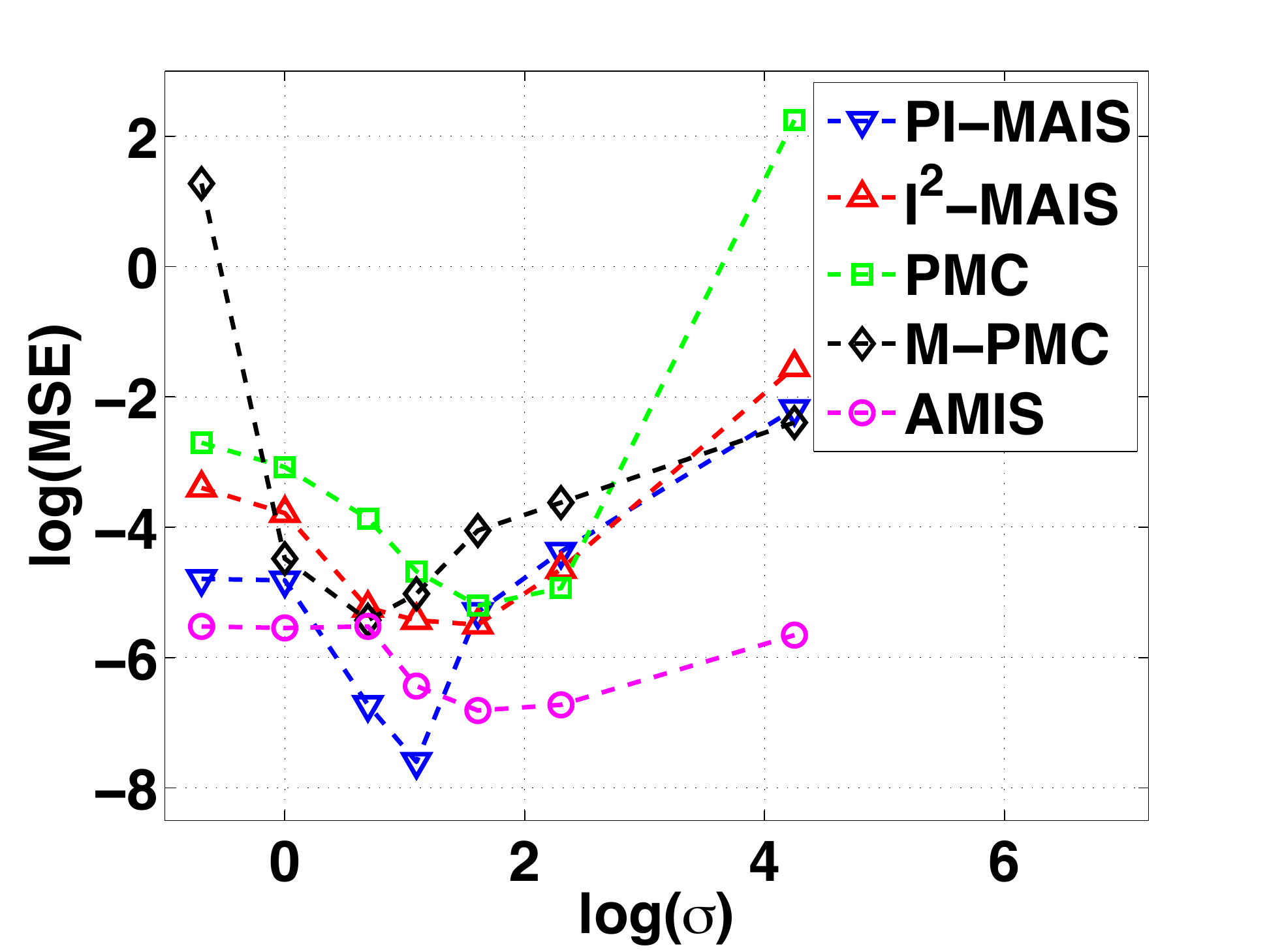}}
  % \subfigure[Possible adapt. in PI-MAIS]{\includegraphics[width=5cm]{AlterMCMC.png}}  
    \hspace{0.3cm}
     \subfigure[Best results.]{\includegraphics[width=8cm]{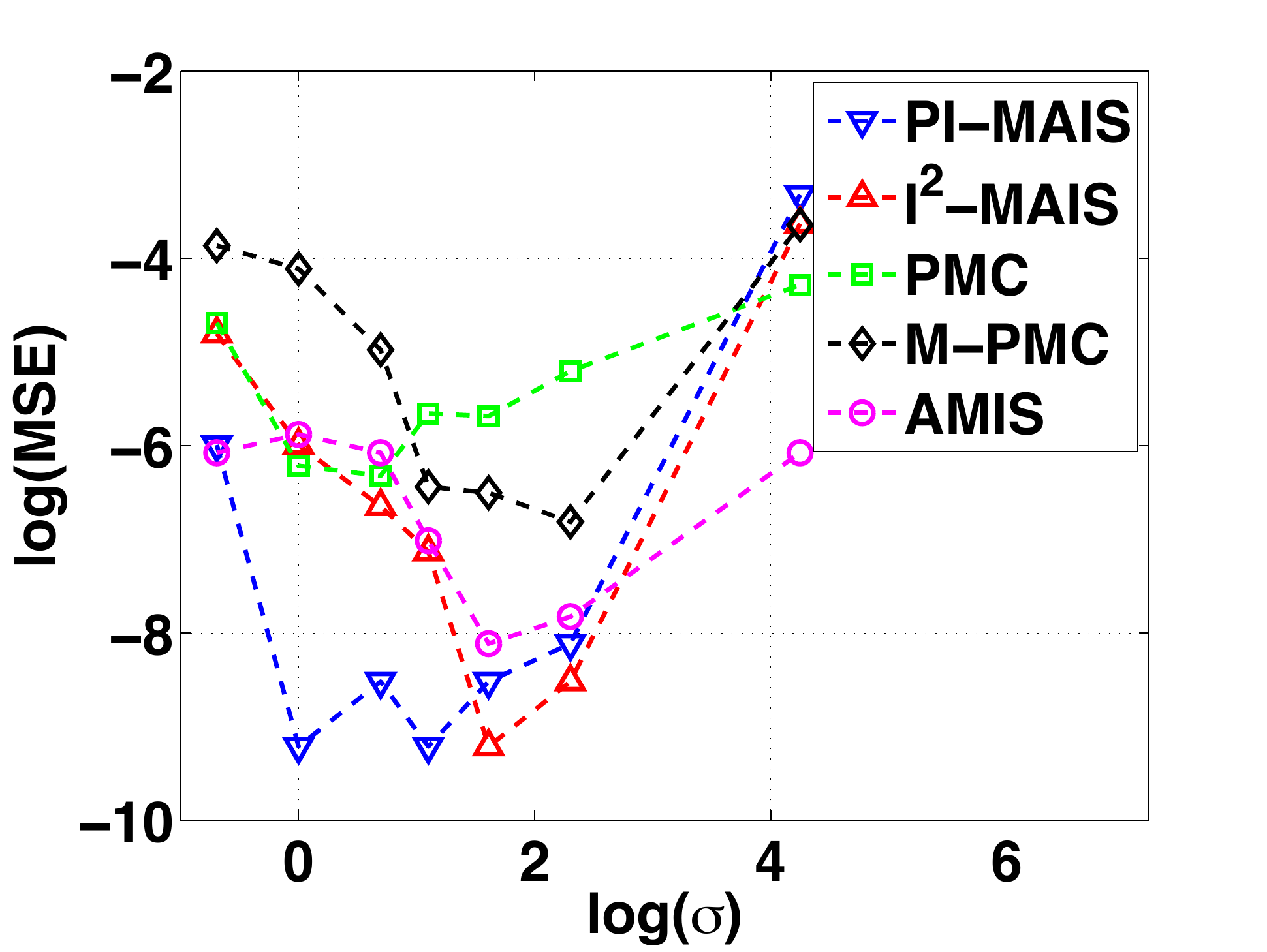}}  
  }
  \caption{{\bf (Ex-Section-\ref{BananaExSect})} Graphical representation of the results in Table \ref{TableResults2} (except for the last column): the curve $\log(\mbox{MSE})$ versus $\log(\sigma)$ with $\sigma\in\{0.5,1,2,3,5,10,70\}$ for the different methods, {\bf (a)} worst and {\bf (b)} best results.  }
\label{figSIMUafterR1}
\end{figure*}

\begin{figure*}[!tb]
\centering 
\centerline{
   \subfigure[$N=100$ and $E=2\cdot 10^5$.]{\includegraphics[width=7.81cm]{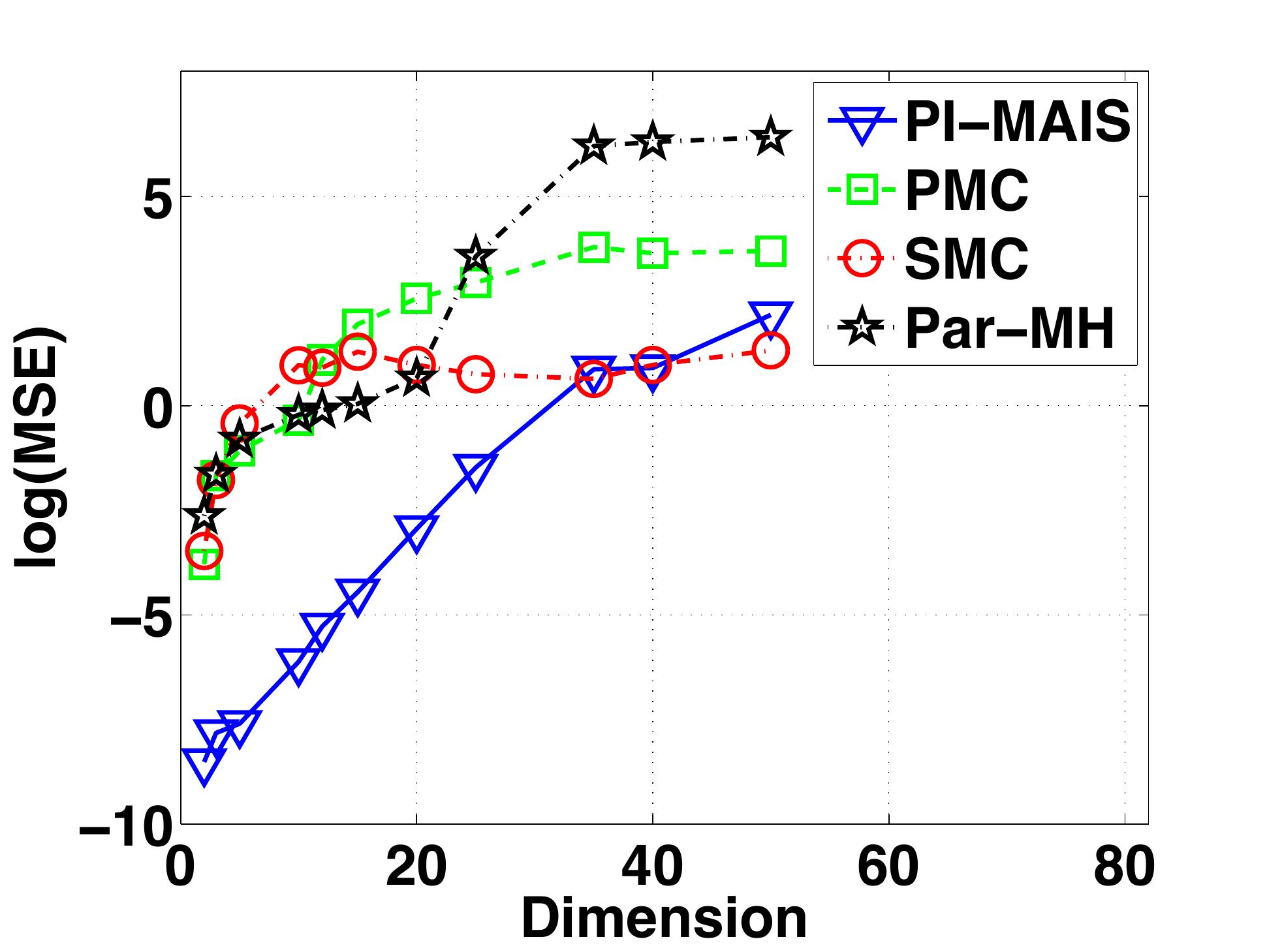}}
  % \subfigure[Possible adapt. in PI-MAIS]{\includegraphics[width=5cm]{AlterMCMC.png}}  
    \hspace{0.3cm}
     \subfigure[$N=500$, keeping $E=2\cdot 10^5$.]{\includegraphics[width=7.81cm]{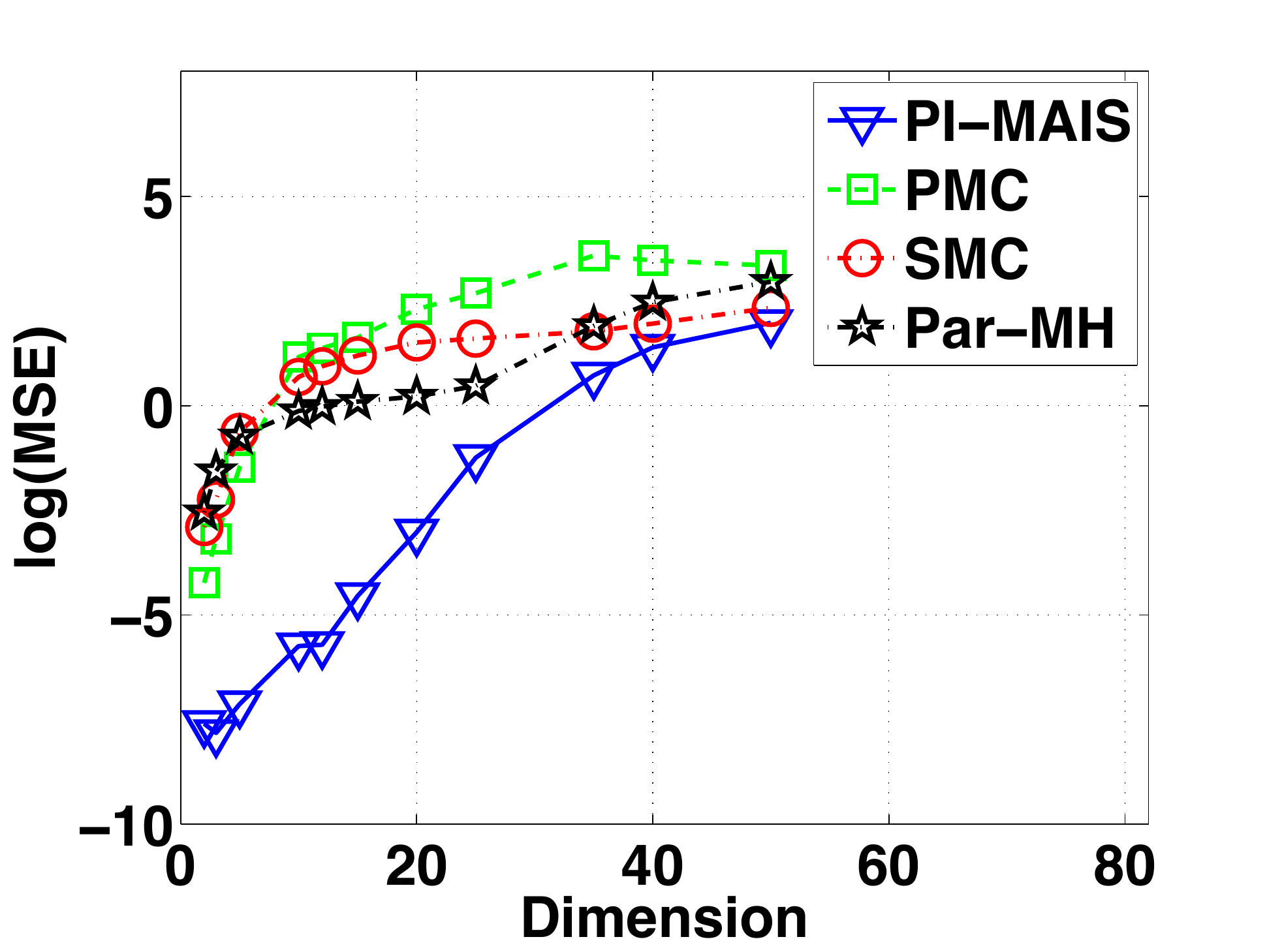}}  
  }
  \caption{ {\bf (Ex-Section-\ref{sec:SIMU_HD})} The curve $\log(\mbox{MSE})$ as function of dimension of the problem, $D_x\in\{2,3,5,10,12,15,20,25,35,40,50\}$, for different methods. We test {\bf (a)} $N=100$ and  {\bf (b)}  $N=500$, keeping fixed the same number of evaluation of the target $E=2\cdot 10^5$. Hence the total number of iterations (of the different algorithms) is greater in Fig. {\bf 9(a)} than in  Fig. {\bf 9(b)}.  }
\label{figSIMU_HD}
\end{figure*}

\end{document}